\documentclass[fleqn,usenatbib,useAMS,usenatbib]{mnras}
\usepackage{graphicx}
\usepackage{natbib}
\usepackage{amsmath}
\usepackage{mathtools}
\usepackage{amssymb}
\usepackage{wasysym}
\usepackage{verbatim}
\usepackage{txfonts}
\usepackage[T1]{fontenc}
\usepackage{ae, aecompl}
\usepackage{hyperref}

\usepackage{latexsym,longtable,epsf,ulem}
\usepackage{cases} 
\usepackage{enumerate} 

\newcommand{\rmgal}{{\rm RM_{gal}}}
\newcommand{\rmqso}{{\rm RM_{qso}}}

\newcommand{\rmc}{{\rm RM_{c}}}
\newcommand{\rmt}{{\rm RM_{t}}}

\newcommand{\zgal}{{z_{\rm gal}}}
\newcommand{\zqso}{{z_{\rm qso}}}

\title[{\rm RM} statistics of intervening galaxies]
{Statistical properties of Faraday rotation measure in external galaxies -- I: intervening disc galaxies}
\author[A. Basu et al.]{Aritra Basu,$^{1,2}$\thanks{E-mail: abasu@mpifr-bonn.mpg.de. \newline Current e-mail: aritra@physik.uni-bielefeld.de} 
S. A. Mao,$^1$ Andrew Fletcher,$^3$ Nissim Kanekar,$^4$ Anvar Shukurov,$^3$
\newauthor Dominic Schnitzeler,$^{1,7}$ Valentina Vacca$^5$ and Henrik Junklewitz$^6$
	\\
$^1$Max-Planck-Institut f{\"u}r Radioastronomie, Auf dem H{\"u}gel 69, D-53121 Bonn, Germany\\
$^2$Fakult{\"a}t f{\"u}r Physik, Universit{\"a}t Bielefeld, Postfach 100131, 33501 Bielefeld, Germany\\
$^3$School of Mathematics and Statistics, Newcastle University, Newcastle-upon-Tyne, NE13 7RU, UK\\
$^4$National Centre for Radio Astrophysics, TIFR, Post Bag 3, Ganeshkhind, Pune 411007, India\\
$^5$INAF-Osservatorio Astronomico di Cagliari, Via della Scienza 5, I-09047 Selargius (CA), Italy\\
$^6$Argelander Institut f{\"u}r Astronomie, Universit{\"a}t Bonn, Auf dem H{\"u}gel 71, 53121 Bonn, Germany\\
$^7$Bendenweg 51, 53121 Bonn, Germany\\
}

\begin{document}

\date{Accepted to be published in MNRAS: 2018 March 19}

\pagerange{\pageref{firstpage}--\pageref{lastpage}} \pubyear{2017}

\maketitle

\label{firstpage}

\begin{abstract}

Deriving the Faraday rotation measure (RM) of quasar absorption line systems,
which are tracers of high-redshift galaxies intervening background quasars, is
a powerful tool for probing magnetic fields in distant galaxies.
Statistically comparing the RM distributions of two quasar samples, with and
without absorption line systems, allows one to infer magnetic field properties
of the intervening galaxy population. Here, we have derived the analytical form
of the probability distribution function (PDF) of RM produced by a single
galaxy with an axisymmetric large-scale magnetic field. We then further
determine the PDF of RM for one random sight line traversing each galaxy in a
population with a large-scale magnetic field prescription. We find that the
resulting PDF of RM is dominated by a Lorentzian with a width that is directly
related to the mean axisymmetric large-scale field strength $\langle B_0
\rangle$ of the galaxy population if the dispersion of $B_0$ within the
population is smaller than $\langle B_0 \rangle$. Provided that RMs produced by
the intervening galaxies have been successfully isolated from other RM
contributions along the line of sight, our simple model suggests that $\langle
B_0 \rangle$ in galaxies probed by quasar absorption line systems can be
measured within $\approx50$ per cent accuracy without additional constraints on
the magneto-ionic medium properties of the galaxies. Finally, we discuss quasar
sample selection criteria that are crucial to reliably interpret observations,
and argue that within the limitations of the current database of absorption
line systems, high-metallicity damped Lyman-$\alpha$ absorbers are best suited
to study galactic dynamo action in distant disc galaxies.

\end{abstract}


\section{Introduction}

Magnetic fields in star-forming disc galaxies in the local Universe ($\lesssim
50$~Mpc) have been studied in detail, with most studies indicating the
existence of a galactic dynamo \citep[see][for reviews]{beck96, beck16}.
However, only a handful of studies exist to date on magnetic fields in high
redshift ($z \gtrsim 0.1$) galaxies \citep[e.g][]{kronb92, oren95, berne08,
joshi13, farne14, kim16, mao17}. The evolution of magnetic fields in disc
galaxies over cosmological timescales has been studied through
magnetohydrodynamic simulations \citep[e.g.][]{deavi05, arsha09, hanas09,
gent13a, gent13b, chama13, burkh13, pakmo14, rodri15}. These simulations have
shown that fields on scales $\ll 1$~kpc can be amplified on a relatively short
timescale (a few hundred Myrs) via a fluctuation dynamo \citep[see
e.g.,][]{kandu99, feder11}. However, the amplification and ordering of
fields on larger, $\gtrsim 1$~kpc, scales via a large-scale dynamo requires
billions of years \citep[e.g.,][]{arsha09, chama13, pakmo14}. Observational
constraints on the amplification timescale of the large-scale magnetic fields
in galaxies are lacking and therefore, the evolution of magnetic fields in
galaxies still remains an important open question in astronomy.

Directly mapping the magnetic field structure and strength in high-$z$
galaxies, as has been done for nearby galaxies, will be a challenging
proposition even for next-generation radio telescopes like the Square Kilometre
Array (SKA) and its pathfinders. However, advances have been made in inferring
the strength of magnetic fields through statistical studies of excess Faraday
rotation measure (RM) introduced by a sample of intervening galaxies when
observed against background polarized quasars \citep[see, e.g.,][]{oren95,
berne08, berne13} --- the so-called ``backlit-experiment''.  Intervening
galaxies are identified through the presence of absorption lines at a redshift
significantly lower than that of the quasar. Since RM is the integral of the
magnetic field component weighted by the free electron density along the entire
line of sight, such studies are faced with the challenge of disentangling the
combined RM of the quasar itself, the intergalactic medium (IGM) and the Milky
Way, from the RM originating in the interstellar medium (ISM) of the
intervening galaxy. The first three terms can be statistically estimated by
measuring the RM towards a sample of quasars without intervening absorption
lines. Comparing the RMs of quasar samples with intervening absorption with
those without intervening absorption then allows one to statistically determine
the excess RM produced by the intervening galaxies. We emphasize that this
approach only allows one to study {\it statistical} properties of magnetic
fields in a sample of high-$z$ galaxies.

To study magnetic fields in individual distant galaxies, \citet{mao17} made use
of gravitational lensing of a polarized background quasar lensed by a
foreground galaxy. The authors inferred the large-scale magnetic field
properties in a galaxy at $z = 0.439$. The RM contributions by the quasar, the
IGM, and the Milky Way are approximately the same for the lensed images of the
quasar. Therefore, by comparing the RMs of the lensed images, it is possible to
measure the magnetic field strength and geometry in the intervening galaxy.
Suitable sightlines for such studies, with a foreground galaxy lensing a
background polarized radio-loud quasar, is limited at present. Further, known
radio-bright lens systems mostly have $\sim$arcsecond-scale image separations,
i.e. to lensing galaxies at the upper end of the galaxy mass function with halo
masses $\gtrsim 10^{10}~\rm M_\odot$. Hence, in order to trace the cosmic
evolution of magnetic fields across a wide range of galaxy types, including
lower mass galaxies that are believed to dominate the galaxy population at
early cosmic times, statistical backlit approaches are needed.

In this paper, we first derive an analytical expression of the probability
distribution function (PDF) of the RM for a galaxy with an axisymmetric
large-scale magnetic field when viewed at an arbitrary inclination. To enable
the derivation of the analytical solution for the PDF of the RM, we have
adopted a simple set of assumptions for the magneto-ionic medium of the target
galaxies. The analytical solutions allow one to explore the dependence of the
PDF of the RM on various magnetic field parameters, and thereby understand how
realistic, often more complicated, assumptions can affect the inferred field
properties. Using the analytical expression we gain insights into the
distribution of RMs measured along lines of sight passing through a sample of
disc galaxies with random distributions of inclination angle, radius and
azimuthal angle of intersection, strength and pitch angle of the magnetic
field, and free electron density.

\begin{figure*}
\begin{centering}
\begin{tabular}{ccc}
{\mbox{\includegraphics[width=5.5cm]{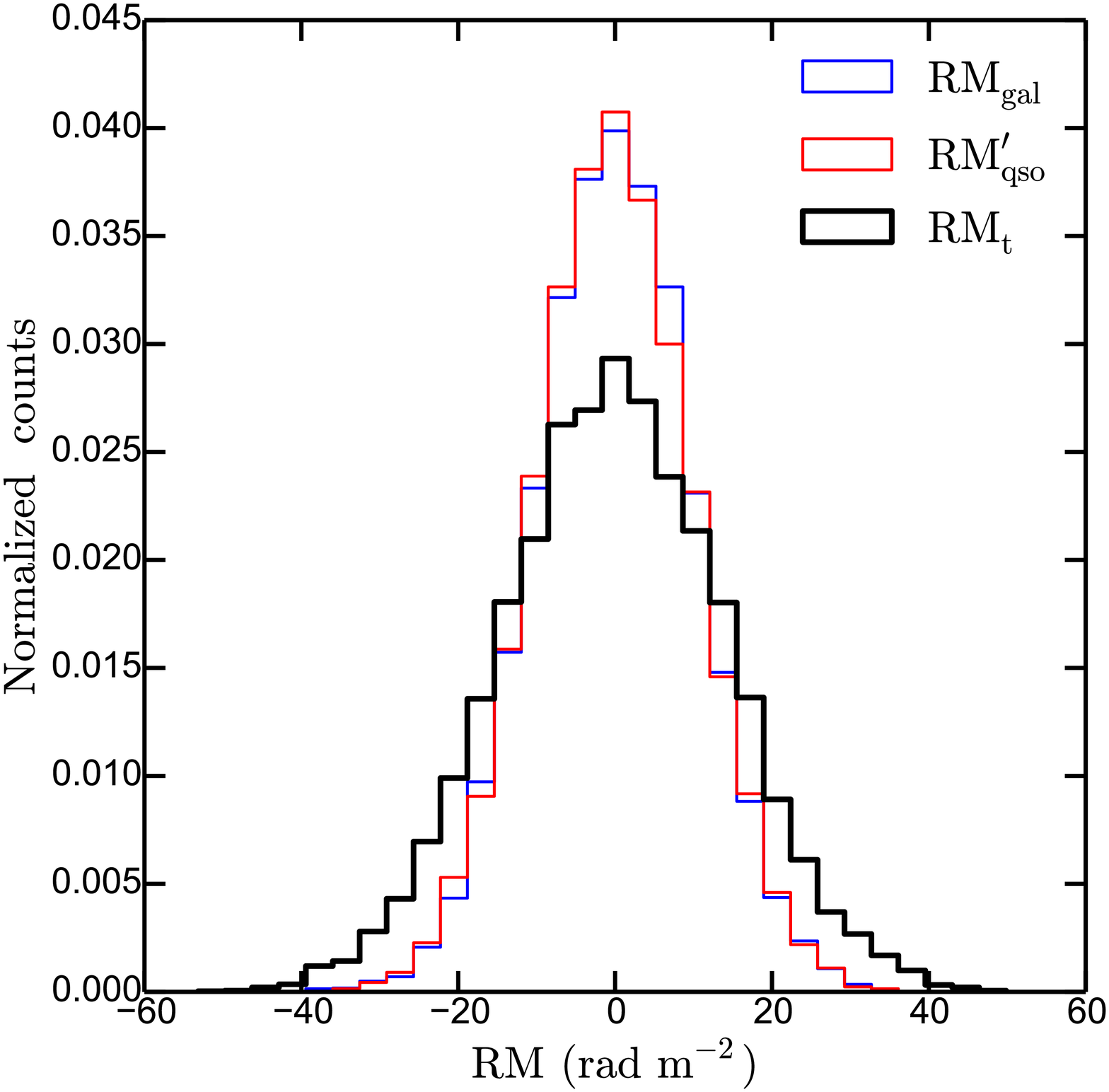}}}&
{\mbox{\includegraphics[width=5.5cm]{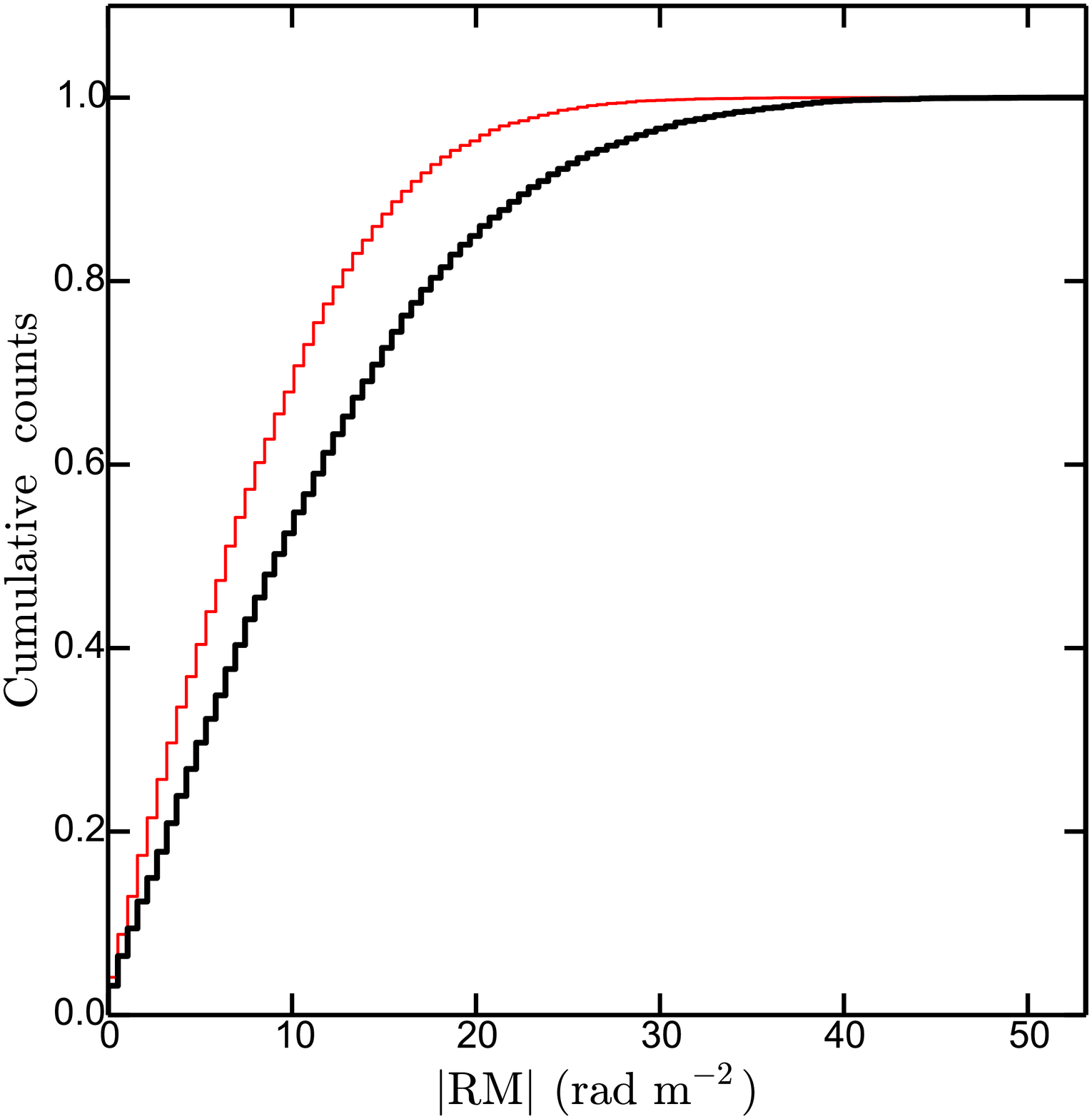}}}&
{\mbox{\includegraphics[width=5.5cm]{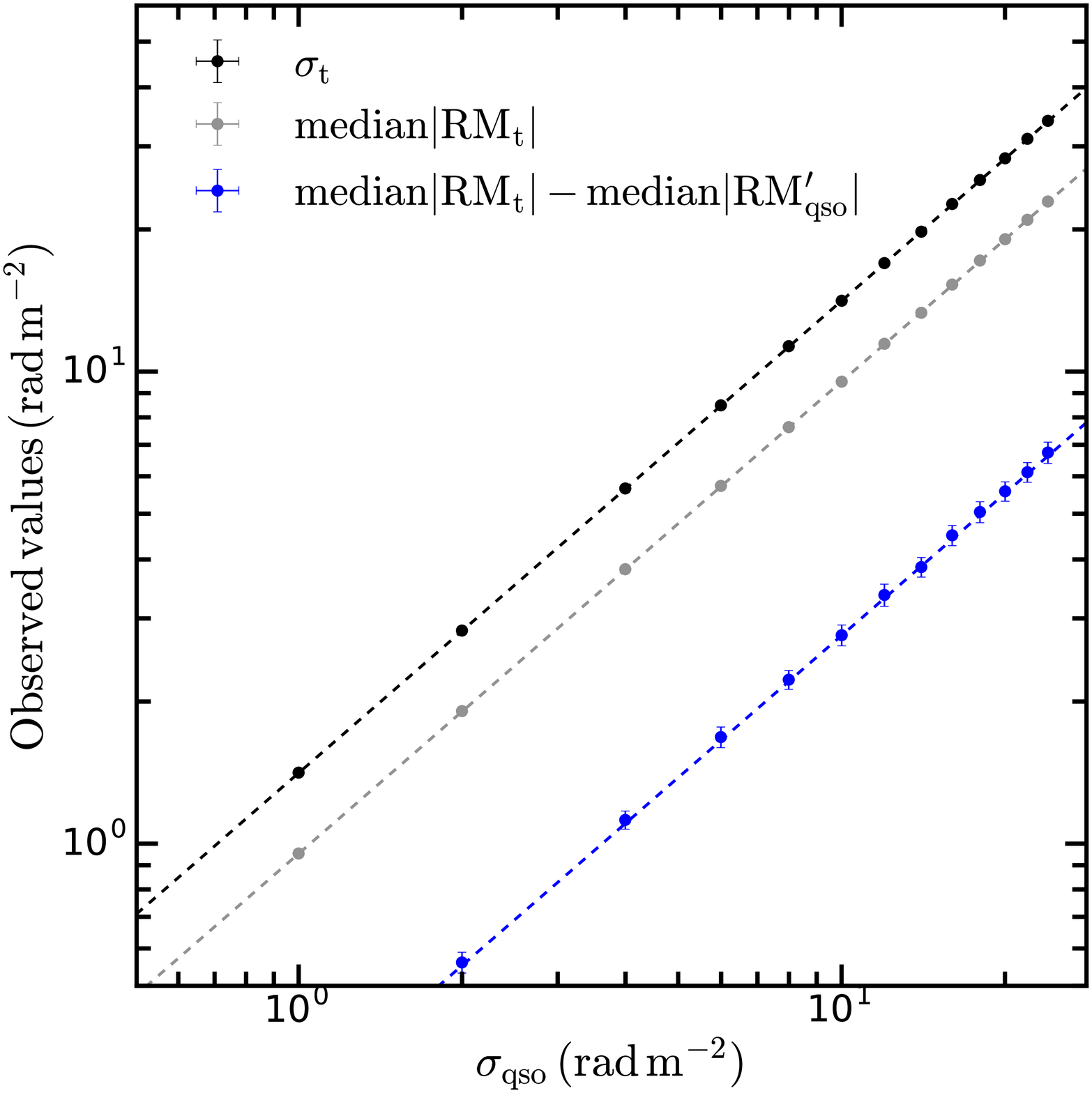}}}\\
\end{tabular}
\end{centering}
\caption{{\it Left}: Distribution of $\rmt=\rmgal+{\rm RM_{qso}^\prime}$ (black
histogram), where $\rmgal$ (blue histogram) and $\rm RM_{qso}^\prime$ (red
histogram) are Gaussian random variables with zero mean and $\sigma=10$ rad
m$^{-2}$. {\it Middle}: Cumulative distributions of $|\rmt|$ (in black) and
$|{\rm RM_{qso}^\prime}|$ (in red). {\it Right}: Variation of the observed
quantities as a function of the $\sigma_{\rm qso}$ assuming both $\rmgal$ and
$\rm RM_{qso}^\prime$ follow the same Gaussian distributions.}
\label{randomRM}
\end{figure*}

In Section~\ref{sec:data}, we describe the general setup for a
backlit-experiment that one uses to derive the RMs of high-redshift galaxies.
We identify three sources of bias in traditional analyses that can give rise to
ambiguity in the measured magnetic field strengths. In
Section~\ref{sec:method}, we present the methodology and assumptions used in
this paper. For simplicity, we only consider magnetic field in galactic discs.
The properties of the PDF of the RM for an intervening galaxy and that for a
sample of intervening galaxies are presented in Section~\ref{sec:result}.
We describe in Section~\ref{sec:sample} the selection criteria of target
absorber sample that satisfies the assumptions we have made. We discuss future
work necessary towards a complete understanding of the evolution of magnetic
fields in galaxies in Section~\ref{sec:future}, and summarize our results in
Section~\ref{sec:summary}.

\section{The data and various pitfalls in the analysis} \label{sec:data}

The presence of a galaxy along, or close to, the line of sight to a
background quasar can be deduced from the presence of ``damped'' or
``sub-damped'' neutral hydrogen Lyman-$\alpha$ absorption or strong metal-line
absorption (arising from transitions in Mg{\sc ii}, Fe{\sc ii}, Si{\sc ii},
etc.) imprinted on the optical and/or ultraviolet (UV) spectra of the quasar.
Damped Lyman-$\alpha$ absorbers (DLAs) are the highest neutral hydrogen (H{\sc
i}) column density absorbers in quasar absorption spectra, with H{\sc i} column
densities $\geq 2 \times 10^{20}$~cm$^{-2}$, and are believed to arise from
sightlines passing through galaxy discs \citep[e.g.][]{wolfe05}. Sub-DLAs have
somewhat lower H{\sc i} column densities, $\approx 10^{19} -
10^{20}$~cm$^{-2}$, and along with Lyman-limit systems and most strong
metal-line absorbers, are likely to mostly arise in the circumgalactic medium
\citep[CGM; e.g.][]{zibet07,nesto07,neele16}. Today, the largest absorber
samples have been obtained from searches through quasar absorption spectra from
the Sloan Digital Sky Survey (SDSS) \citep[e.g.][]{prochaska05, noter09,
noter12, zhu13}, with $\gtrsim 10,000$ DLAs currently known at $z \gtrsim 2$
\citep[e.g.][]{noter12} and $\gtrsim 40,000$ strong Mg{\sc ii} absorbers known
at $z \gtrsim 0.4$ \citep[e.g.][]{zhu13}. The sample of quasars with
foreground absorbing galaxies will henceforth be referred to as the ``target''
sample. We emphasize that our aim is to determine the magnetic fields of the
intervening absorber galaxies, and that the quasars are used only as background
torches.

The net RM ($\rmt$) of the background quasar and the absorption system in the
observer's frame for a single line of sight is given by,
\begin{equation}
\rmt = \dfrac{\rmgal}{(1+z_{\rm gal})^2} + \dfrac{\rmqso}{(1+z_{\rm qso})^2} + \Delta_{\rm RM}.
\label{rmobs}
\end{equation}
Here, $\rm RM_{gal}$ and $\rm RM_{qso}$ are the RM of the foreground absorber
galaxy and the intrinsic RM of the background quasar in their rest frames,
respectively, and $z_{\rm gal}$ and $z_{\rm qso}$ are their redshifts.
$\Delta_{\rm RM}$ is the additional contributions to the RM along the line
of sight and is defined as $\rm \Delta_{RM} = RM_{IGM} + RM_{MW} + \delta_{\rm
RM}$, where $\rm RM_{IGM}$ and $\rm RM_{MW}$ are the contributions from the IGM
and the Milky Way, respectively, and $\delta_{\rm RM}$ is the measurement
noise. We define $\rm RM_{qso}^\prime = \rmqso\,{(1 + \zqso)^{-2}} +
\Delta_{\rm RM}$, as the RM that would have been measured towards the quasar in
the absence of the intervening galaxy; thus, $\rmt = \rmgal\,(1+z_{\rm
gal})^{-2} + {\rm RM_{qso}^\prime}$. 

It is not possible to determine $\rmgal$ from a single measurement of $\rmt$,
and therefore measurements along lines of sight towards quasars {\it without}
intervening absorbers are required to statistically infer $\rm
RM_{qso}^\prime$. These can be compared with the measurement of $\rmt$ towards
quasar sightlines with absorbing galaxies to estimate the statistical
properties of $\rmgal$, and to subsequently derive the strength of the magnetic
field in the absorber galaxy. The sample of quasars without foreground absorber
galaxies, which are used to estimate $\rm RM_{qso}^\prime$, are referred to as
the ``control'' sample. The RM values of the quasars in the control sample are
given by,
\begin{equation} 
{\rm RM_{c}} = \dfrac{\rm RM_{qso,c}}{(1+z_{\rm qso, c})^2} + \Delta_{\rm RM}.
\label{rmcontrol}
\end{equation}
In order to infer $\rmgal$ produced by the absorber galaxies in our target
sample, we will assume that $\rmc$ and $\rm RM_{qso}^\prime$ have the same
statistical properties. Thus, a comparison between the distributions of $\rmt$
and $\rmc$ would then yield the excess contribution from ${\rm
RM_{gal}}\,{(1+z_{\rm gal})^{-2}}$ in the former quantity.

When interpreting the distribution of $\rmt$, it is important to consider the
effects of each of the variables, i.e., $\rm RM_{gal}$, $\rm RM_{qso}$, $\rm
RM_{qso,c}$, $\Delta_{\rm RM}$, $z_{\rm gal}$, $z_{\rm qso}$, and $z_{\rm qso,
c}$, where subscript `c' refers to the control sample. In the following, we
discuss the potential sources of bias that are introduced by standard analysis
procedures used in the literature to date. In Table~\ref{variables}, we list
and describe in brief the variables and notations used in this paper and in the
literature.

We note, that the distribution of $\rmt$ is the convolution of the
distributions of the terms, $\rmgal\,(1+\zgal)^{-2}$ and $\rm RM_{qso}^\prime$,
in Eq.~\eqref{rmobs}, i.e., $\mathrm{PDF}(\rmt) =
\mathrm{PDF}\left(\rmgal\,(1+\zgal)^{-2}\right) \otimes \mathrm{PDF}\left(\rm
RM_{qso}^\prime\right)$. Therefore, a formal approach to study the distribution
of $\rmgal$ is to deconvolve the distribution of $\rm RM_{qso}^\prime$ (using
the distribution of $\rmc$ as its proxy), from that of $\rmt$. Unfortunately,
we do not {\it a priori} know the distributions of $\rmt$ and $\rmc$, and there
is no formal mathematical procedure in the literature to perform a
deconvolution on arbitrary statistical distributions. One hence usually simply
compares the distributions of $\rmt$ and $\rmc$ to test whether or not the
contribution from $\rmgal$ produces a statistical difference between the two
distributions. It is beyond the scope of this paper to investigate a formal
deconvolution method, and how such a procedure might affect the results. In
this work, we study the properties of the distribution of $\rmgal$ through
simulations, and how this distribution is related to the strength of the
large-scale magnetic fields in high-$z$ galaxies.

\subsection{The bias from using $|\rm RM|$} \label{sec:absRM}

To determine the contribution of $\rmgal$ to $\rmt$, the empirical cumulative
distribution function (CDF) of $|\rmt|$ has been compared to that of $|\rmc|$
in the past \citep[see, e.g.,][]{berne08, farne14}. The difference in their
median values has been used to estimate the magnetic field strengths
\cite[e.g.,][]{farne14}. In this section, we demonstrate that using the
absolute value of RM, rather than the value of RM itself, introduces a
systematic bias in the results.

To demonstrate this, we consider the simple case where $\zgal\approx0$, so that
$\rmt = \rmgal + \rm RM_{qso}^\prime$. We assume Gaussian random distributions
for both $\rmgal$ and $\rm RM_{qso}^\prime$, i.e. their PDFs are, respectively,
given by $X = \mathcal{N}\left(\langle \rmgal \rangle, \sigma_{\rm gal}
\right)$ and $Y = \mathcal{N}\left( \langle \rm RM_{qso}^\prime \rangle,
\sigma_{\rm qso} \right)$.  This is equivalent to the case where the
probability distribution of $\rmgal\,(1+ \zgal)^{-2}$ is Gaussian. In this
case, the PDF of the sum $X + Y$ is the convolution of the two PDFs and is
given by $X + Y = \mathcal{N}\left(\langle \rmgal \rangle, \sigma_{\rm
gal}\right) \otimes \mathcal {N}\left(\langle \rm RM_{qso}^\prime \rangle,
\sigma_{\rm qso}\right) = \mathcal{N}\left(\langle \rmgal \rangle + \langle \rm
RM_{qso}^\prime \rangle, \sqrt{\sigma_{\rm gal}^2 + \sigma_{\rm
qso}^2}\right)$. We emphasize that the statistical properties of $|X+Y|$ are
very different, and significantly more complicated, than those of $X+Y$. 
The absolute value of a Gaussian is known as the folded normal distribution.
The median, mean and variance of this distribution cannot be written in closed
form, unless the mean is zero, and its shape is determined by both its mean and
variance.

We find that the difference between the median values of $|\rmt|$ and $|\rm
RM_{qso}^\prime|$ increases with increasing $\sigma_{\rm qso}$. In
Fig.~\ref{randomRM} (left panel), we simulate the distributions\footnote{All
distributions presented in this paper are normalized to their areas.} of
$\rmt$, $\rmgal$ and $\rm RM_{qso}^\prime$. We consider similar Gaussian
distributions for $\rmgal$ and $\rm RM_{qso}^\prime$ with zero mean and the
same dispersion $\sigma_{\rm gal} = \sigma_{\rm qso}$. For such a case,
$\langle \rmt \rangle = \langle \rmgal \rangle + \langle \rm RM_{qso}^\prime
\rangle = 0$~rad\,m$^{-2}$, and the observed dispersion of $\rmt$ ($\sigma_{\rm
t}$) is given by $\sigma_{\rm t}^2 = \sigma_{\rm gal}^2 + \sigma_{\rm qso}^2$.
However, if we compare the absolute values of $\rmt$ and $\rm RM_{qso}^\prime$
(as shown in the middle panel of Fig.~\ref{randomRM}), we find that the CDFs
are different, such that the median value of $|\rmt|$ is greater than the
median value of $|\rm RM_{qso}^\prime|$. Such a difference in the CDF of $|\rm
RM|$ between the target and control samples has been interpreted in the past as
implying that intervening galaxies show excess median RM and has been
used to derive the strength of the large-scale field \citep{farne14}. Our
simulated data indicate that such a difference in the CDFs may simply arise
from the increased dispersion in the RM of sightlines that host an intervening
galaxy, and does not necessarily imply an excess in median value of RM
originating from large-scale fields in the absorbers.

Figure~\ref{randomRM} (right panel) shows how the inferred difference of the
statistics of absolute RM values varies as a function of $\sigma_{\rm qso}$
(which is the same as $\sigma_{\rm gal}$ in this example).  The black points
show the dispersion of the observed RM, $\sigma_{\rm t}$; this is related to
$\sigma_{\rm qso}$ as $\sigma_{\rm t} = \sqrt{2}\,\sigma_{\rm qso}$ in this
example. The grey points show how the median value of $|\rmt|$ varies as a
function of $\sigma_{\rm qso}$. Clearly, the median value of $|\rmt|$ has a
linear relation with $\sigma_{\rm qso}$, whereas the median value of $\rmt$ is
zero.  The blue points show the variation of the difference between the median
values of $|\rmt|$ and $|\rm RM_{qso}^\prime|$. This difference is just an
artefact of the increased dispersion, and does not necessarily imply a true
difference between the medians of the two population.  For $\sigma_{\rm qso}
\sim 6\text{--}7$ rad m$^{-2}$ \citep{schni10, opper15}, a typical value for
the dispersion of the RM of extragalactic sources in the observer's frame, the
difference of medians can be up to $\sim 2$~rad\,m$^{-2}$. This is a
significant fraction of the differences in median $|\rm RM|$ reported earlier
in studies based on Mg{\sc ii} absorption systems \citep[e.g.][]{farne14}.
Using $|\rm RM|$ instead of RM introduces a similar bias also for other choices
of the means and variances of $\rmgal$, $\rm RM_{qso}^\prime$ and $\rmc$.
Therefore we believe, RM, rather than $\rm |RM|$, is perhaps a better
quantity to compare between the target and the control sample to infer the
magnetic field properties of the intervening galaxies.

\subsection{The redshift bias} \label{sec-redshift}

A comparison between the distributions of RM of the target and control sample,
$\rmt$ and $\rmc$, is only meaningful if the distributions of $\rmc$ and $\rm
RM_{qso}^\prime$ are similar. For this, a comparison between Eqs.~\eqref{rmobs}
and \eqref{rmcontrol} shows that quasars in the target and control samples
should be selected in such a way that $\rmqso\,(1+ \zqso)^{-2}$ and ${\rm
RM_{qso,c}}\,(1+z_{\rm qso,c})^{-2}$ follow similar statistical distribution.
Even if $\rmqso$ and $\rm RM_{qso, c}$ follow the same distribution, a
difference in the redshift distributions of the target and control samples can
introduce a systematic bias, due to the $(1 + z)^{-2}$ factor in the expression
for RM in the observer's frame. In fact, in such a situation, the contribution
of $\rm RM_{IGM}$ in the $\Delta_{\rm RM}$ term will also be different for the
two sets of lines of sight, introducing additional bias
\citep[e.g.][]{akaho16}.

\begin{figure}
\begin{centering}
\begin{tabular}{c}
{\mbox{\includegraphics[width=8cm]{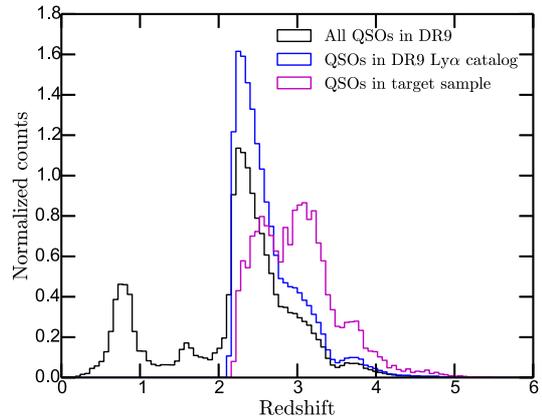}}}\\
\end{tabular}
\end{centering}
\caption{Redshift distribution of the quasars in the SDSS DR9 catalog are shown
as the black histogram and for background quasars in the target DLA sample as
the magenta histogram. Blue histogram show the redshift distribution of quasars
in the BOSS Lyman-$\alpha$ forest catalog.}
\label{zdistr}
\end{figure}

Here, we assess the systematic bias produced by different redshift
distributions of the quasars in the target and the control samples, using DLAs
as the intervening galaxies. Fig.~\ref{zdistr} shows the redshift distribution
of various quasar sub-samples drawn from the SDSS DR9 catalogue \citep{ahn12}.
The redshift distribution of all quasars in the SDSS DR9 catalog
\citep{paris12} is shown in black. The redshift distribution of the quasars in
the BOSS Lyman-$\alpha$ forest sample from the SDSS~DR9 \citep{lee13} is shown
in blue, while that of SDSS~DR9 quasars with foreground DLAs
\citep{noter12,lee13} is shown in magenta.  Thus, the quasars in the control
sample have $z_{\rm qso, c} \approx 0.5 - 4$, while those in the target sample
have $\zqso \approx 2 - 4$. It is clear that a simple use of the full SDSS DR9
quasar catalogue (i.e. without redshift cuts to ensure similar quasar redshift
distributions in the target and control samples) to select polarized quasars
for studies of intervening galaxies identified as DLAs would yield $\langle
\zqso \rangle \gtrsim \langle z_{\rm qso, c} \rangle$.

\begin{figure}
\begin{centering}
\begin{tabular}{c}
{\mbox{\includegraphics[width=7.0cm, trim=.25cm 0.5cm 0cm 1cm, clip]{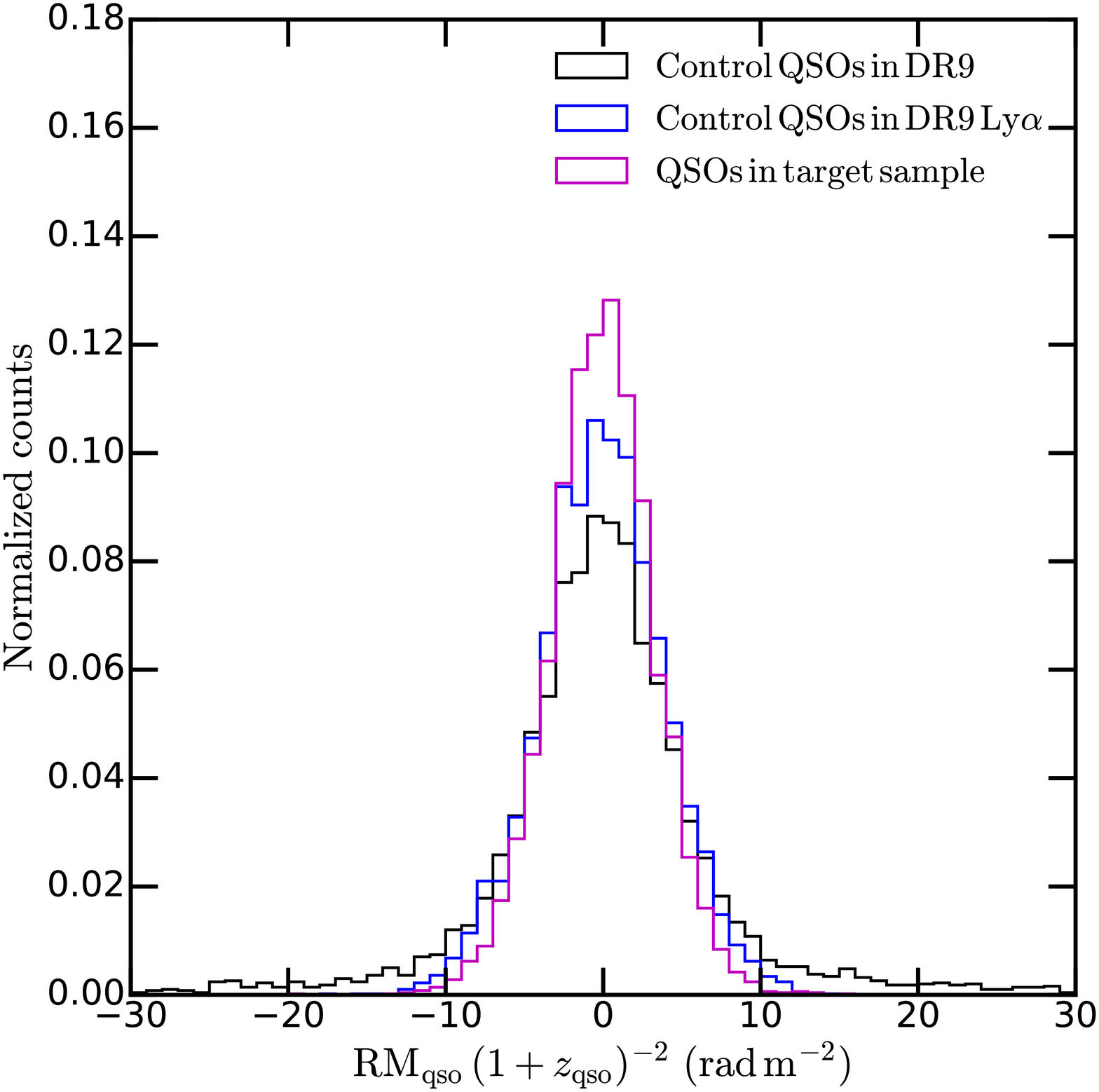}}}\\
{\mbox{\includegraphics[width=7.0cm, trim=.25cm 0.5cm 0cm 1cm, clip]{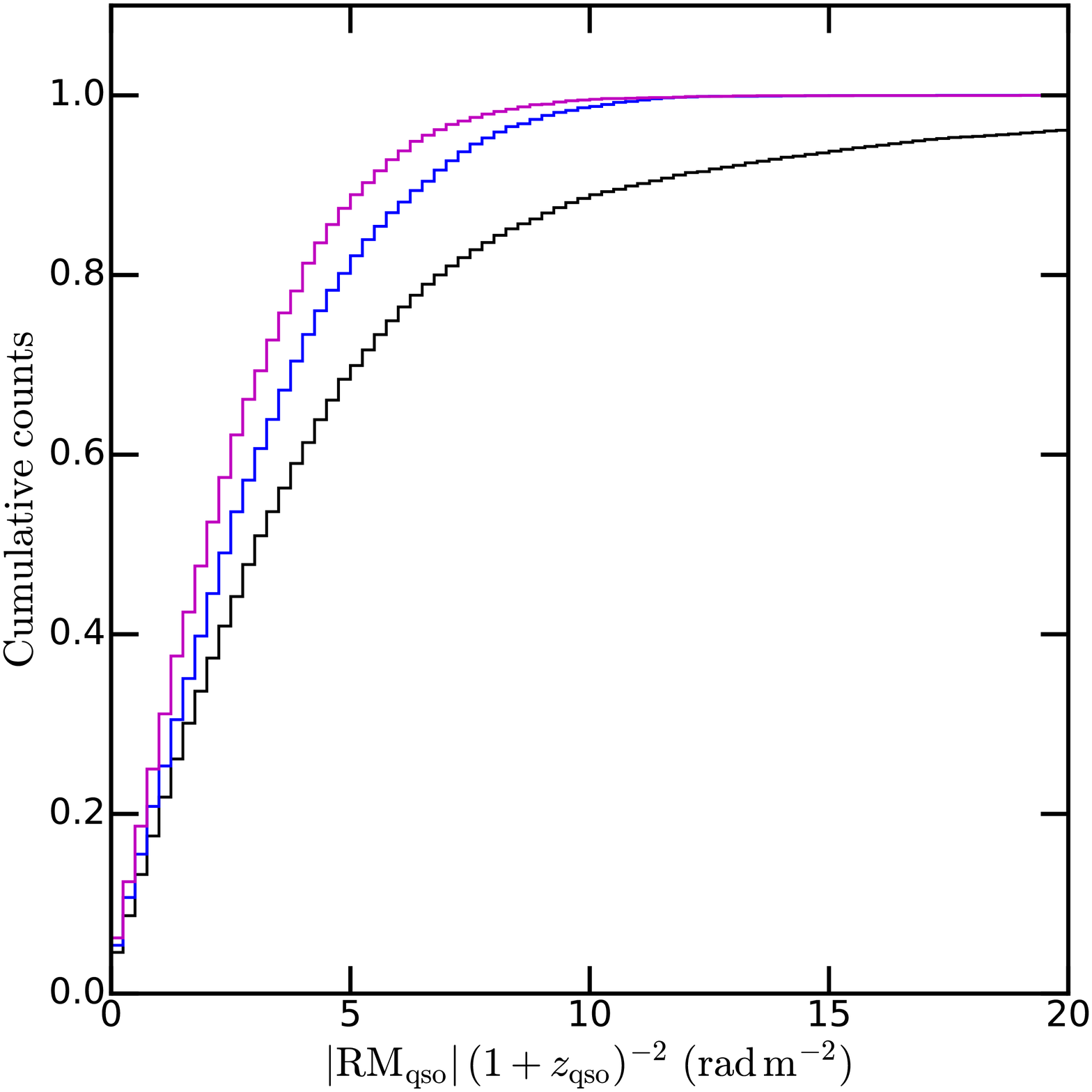}}}\\
\end{tabular}
\end{centering}
\caption{{\it Top}: Distribution of $\rmqso\,(1+\zqso)^{-2}$ for different
redshift coverages given in Fig.~\ref{zdistr}. {\it Bottom}: Cumulative
distribution of $|\rmqso|\,(1+\zqso)^{-2}$. We assume that the intrinsic
(rest-frame) RMs of the quasars follow a Gaussian distribution with zero mean
and a standard deviation of $50\,\rm rad\,m^{-2}$. The distribution in black is
for control quasars chosen from the entire SDSS DR9 catalog, blue is for
control quasars chosen from BOSS Lyman-$\alpha$ forest catalog and magenta is
for quasars in the target DLA sample.} 
\label{zrmqso}
\end{figure}

The top and bottom panels of Fig.~\ref{zrmqso} show the effect of different
redshift distributions for the control and target quasar samples on the PDF of
RM$_{\rm qso}(1+z_{\rm qso})^{-2}$ (top panel) and the CDF of $|{\rm
RM_{qso}}|(1+z_{\rm qso})^{-2}$ (bottom panel). Here, we have assumed that
quasars at all redshifts have an intrinsic RM distribution with zero mean and
$\sigma_{\rm qso} = 50$~rad\,m$^{-2}$ in their rest-frame.\footnote{We chose
$\sigma_{\rm qso} = 50$~rad\,m$^{-2}$ because the typical observed $\sigma_{\rm
RM}$ for extragalactic sources (in the observer's frame) is $\approx
6\text{--}7$~rad\,m$^{-2}$ \citep{schni10}. The sources are expected to be
mostly active galactic nuclei (AGNs), with a redshift distribution that peaks
at $z \approx 2$ (see Fig.~\ref{zdistr}).  Hence, assuming a typical AGN
redshift of $z \approx 2$, we expect $\sigma_{\rm qso} \approx 50
$~rad\,m$^{-2}$.} Since $(1+\zqso)^{-2}$ is non-linear in $\zqso$, the
distribution of $\rmt$ is non-Gaussian. If we do not impose redshift
constraints on the control and target samples, the redshifts of the control and
target samples would be drawn from the distributions shown, respectively, in
black and magenta in Fig.~\ref{zdistr}. This implies that the control sample
can have $\sigma_{\rm qso}$ up to $\approx 5$~rad\,m$^{-2}$ larger than that of
the target sample in the observer's frame (comparing the black and magenta
histograms in the top panel of Fig.~\ref{zrmqso}). This would yield a median
value of $|\rmc|$ lower by $\approx 1.5$~rad\,m$^{-2}$ than the median value of
$|\rm RM_{qso}^\prime|$.

The above differences can be mitigated if the quasars of the control and target
samples are chosen to have the same redshift distribution. In the case of the
SDSS~DR9, it would appear that this might be achieved by selecting both the
control and the target quasars from the BOSS Lyman-$\alpha$ forest sample,
i.e., the redshift distribution shown in blue in Fig.~\ref{zdistr}. We note,
however, that such a selection from the BOSS catalogue actually does {\it not}
yield similar statistical distributions of $\rmc$ and $\rm RM_{qso}^\prime$
(see the blue and magenta curves in Fig.~\ref{zrmqso}), despite similar range
of their redshift coverage. The difference arises due to the differences in the
actual redshift distributions of the quasars of the two samples (see the blue
and magenta curves in Fig.~\ref{zdistr}), as the redshift distribution of
quasars in the target sample contains an additional dependence on the
probability of finding an absorbing galaxy at a given redshift. It is hence
critical to carry out statistical tests to ensure statistically identical
redshift distributions for the radio-bright quasars in the target and the
control samples.

\subsection{Bias from incomplete absorber redshift coverage}

An additional bias arises due to the restricted wavelength coverage of the
spectrographs used for the absorption surveys, which implies that they are only
sensitive to absorbers lying in a limited redshift range. For example,
ground-based optical surveys for DLAs are sensitive to Lyman-$\alpha$
absorption only at $z \gtrsim 1.7$, even when using UV-sensitive spectrographs;
the SDSS is only sensitive to DLAs at $z \gtrsim 2$ \citep[e.g.][]{noter12}.
However, the redshift line number density of DLAs is $\approx 0.25$ per unit
redshift at $z\gtrsim2$ \citep{prochaska05}, and $\approx 0.1$ per unit
redshift at $z\approx 1$ \citep{rao06}. This implies that it is not trivial to
generate a ``clean'' high-redshift quasar control sample from a survey such as
the SDSS~DR9, because there is $\approx 25$ per cent probability that a quasar
at $z>2$ would have an absorbing galaxy at $z < 2$ that is undetected simply
because the relevant wavelength range has not been covered in the survey. 

The $(1+z)^{-2}$ dilution of the rest-frame $\rmgal$ makes this a serious
issue, because the RM contribution in the observer's frame from an absorber at
$z\lesssim 1$ is significantly larger (typically by factor $\sim2\text{--}4$)
than that at $z\gtrsim2$. The presence of a DLA at $z \lesssim 1$ towards a
high-$z$ quasar would imply a higher $|\rm RM|$ than if the DLA were absent. 

The effect of such putative undetected absorbers on the derived distribution of
$\rmt$ depends critically on the redshift distributions of the target and the
control quasar samples.  For the simple SDSS~DR9 case discussed in the previous
section, where the quasar control sample has a lower median redshift, quasars
in the target sample are more likely to have undetected absorbers at low
redshifts, and should hence have, statistically, a higher $\rmt$ than the
quasars of the control sample. If this effect is not corrected for, it would
yield higher $\rmgal$ values for the DLAs in the sample than the true ones. We
note that this bias goes in the opposite direction as the bias discussed in the
previous section.

The best way to remove this bias is again to ensure that the target and control
quasar samples have the same redshift distribution. If so, the issue of
undetected absorbers at low redshifts should affect the quasars of both the
target and the control samples in the same manner, implying no systematic bias
in the derived $\rmgal$ values for the target sample.

\section{Basic equations} \label{sec:method}

As discussed earlier, our approach to estimating the large-scale magnetic field
is based on RM measurements towards a large number of quasars with foreground
absorbers; the RM estimates contain information on the magnetic field component
along the line of sight for each intervening galaxy. Since $|\rm RM|$
introduces a systematic bias (Section~\ref{sec:absRM}), one should work
directly with the distribution of RM. In this section, we derive an analytical
form of the PDF of the RM for an intervening disc galaxy with an axisymmetric
spiral magnetic field geometry, observed along random lines of sight through
the disc. This will be used to infer the magnetic field properties of high-$z$
disc galaxies from the observed RM distribution.

Note that we will, in later sections, focus on the properties of the
distribution of the RM originating only from the large-scale fields in the
intervening galaxies in their rest frame. This is equivalent to the case --- in
an observed backlit-experiment --- that the contribution of RM from the quasar,
the IGM, the Milky Way and noise, i.e. $\rm RM_{qso}^\prime$, is either
negligible compared to the observed RM or $\rmgal$ has already been isolated
from the contribution of $\rm RM_{qso}^\prime$ (see the discussion in
Section~\ref{sec:data}).

\subsection{Assumptions on the Faraday rotating medium}

We model an intervening galaxy as a disc with an axisymmetric spiral magnetic
field and a radially decreasing free electron density ($n_{\rm e}$). We assume
that the amplitude and geometry of the magnetic field, as well as the electron
density, do not vary with distance from the mid-plane of the galaxy. 
Faraday rotation in galaxies is produced from both turbulent and large-scale
magnetic fields. The distribution of RM due to isotropic turbulent fields is
expected to be a Gaussian with zero mean and its standard deviation is a
measure of the strength of the turbulent field (see Section~\ref{sec:turb_field}). 
Thus, for characterizing turbulent magnetic fields the dispersion of observed
RM is sufficient. Since we are interested in studying the properties of RM
produced by large-scale fields, we will work under the assumption that the RM
originating from turbulent magnetic fields is insignificant within the
three-dimensional illumination beam passing through the magneto-ionic medium of
the galaxy. This can be achieved when the spatial extent of the polarized
emission from the background quasar, as seen by the foreground galaxy, is large
enough to encompass several turbulent cells, but small enough so that the RM
contributed by the large-scale field does not vary significantly across the
beam (see Section~\ref{sec:turb_field}).

The magnetic field component along the line of sight ($B_\|$) at each point in
a galaxy can be calculated via \citep[e.g.][]{berkh97},
\begin{equation}
B_\| = -(B_r \sin \theta + B_\theta \cos \theta) \sin i + B_z \cos i \;.
\label{eq_bparallel}
\end{equation}
Here, $B_r$ and $B_\theta$ are, respectively, the radial and azimuthal
components of the magnetic field, $\theta$ is the azimuthal angle, and $i$ is
the inclination angle with respect to the plane of the sky. For an axisymmetric
spiral magnetic field the pitch angle is defined as, $p =
\arctan\left(B_r/B_\theta\right)$. $B_z$ is the magnetic field component
perpendicular to the disc. In this paper we assume that $B_z=0$, to simplify
our calculations, as it is significantly smaller in magnitude than $B_r$ and
$B_\theta$ \citep{mao10,chama16}.

We assume that the total large-scale magnetic field ($B = \sqrt{B_r^2 +
B_\theta^2}$) varies exponentially with radius $r$ \citep[e.g.][]{beck07,
beck15}, such that, $B(r) = B_{\rm 0}\, {\rm e}^{-r/r_B}$, where $r_B$ is the
radial scale-length and $B_0$ is the large-scale field strength at the centre
of the galaxy.\footnote{In this formulation, the radial variations of $B_r$ and
$B_\theta$ are given by $B_r(r) = B_0\, \sin p \, {\rm e}^{-r/r_B}$ and
$B_\theta(r) = B_0\, \cos p \, {\rm e}^{-r/r_B}$.} The radial scale-length of
the ordered magnetic field in spiral galaxies is typically $\approx 15 -25$~kpc
\citep{beck15, berkh16}; we hence adopt $r_B = 20$ kpc.

The radial variation of $n_{\rm e}$ is modelled as $n_{\rm e}(r) = n_0\, {\rm
e}^{-r/r_{\rm e}}$. Here, $n_0$ is the electron density at the centre of the
galaxy and $r_{\rm e}$ is the radial scale-length. We assume that $n_{\rm
e}$ does not vary with distance from the midplane, and that the electrons are
in a thick disc of uniform thickness, $h_{\rm ion}= 500$ pc, centred at the
midplane. For a sightline of inclination angle $i$ with respect to the plane of
the sky, the path length through the disc is then $L = h_\mathrm{ion}/\cos i$.
Finally, according to the NE2001 model \citep{ne2001}, the ionized thick disc
of the Milky Way shows a gradual decrease of $n_{\rm e}$ with Galactocentric
radius, with $r_{\rm e} \gtrsim 15$~kpc, i.e. comparable to $r_B$.  We hence
adopt $r_{\rm e} \approx r_B \approx 20$~kpc for our calculations.

With these assumptions, the RM can be written as,
\begin{equation} \label{eq_rm}
\begin{split}
\mathrm{RM} & = 0.81\, \left(\dfrac{\langle n_{\rm e}\rangle}{\rm cm^{-3}} \right)\, \left(\dfrac{\langle B_\parallel\rangle}{\rm \umu G}\right)\, \left(\dfrac{L}{\rm pc}\right) ~ {\rm rad\,m^{-2}}  \\
 & = -0.81\, n_0\, B_0\, {\rm e}^{-r/r_0^\prime}\,\cos (\theta - p) \, h_{\mathrm{ion}} \tan i \;, 
\end{split}
\end{equation}
\noindent where $1/r_0^\prime = 1/r_B + 1/r_{\rm e}$.

\subsection{Line of sight approximation}

To calculate $B_\|$ and RM along a single line of sight, we have used a
constant value of $B_\|$, $n_{\rm e}$ and $\theta$ throughout the ionized
medium of the galactic disc that gives rise to Faraday rotation. This allows us
to compute analytical solutions for the quantities of interest. Strictly
speaking, such a simplification is inadequate for an inclined disc. Sightlines
with $i\lesssim 45^\circ$ probe relatively narrow ranges of both galactocentric
radii and $B$ values. For example, for our assumed $h_{\rm ion} = 500$~pc and
$r_B \approx 20$~kpc, and sightlines with $i \lesssim 45^\circ$, $B$ varies by
less than $2$ per cent at the near and the far sides of the disc, with respect
to its value at the mid-plane, while $\theta$ varies by less than $1.5^\circ$.
Such small variations will not significantly affect our derived values of
RM and subsequent results.

At higher inclinations, e.g. $i \approx 75^\circ$, $B$ varies by $\lesssim 10$
per cent along the sightline, and $\theta$ by $\lesssim 5^\circ$.  This will
affect the estimated values of RM at $\lesssim15$ per cent level.  At even
larger inclinations $i > 75^\circ$, while the errors due to the assumptions of
constant $B$ and $\theta$ will be significant ($>20$ per cent), the probability
of finding a galaxy-quasar pair will also be low, as the projected area of the
galaxy on the sky is low for high $i$.  Only a small fraction of the sightlines
in the target sample would hence be at such high inclinations, implying that
our approximations are unlikely to significantly affect the final results.

We note, that the inclination angle does not have a marked effect on $B$
primarily because $r_B \gg h_{\rm ion}$, and therefore the magneto-ionic disc
essentially behaves like a thin disc. Similarly, because $r_{\rm e}$ is large,
the variation of $n_{\rm e}$ through the disc at a particular radius is small.
Overall, our simplifications will affect the derived values of RM at
$\lesssim15$ per cent for the extreme case, where the galaxy is inclined at
$\approx 75^\circ$.

\subsection{Distribution functions of the random parameters: azimuthal angle,
inclination angle and impact radius} \label{rand_distr}

The quasar sightlines can intersect the intervening galaxies at any impact
radius, inclination angle, and azimuthal angle. To model this, we assume
uniform distributions for $\theta$ and $i$ where all values are equally likely,
i.e., the probability densities have the form,
\begin{equation}
f_\Theta(\theta) =
\begin{cases}
~\dfrac{1}{2\upi}, & \text{for } 0 \leq i \leq {2\upi},
\\
~0, & \text{otherwise. }
\end{cases}
\end{equation}
\begin{equation}
f_I(i) =
\begin{cases}
~\dfrac{2}{\upi}, & \text{for } 0 \leq i \leq \dfrac{\upi}{2},
\\
~0, & \text{otherwise. }
\end{cases}
\end{equation}

We note, however, that the distribution of $i$ may not be strictly uniform
even for the case of random lines of sight, with no preferred inclination
angles of the intervening galaxy with respect to the observer and the
background quasar.  This is due to two competing reasons. On one hand, the
probability of having a quasar behind a highly inclined galaxy is lower as such
a galaxy would have a far smaller projected area in the plane of the sky than a
relatively face-on galaxy. In Appendix~\ref{sec:inclination}, we discuss how
this may affect our results. On the other hand, for galaxies selected through
absorption, the inclination angle and radius of intersection are related
because, at large impact radii, a sightline through a highly inclined galaxy
could yield a higher H{\sc i} column density and hence stronger absorption
compared to a face-on galaxy. Modelling these effects simultaneously is
difficult due to the lack of a large sample of absorber galaxies with
information on the impact radii, the column density of the absorbing gas, and
the inclination angle \citep[see e.g.][]{kacpr11}.

\subsubsection{Distribution function of the impact radius} \label{sec:rpdf}

The projected distance between the background quasar and the center of the
absorbing object, known as the impact parameter, is used to quantify the
distance at which the line of sight intersects an absorber galaxy. For
simplicity, we use the deprojected radial distance from the centre, $r$,
instead of the impact parameter in this study. Using the radius in our
calculations has the advantage of modelling the variations of physical
parameters relatively easily. In contrast, the impact parameter depends on both
the radii of impact and the inclination angle.

\begin{figure}
\begin{tabular}{c}
{\mbox{\includegraphics[width=8cm]{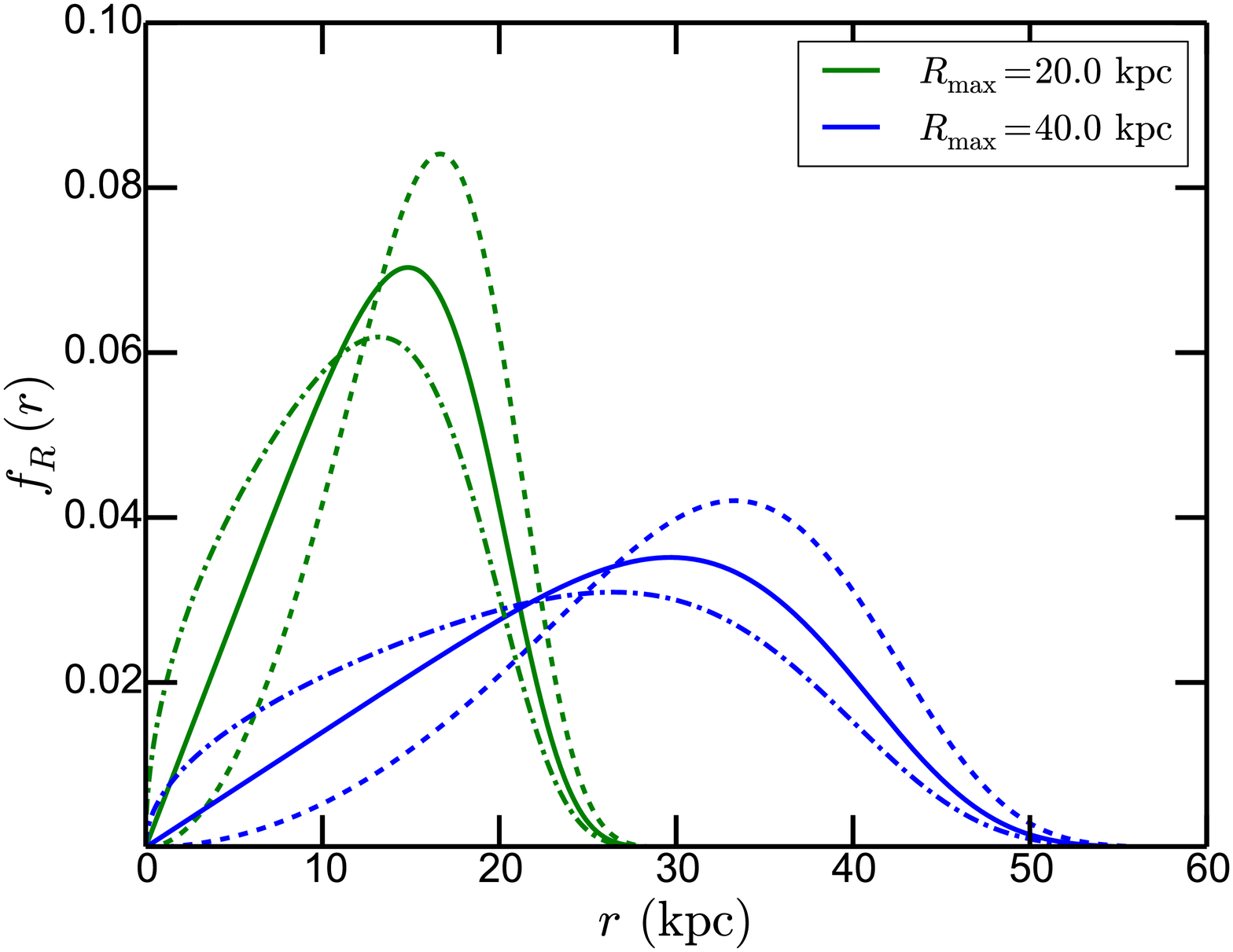}}} \\
\end{tabular}
\caption{Assumed distribution function of $r$ as described by Eq.~\eqref{r_pdf2}
for $\eta=0.5$ (dashed-dot lines), $\eta = 1$ (solid lines) and $\eta=2$
(dashed lines). The green and the blue curves are for $R_{\rm max} = 20$ and 40
kpc, respectively.}
\label{fig:r_pdf2}
\end{figure}

In general, we assume that a particular species of absorption line probes the
radius range $R_{\rm min} \le r \le R_{\rm max}$ in the intervening galaxies.
For example, $R_{\rm max}$ can be the transition radius at which the H{\sc i}
column densities drop below the DLA threshold column density of $2\times
10^{20}$~cm$^{-2}$ \citep{wolfe05}. Or, for Mg{\sc ii} absorbers, it could be
the radius at which the Mg{\sc ii}$\lambda$2796\AA\ rest equivalent width falls
below a threshold width \citep[e.g. 0.5~\AA; ][]{rao06}. For the distribution
of $r$, we assume the following two cases:
\begin{itemize}
\item{{\bf Case 1:} For a sample of galaxies, the probability of finding a
background quasar at a radius $r$ within the range $r\pm \Delta r/2$ increases
as $r\,\Delta r$. However, for an absorption selected sample of intervening
galaxies, due to inclination effects such a linear behaviour with $r$ can be
different and therefore we assume that the probability of finding a quasar at a
radius $r$ scales as $r^\eta$ where $\eta$ could lie between 0 and 2. Since the
absorber species probes up to a maximum radius $R_{\rm max}$, we truncate the
$r^\eta$ function with an exponential of the form ${\rm e}^{-(r/R_{\rm
max})^6}$. In this case, the PDF of $r$ is described by,
\begin{equation}
f_R(r) = \dfrac{6}{R_{\rm max}^{\eta + 1}\, \Gamma\left(\dfrac{\eta + 1}{6}\right)} \, r^\eta\, {\rm e}^{-(r/R_{\rm max})^6}, ~~~  0<r<\infty.
\label{r_pdf2}
\end{equation}
The PDF $f_R(r)$ is shown in Fig.~\ref{fig:r_pdf2} for different values of
$\eta$. It can easily be shown that, for the case where $r$ has a minimum
radius $R_{\rm min}$, the $R_{\rm max}^{\eta+1}$ term in the denominator is
replaced by $R_{\rm max}^{\eta+1} - R_{\rm min}^{\eta+1}$ and the gamma
function is modified to $\Gamma\left({(\eta + 1)}/{6}, {R_{\rm min}^6}/{R_{\rm
max}^6} \right)$. For $R_{\rm min} \ll R_{\rm max}$, $\Gamma\left({(\eta +
1)}/{6}, {R_{\rm min}^6}/{R_{\rm max}^6} \right) \approx \Gamma\left({(\eta +
1)}/{6}, 0 \right)$ and $R_{\rm max}^{\eta+1} - R_{\rm min}^{\eta+1} \approx
R_{\rm max}^{\eta+1}$. Hence, we do not additionally invoke a $R_{\rm min}$. }

\item{{\bf Case 2:} It is plausible that the database of absorption line samples 
is incomplete in terms of the range of radii probed by the background quasars. 
To account for such a situation, we consider a simplified scenario where the 
radii at which the background quasars probe the absorber galaxies are uniformly 
distributed:
\begin{equation}
f_R(r) =
\begin{cases}
~\dfrac{1}{R_{\rm max} - R_{\rm min}}, & \text{for } R_{\rm min} \leq r \leq R_{\rm max},
\\
~0, & \text{otherwise. }
\end{cases}
\end{equation}

}
\end{itemize}

\begin{figure*}
\begin{tabular}{cc}
{\mbox{\includegraphics[width=8cm]{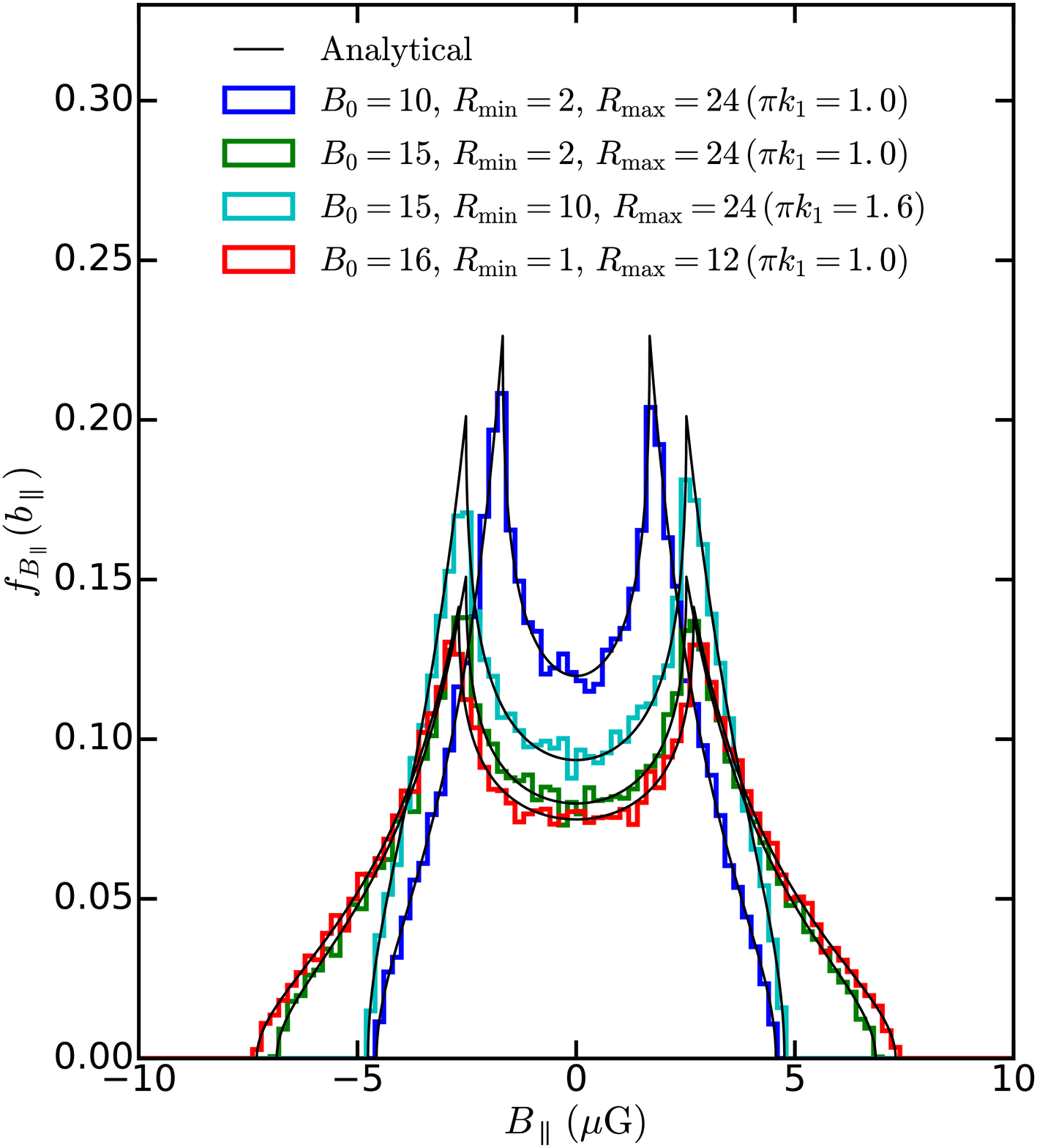}}} &
{\mbox{\includegraphics[width=8cm]{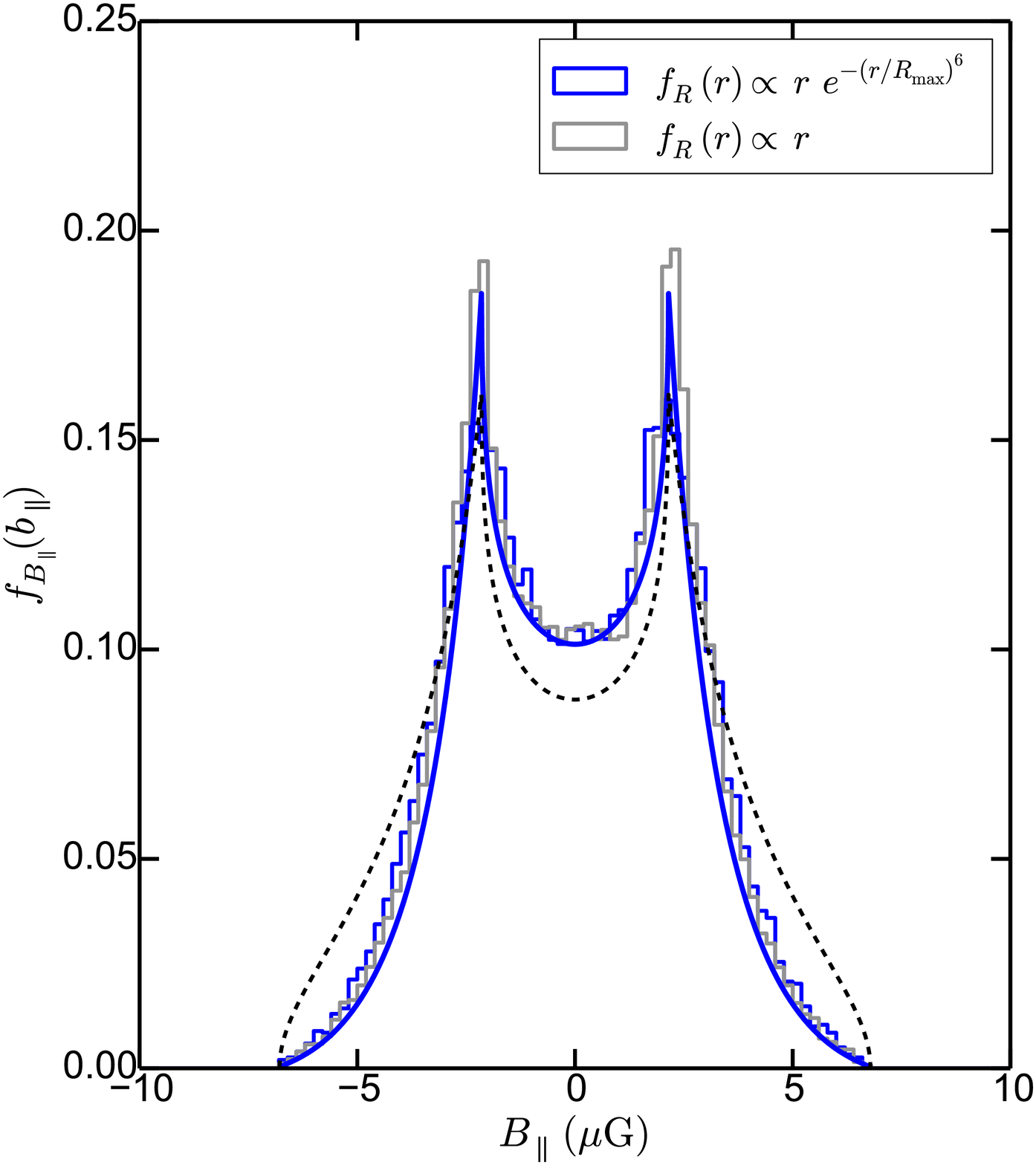}}} \\
\end{tabular}
\caption{
{\it Left:} Distribution of $B_\|$ for a galaxy inclined at $i = 30^\circ$ and
uniform distribution of radius (Case 2). The black curves are the analytical
PDF of a galaxy in Eq.~\eqref{eq:pdf_b}. {\it Right:} Distribution of $B_\|$
for the case when radii are distributed as per Eq.~\eqref{r_pdf2} with $\eta=1$
(Case 1) is shown as the blue histogram. The parameters are listed in
Table~\ref{variables}. The blue curve is the approximated analytical PDF given
in Eq.~\eqref{eq:pdf_b1} (see text for details). For comparison, the analytical
distribution of $B_\|$ for uniform distribution of radii with the same
parameters is shown as the dashed line. The grey histogram is the distribution
of $B_\|$ for the case when the distribution of impact radii given by
Eq.~\eqref{r_pdf2} is approximated as Eq.~\eqref{r_pdf2_approx}.}
\label{distr_bparallel}
\end{figure*}

\section{Results} \label{sec:result}

\subsection{PDF of $B_{\parallel}$ and RM for a single galaxy} \label{pdf_single}

We first consider the case of a single galaxy with inclination angle $i$, assuming
Case~2 above for the distribution of the radii of the intersecting lines of sight.
Applying standard laws pertaining to the distribution functions of random variables 
\citep[e.g.][]{svesh68} and to the product of two continuous random functions
\citep[e.g.][]{rohat76, glen04} to Eq.~\eqref{eq_bparallel} (also given by
Eq.~\eqref{eq_bparallel_app} under our assumptions), we obtain the PDF of
$B_\parallel$ ($f_{B_\parallel} (b_\parallel)$) for this situation (see
Appendix~\ref{pdfcalc}) as, 
\begin{equation}
f_{B_\parallel}(b_\parallel) =
\begin{cases}
~\dfrac{k_1}{|b_\parallel|}\left[ \arcsin \left(\dfrac{|b_\parallel|}{a} \right)-  \arcsin\left(\dfrac{|b_\parallel|}{b}\right) \right], &  -a\le b_\parallel \le a ,
\\
~\dfrac{k_1}{|b_\parallel|}\arccos\left(\dfrac{|b_\parallel|}{b} \right), & b_\parallel \in [-b,-a) \cup (a,b]. 
\\
\end{cases}
\label{eq:pdf_b}
\end{equation}
Here, $k_1=r_B/[\upi(R_{\rm max} - R_{\rm min})]$, $a=B_0 \sin i\, {\rm
e}^{-R_{\rm max}/r_B}$ and $b=B_0 \sin i\, {\rm e}^{-R_{\rm min}/r_B}$. In left
panel of Fig.~\ref{distr_bparallel}, we compare the analytical PDF of
$B_\parallel$ to that of simulated distributions to verify the results of the
calculation. The simulated distributions were carried out for 10,000 lines
of sight, each passing through a galaxy with $B_\|$ computed from
Eq.~\ref{eq_bparallel_app}. The random variables $r$ and $\theta$ were drawn
from the distributions described in Section~\ref{rand_distr}.

It is evident from Fig.~\ref{distr_bparallel} that the PDF of $B_\|$ for one
galaxy has characteristic features that can be directly used to estimate the
strength of the large-scale magnetic field, $B_0$. The location of the cusp in
the PDF at $a$ is the magnitude of the magnetic field vector projected along
the line of sight at the outer edge, while the location of the truncation of
the PDF, $b$, is the same but for the inner radius of the ionized disc.
$(\upi\,k_1)^{-1}$ represents the number of radial scale lengths of the
magnetic field across the ionized disc. For a galaxy with known values of
$R_{\rm min}$ and $R_{\rm max}$, the parameters $a$ and $b$ can be used to
estimate $B_0$ and $r_B$. It can be easily shown that,
\begin{equation}
r_B = \frac{R_{\rm max} - R_{\rm min}}{\ln(b/a)}.
\end{equation}
Having determined $r_B$, $B_0 \,\sin i$ can be directly evaluated.

For the case when the radii of the intersecting lines of sight are distributed as
in Case~1, with $\eta=1$, the PDF of $B_\|$ can be approximated as:
\begin{equation}
f_{B_\parallel}(b_\parallel) \approx 
\begin{cases}
~\dfrac{k_2}{|b_\||}\left(\dfrac{R_{\rm max}}{r_B}\right)\left[ \arcsin \left(\dfrac{|b_\||}{a} \right)-  \arcsin\left(\dfrac{|b_\||}{b}\right) \right], &~\\
~~~~~~~~~~~~~~~~~~~~~~~~~~~~~~~~~~~~~ -a \le b_\| \le a, &\\
~\dfrac{k_2}{|b_\||}\ln\left(\dfrac{B_0\,\sin i}{|b_\||}\right)\left[\arccos\left(\dfrac{|b_\||}{b} \right)\right], &~\\
~~~~~~~~~~~~~~~~~~~~~~~~~~~~~\forall\,\, b_\| \in [-b,-a) \cup (a,b].&
\end{cases}
\label{eq:pdf_b1}
\end{equation}
Here, $k_2 = r_B/(\upi \, R_{\rm max})$. The right panel of
Fig.~\ref{distr_bparallel} shows the simulated distribution of $B_\|$ (again
assuming Case~1 for the distribution of impact radii), and compare this to the
analytical results. The location of the cusps in Fig.~\ref{distr_bparallel}
(left panel) are the same as for Case~2. However, the distributions are
slightly wider between the cusps due to the exponential turn-over of the radii
distribution in Case~1, while the wings of the distribution of $B_\|$ are
narrower than those for the case when the impact radii are uniformly
distributed.

The approximate analytical form of the PDF given by Eq.~\eqref{eq:pdf_b1} was
derived by ignoring the exponential cut-off and using a radius range of
$0\text{--} {\rm R_{max}}$, i.e., a sharp cut-off at $\rm R_{max}$. In this
approximation, the distribution of $r$ is, 
\begin{equation}
f_R(r) = \frac{\eta + 1}{R_{\rm max}^{\eta + 1}} \, r^\eta, ~~~  0<r<R_{\rm max} \; .
\label{r_pdf2_approx}
\end{equation}
For comparison, the simulated distribution of $B_\|$ for the above distribution
of $r$ is shown as the grey histogram in Fig.~\ref{distr_bparallel} (right
panel). The grey and blue histograms show fairly good agreement. Further, in
Eq.~\eqref{eq:pdf_b1}, we have neglected the imaginary terms arising while
performing the integrals in Eq.~\eqref{eq:pdf_2func}. These approximations
cause the analytical function in Eq.~\eqref{eq:pdf_b1} to underestimate the
true distribution of $B_\|$ by $\lesssim10$ per cent, which results in a
narrowing of the wings of the analytical function by $\lesssim5$ per cent.

\begin{figure}
\begin{centering}
\begin{tabular}{c}
{\mbox{\includegraphics[width=8cm]{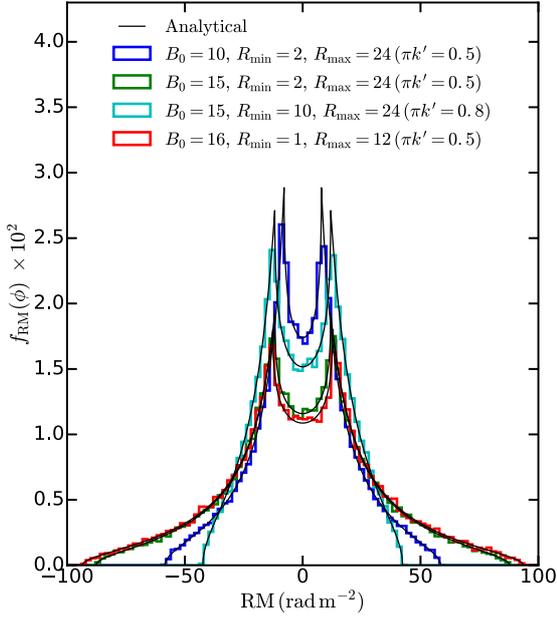}}} \\
\end{tabular}
\end{centering}
\caption{Distribution of RM for a galaxy inclined at $i = 30^\circ$
for uniform distribution of radius. The black lines are the analytical
PDF of RM as given in Eq.~\eqref{eq:pdf_rm}.}
\label{distr_rm}
\end{figure}

Under our assumptions for the magneto-ionic medium, the RM along each line of
sight through a galaxy (given by Eq.~\eqref{eq_rm}) has a form similar to
Eq.~\eqref{eq_bparallel_app}. Therefore, the PDF of RM for Case 2 can be
obtained by simply replacing $r_B$ by $r_0^\prime$ and $B_0\, \sin i$ by
$0.81\,B_0\, n_0\,h_{\rm ion}\, \tan i$ in Eq.~\eqref{eq:pdf_b}. The
probability distribution function of RM, $f_{\rm RM}(\phi)$, is then given as,
\begin{equation} 
f_{\rm RM}(\phi) = 
\begin{cases}
~\dfrac{k^\prime}{|\phi|}\left[ \arcsin \left(\dfrac{|\phi|}{a^\prime} \right)-  \arcsin\left(\dfrac{|\phi|}{b^\prime}\right) \right], &  -a^\prime \le \phi \le a^\prime,
\\
~\dfrac{k^\prime}{|\phi|}\,\arccos \left(\dfrac{|\phi|}{b^\prime} \right), & \phi \in [-b^\prime,-a^\prime) \cup (a^\prime,b^\prime].
\\
\end{cases}
\label{eq:pdf_rm}
\end{equation}
Here, $k^\prime=r^\prime_0/[\upi(R_{\rm max} - R_{\rm min})]$. The parameters
$a^\prime$ and $b^\prime$ are given as,
\begin{equation}
\begin{split}
a^\prime & = 0.81\,B_0\,n_0\,h_{\rm ion} \tan i\, {\rm e}^{-R_{\rm max}/r^\prime_0}\\
b^\prime & = 0.81\,B_0\,n_0\,h_{\rm ion} \tan i\, {\rm e}^{-R_{\rm min}/r^\prime_0}.
\end{split}
\label{eq:abprime}
\end{equation}
The RM distribution for one galaxy inclined at $i = 30^\circ$ with different
values for the fixed parameters is shown in Fig.~\ref{distr_rm}.

\begin{figure*}
\begin{tabular}{cc}
{\mbox{\includegraphics[width=8cm]{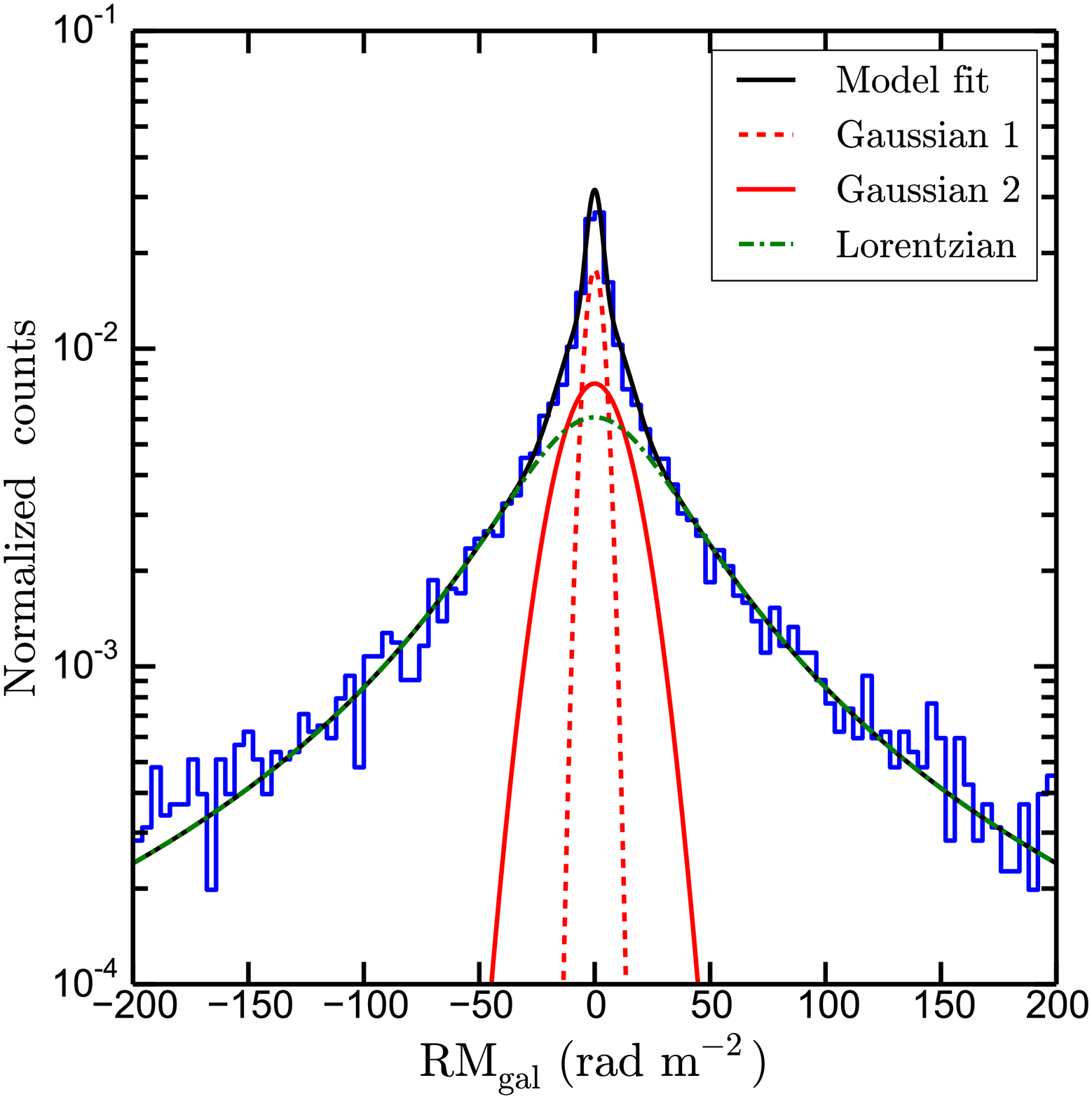}}} &
{\mbox{\includegraphics[width=8cm]{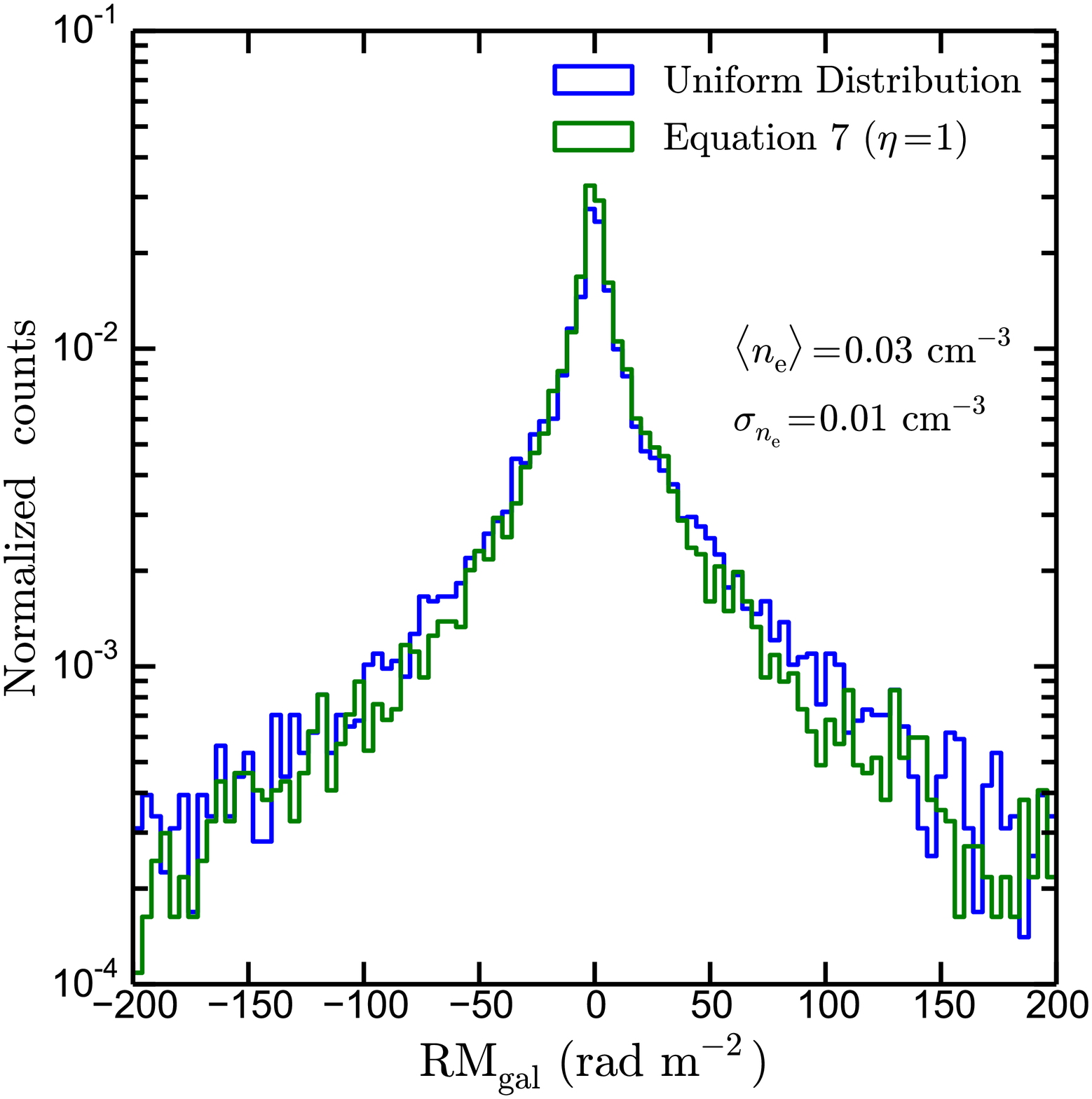}}} \\
\end{tabular}
\caption{{\it Left}: Distribution of $\rmgal$ for a sample of 10,000 galaxies
with uniform distribution of $i$, $\theta$ and $r$ as described in
Section~\ref{rand_distr} and, $B_0$ and $n_0$ of the sample have Gaussian
distributions. The parameters used are listed in Table~\ref{variables}. The
distribution is modelled as a sum of one Lorentzian and two Gaussian components
and is shown as the black solid line. The individual components are shown as
green dot-dashed, red solid and dashed lines. {\it Right}: Comparison of
distribution of $\rmgal$ when the radius at which the lines of sight intersect
the galaxies are distributed uniformly same as left-hand panel (blue histogram)
and for the case when they are distributed as in Eq.~\eqref{r_pdf2} with
$\eta=1$ (green histogram).}
\label{distr_rm_full}
\end{figure*}

The probability distribution of $B_\parallel$, and therefore RM, over all
azimuthal angles, does not depend on the pitch angle $p$ of the magnetic field.
This is because of the $2\upi$-periodicity of $\cos (\theta - p)$. 
Therefore, it is not possible to determine the pitch angle of the large-scale
magnetic field from backlit experiments. However, within a sector of a galaxy,
$\theta_{\rm min} \le \theta \le \theta_{\rm max}$, where $\theta_{\rm max} -
\theta_{\rm min} < \upi$, the function ceases to be periodic. In this case, the
PDFs of $B_\parallel$ and RM within a segment depend on $p$; this situation is
discussed in Appendix~\ref{appendix_RM}.

\subsection{The PDF of RM for a sample of galaxies} \label{sec:RMsample}

To simulate a realistic RM distribution for a sample of 10,000 disc galaxies,
we computed the RM assuming a single line of sight per galaxy with random
distributions of the inclination angle, the azimuthal angle, and the radius of
intersection. These parameters have the probability distributions described in
Section~\ref{rand_distr}. We adopted a normal probability distribution for the
strength of the large-scale magnetic field $B_0$ of the sample of galaxies,
with sample mean $\langle B_0 \rangle$ and standard deviation
$\sigma_{B_0}$.\footnote{Note that $\sigma_{B_0}$ is not to be confused
with the strength or root-mean-square (rms) of the turbulent magnetic field.}
Similarly, the free electron density was assumed to have sample mean $\langle
n_{\rm e}\rangle$ and standard deviation $\sigma_{n_{\rm e}}$. Thus, the
distributions of $B_0$ and $n_{\rm e}$ in the sample of galaxies are given as:
\begin{equation}
\begin{split}
{\rm PDF}(B_0) & = \dfrac{1}{\sqrt{2\,\upi\,\sigma_{B_0}^2}}{\rm exp}\left[\dfrac{-\left(B_0 - \langle B_0 \rangle\right)^2}{2\sigma_{B_0}^2}\right]\\
{\rm PDF}(n_{\rm e}) & = \dfrac{1}{\sqrt{2\,\upi\,\sigma_{n_{\rm e}}^2}}{\rm exp}\left[\dfrac{-\left(n_{\rm e} - \langle n_{\rm e} \rangle\right)^2}{2\sigma_{n_{\rm e}}^2}\right].
\end{split}
\label{eq:b0_distr}
\end{equation}

The left panel of Fig.~\ref{distr_rm_full} shows the simulated distribution of
$\rmgal$ as the blue histogram for the case where the line of sight passes
through a sample of absorber systems with a uniform distribution of impact
radii (Case~2 of Section~\ref{sec:rpdf}). The distinctive features of the RM
PDF for a single galaxy described in Section~\ref{pdf_single} are washed out,
and we empirically model the PDF of $\rmgal$ as the sum of one Lorentzian and
two Gaussian functions:
\begin{eqnarray}
f_\rmgal(x) = a_{1}\left[ \frac{w_{1}^2}{w_{1}^2 + (x - m_{1})^2} \right] 
+ \,a_2 \, \exp\left[\dfrac{-(x-m_2)^2}{w_{2}}\right] \nonumber \\ 
+ \,a_3 \, \exp\left[\dfrac{-(x-m_3)^2}{w_{3}}\right]. 
\label{model_pdf}
\end{eqnarray}
Here, the parameters $w_1$, $w_2$ and $w_3$ are the characteristic widths of
the respective components; $a_1$, $a_2$ and $a_3$ are the amplitude
normalizations; and $m_1$, $m_2$ and $m_3$ are the mean values (which are close
to zero, as expected). In Eq.~\eqref{model_pdf}, $w_1$ is in
rad\,m$^{-2}$, and $w_2$ and $w_3$ are in (rad\,m$^{-2}$)$^2$. $w_3$ is the
wider of the two Gaussian components. The fitted PDF is shown as the solid
black line in Fig.~\ref{distr_rm_full}.

The choice of the functions that we used to empirically model the distribution
of $\rmgal$ is motivated by the shape of the PDF of the RM for a single galaxy.
The Lorentzian is chosen to capture the wings in the distribution of RM as seen
in Fig.~\ref{distr_rm}. Since the cusps at $\pm a^\prime$ for one galaxy
depend on $B_0 \, n_0$, the distribution should be widened approximately by a
Gaussian function if $B_0$ and $n_0$ have a normal distribution for the galaxy
sample. This motivates the choice of the wider of the two Gaussian functions.
Finally, the sharp peak at $\rmgal = 0$~rad\,m$^{-2}$ is a manifestation of the
distribution of the $\sin i$ function for a uniform distribution of $i$ in the
range 0 to $\upi/2$, and the narrower Gaussian accounts for this.

The right panel of Fig.~\ref{distr_rm_full} compares the distributions of
$\rmgal$ in Case~2 and Case~1 (where we again assume $\eta = 1$ for the radius
distribution of the sightlines; see Section~\ref{sec:rpdf}), keeping all the
other variables the same as in the left panel. The two distributions have
negligible overall differences, except that the width of the distribution for
Case~1 is marginally smaller than that for Case~2. This does not significantly
affect the result of the fit using Eq.~\eqref{model_pdf}: the fitted
parameters $w_2$ and $w_3$ are consistent within the errors, with $w_1 \approx
20$ per cent lower for Case~1 than in Case~2. This is also true for other
choices of $\eta$, e.g. $\eta = 0.5$ or $\eta = 2$. We therefore use Case~2 for
the radius distribution of the sightlines in the rest of the paper.

\subsection{Model PDF and physical parameters} \label{sec:modelPDF}

\begin{figure*}
\begin{tabular}{cc}
{\mbox{\includegraphics[width=8cm]{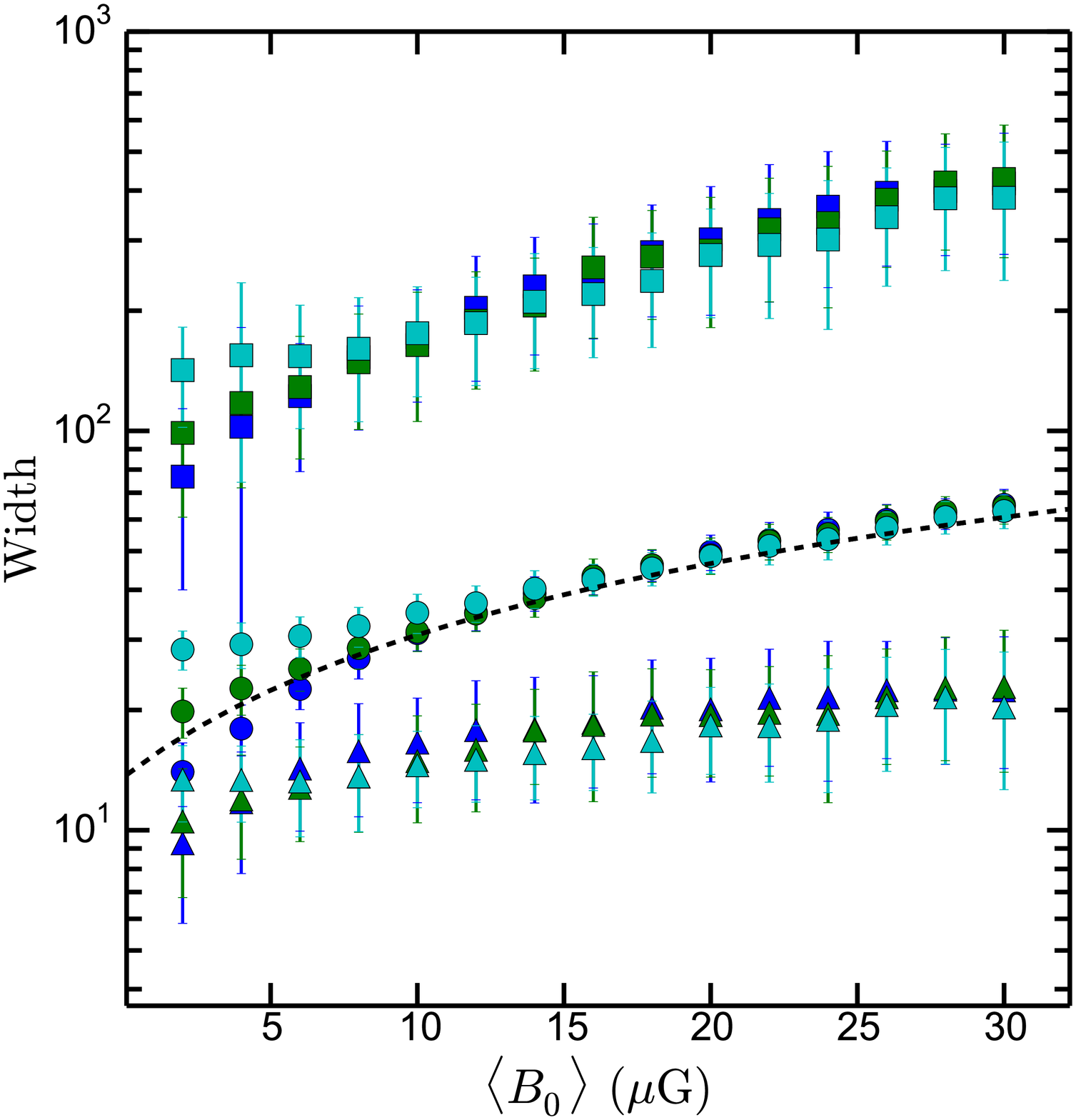}}} &
{\mbox{\includegraphics[width=8cm]{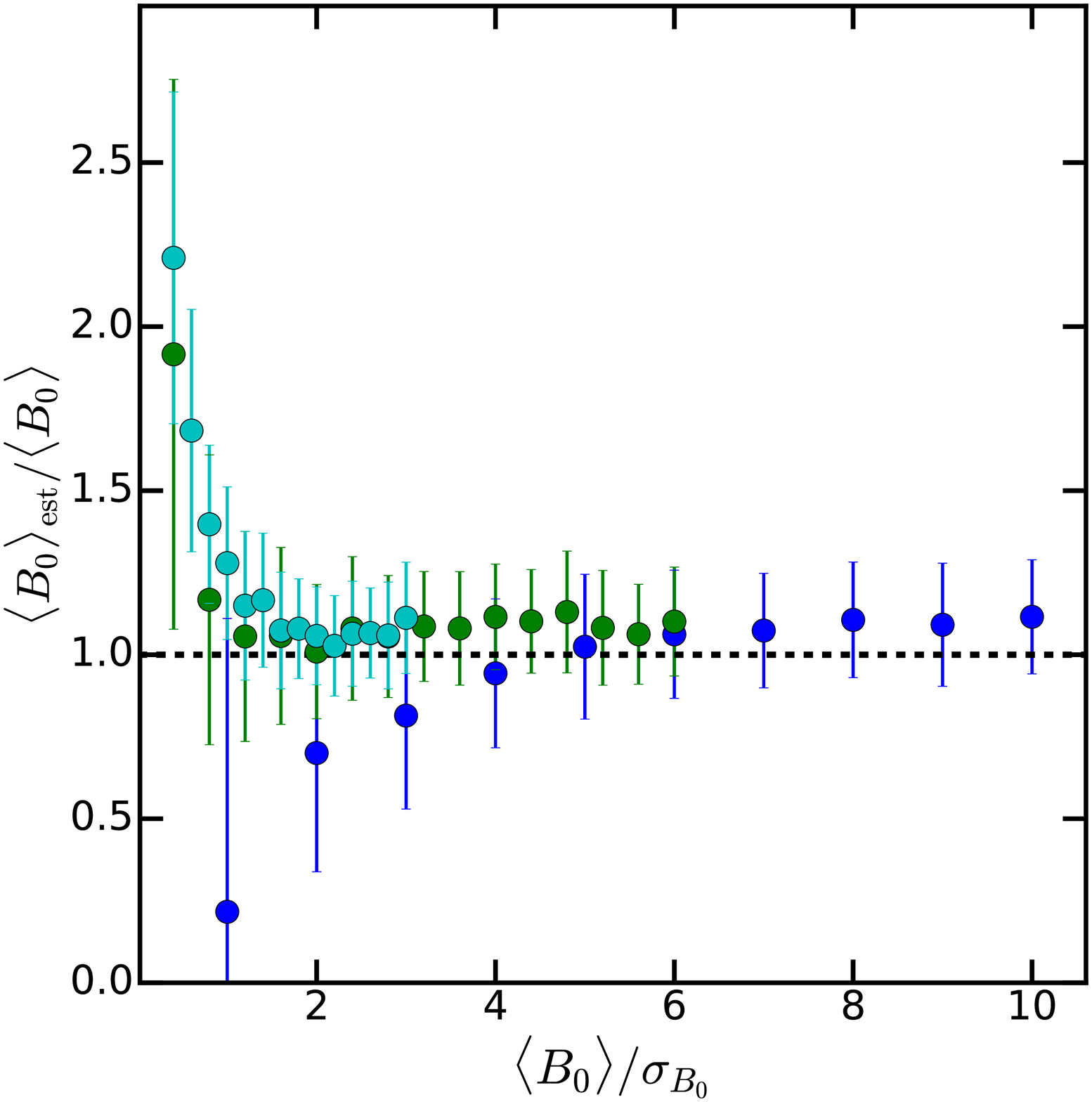}}} \\
\end{tabular}
\caption{{\it Left}: Variation of the widths of the modelled components with
$\langle B_0 \rangle$. Circles, triangles and squares are for $w_1$, $w_2$ and
$w_3$, the widths of the Lorentzian and the two Gaussian components,
respectively. $w_1$ is in rad\,m$^{-2}$ and, $w_2$ and $w_3$ are in
(rad\,m$^{-2}$)$^2$. The different colours are for different values of
$\sigma_{B_0}$ (blue=$2\,\umu$G, green=$5\,\umu$G, cyan=$10\,\umu$G). The
dashed line is the empirical fit to $w_1$ as a function of $\langle B_0
\rangle$ given by Eq.~\eqref{wlor_model}. {\it Right}: Variation of
$\langle B_0 \rangle_{\rm est}/\langle B_0\rangle$ as a function of $\langle
B_0\rangle/\sigma_{B_0}$. Here, $\langle B_0\rangle_{\rm est}$ is the estimated
$\langle B_0\rangle$ using Eq.~\eqref{wlor_model}. The dashed line is for 
$\langle B_0\rangle_{\rm est}/\langle B_0\rangle = 1$. The different colours 
have the same meaning as in the left-hand panel.}
\label{width}
\end{figure*}

For a single galaxy, the width of the PDF of $\rmgal$ depends on two parameters
$a^\prime$ and $b^\prime$, both of which directly depend on $B_0$ (see
Eq.~\eqref{eq:abprime}). We therefore expect the width of the Lorentzian
component of the PDF of $\rmgal$ for a sample of galaxies to depend on $\langle
B_0 \rangle$. In the left-hand panel of Fig.~\ref{width}, we show the variation
of the widths of the different fitted components as a function of $\langle B_0
\rangle$, for values of $\sigma_{B_0}$ in the range $1-10 \umu$G.\footnote{To
compute the errors in the fitted parameters, we performed Monte-Carlo
simulations with 100 realizations for the random variables. For each
realization, the distribution of $\rmgal$ was fitted with
Eq.~\eqref{model_pdf}, and we use the standard deviation of the derived set of
parameters to be their error.} We find that $w_2$ varies only weakly with
$\langle B_0 \rangle$, consistent with being constant and with values of $15 -
20$. However, both $w_1$ and $w_3$ shows significant variation with $\langle
B_0 \rangle$,  with $w_1$ showing the strongest dependence on $\langle B_0
\rangle$ (after taking the errors into account). It should be noted that
$\langle B_0 \rangle$ scales with $\langle n_{\rm e} \rangle$ and $h_{\rm ion}$
as,
\begin{equation}
\left(\dfrac{\langle B_0 \rangle}{\rm \umu G}\right) \equiv \langle B_0 \rangle \left(\dfrac{\langle n_{\rm e}\rangle}{\rm 0.03\,cm^{-3}}\right)\left(\dfrac{h_{\rm ion}}{\rm 500\,pc}\right)
\end{equation}

It is apparent from Fig.~\ref{width} (left-hand panel) that for all values of
$\sigma_{B_0}$, $w_1(\langle B_0 \rangle)$ converges asymptotically when
$\langle B_0 \rangle \gtrsim \sigma_{B_0}$. We empirically model the asymptotic
dependence of $w_1$ (in $\rm rad\,m^{-2}$) on $\langle B_0 \rangle$ (in
$\umu$G) as,
\begin{equation}
w_1(\langle B_0 \rangle) = p_0 + p_1\,\langle B_0 \rangle + p_2\,\langle B_0 \rangle^2.
\label{wlor_model}
\end{equation}
The best-fit values of the parameters $p_0$, $p_1$ and $p_2$ are found to be
13.6, 1.8 and $-0.0076$, respectively. The best fit is shown as the dashed
line in Fig.~\ref{width} (left). Note that we have fixed $w_2 = 18$ (due to its
weak dependence on $\langle B_0 \rangle$) when determining the above empirical
dependence $w_1 (\langle B_0 \rangle)$. The error due to this assumption on
$w_1(\langle B_0 \rangle)$ is $\lesssim 5$ per cent (compared to the case when
$w_2$ is left as a free parameter). The net error on the estimated $\langle B_0
\rangle$ is also $\lesssim 5$ per cent.

When $\langle B_0 \rangle \lesssim \sigma_{B_0}$, $w_1$ changes marginally or
remains roughly constant at a value that depends on $\sigma_{B_0}$. In the
right-hand panel of Fig.~\ref{width}, we show the variation of the ratio of
$\langle B_0 \rangle$ estimated using Eq.~\eqref{wlor_model}, $\langle
B_0\rangle_{\rm est}$, to that of the true $\langle B_0\rangle$ in our
simulations as a function of $\langle B_0\rangle/\sigma_{B_0}$. For $\langle
B_0 \rangle/\sigma_{B_0} \gtrsim 1$, $\langle B_0\rangle_{\rm est}$ agrees well
with the true $\langle B_0\rangle$, except for the blue points for which
$\sigma_{B_0} = 2\,\umu$G. This is because the parameters $p_0$, $p_1$ and
$p_2$ in Eq.~\eqref{wlor_model} are estimated in the regime where $w_1$
converges asymptotically in Fig.~\ref{width} (left), i.e., for $\langle B_0
\rangle > 10\,\umu$G. This causes $\langle B_0\rangle_{\rm est}$ to deviate
significantly from $\langle B_0 \rangle$ for the case where $\sigma_{B_0} =
2\,\umu$G up to $\langle B_0\rangle/\sigma_{B_0} \approx 4$. However, we
believe that it is unlikely that a sample of galaxies would have $\sigma_{\rm
B_0} = 2\,\umu$G \citep[e.g.,][]{fletc10}, and have hence not extended the fit
to account for the above deviation. Thus, in the case where the dispersion of
the magnetic field strengths of the galaxies in the sample is larger than their
mean field strength, determining $\langle B_0 \rangle$ will be difficult. This
demonstrates the importance of careful selection of the absorber sample: a
large variation of galaxy types in the sample is likely to lead to a large
$\sigma_{B_0}$, which would make it difficult to interpret the results.

Some previous works on this topic, using strong Mg{\sc ii} absorbers as
the intervening galaxies, attributed the dispersion in the distribution of RM
to predominantly arise from turbulent magnetic fields, typically with RM
dispersion $\approx150~\rm rad~m^{-2}$ \citep[see, e.g.,][]{berne08, berne13}.
It is interesting to note from Fig.~\ref{width} (left-hand panel), for a sample
of galaxies, the mean large-scale field strength, $\langle B_0\rangle$, and its
dispersion within the sample, $\sigma_{B_0}$, could also give rise to a
significant spread in the RM distribution, of a magnitude comparable to that
found in the previous studies. Also, the PDF of RM in the presence of
large-scale magnetic fields in galaxies deviates from that of a Gaussian. Large
sample size in future backlit-experiments could be used to study this deviation
from Gaussianity and distinguish between dominating large-scale fields or
dominating turbulent fields.

We note that the empirical model given in Eq.~\eqref{wlor_model} does depend on
the assumed values for the fixed parameters. In
Appendix~\ref{appendix_variation}, we discuss in detail how different choices
of the fixed variables $r_B$, $r_{\rm e}$, $R_{\rm min}$, and $R_{\rm max}$,
and the assumed distribution of impact radii, i.e., Cases~1 and 2, affect the
empirical form of Eq.~\eqref{wlor_model}, and hence inferring $\langle B_0
\rangle$. In the absence of additional constraints on the above free
parameters, and considering the typical ranges of values of these parameters,
the true value of $\langle B_0 \rangle$ lies within $\approx50$ per cent
of the value of $\langle B_0 \rangle$ estimated using Eq.~\eqref{wlor_model}.

\section{Sample selection criteria for observations} \label{sec:sample}

In the light of our results on the statistical properties of the Faraday
rotation measure for absorption-selected galaxies, we discuss in this section
the selection criteria of the control and the target samples that are important
for a better estimation of the large-scale magnetic field properties in the
intervening galaxies.

\subsection{Redshift coverage of the quasars}

As has been pointed out in Section~\ref{sec-redshift}, differences in the
redshift distribution of the background quasars in the target and the control
samples can lead to biases when inferring magnetic fields in the intervening
galaxies. In fact, there are clear differences in the redshift distributions of
the target and control samples that have been used in earlier studies in the
literature \citep[e.g.][]{farne14, kim16}. The undesirable bias introduced due
to the mismatch in quasar redshifts can be avoided by ensuring the same redshift
distribution for the quasars in the target and control samples.

Further, a random selection of quasars from the currently available optical
catalogs (e.g. as shown in Fig.~\ref{zdistr}) will introduce artefacts in the
observed PDFs, making it difficult to separate the PDF of the RM contributed by
the intervening galaxies from that of the quasars. It is therefore desirable to
use a sample that is distributed uniformly in redshift, i.e., with roughly
equal number of sources per redshift bin. This would allow the contribution of
the $(1+z)^{-2}$ factors to be analytically filtered out.

\subsection{The background quasars}

Relating the observed PDF of RM to parameters describing the large-scale
magnetic field can be greatly simplified if the RM produced by the large-scale
fields is unaffected by the RM contributed by turbulent fields in the
intervening galaxies and the background polarized quasar remain unresolved
within the telescope beam. The former can be achieved when the projected
linear extent of the polarized emission from the background quasar encompass
several turbulent cells, with the RM contributed by the large-scale field not
varying significantly ($\lesssim 10$ per cent) within the illumination beam.  

The typical sizes of turbulent cells in nearby galaxies are $\approx
50\text{--}100$~pc \citep[e.g.][]{ohno93, lazar00, beck16}. For
intervening galaxies at $z \gtrsim 1$, the angular diameter distance to the
foreground absorber and the background quasar is approximately the same.
Assuming that the turbulent cell sizes in high-$z$ galaxies are similar to that
of the nearby galaxies, we thus require that the spatial extent of the
polarized emission in quasars should be $\approx 100\text{--}250$~pc (see
Section~\ref{sec:turb_field}), i.e.  $\approx 25\text{--}50$~mas for quasars at
$z \gtrsim 1$. The flat- or inverted-spectrum cores of radio quasars or BL~Lac
objects are known to emit polarized synchrotron radiation on such scales
\citep[e.g.][]{bondi96book, zensus97, aller99, lister01, bondi04, lister16},
and would make good targets for such an RM study. Steep-spectrum polarized
sources, e.g. the lobes of radio galaxies, that remain unresolved with the
currently available radio telescopes, should be avoided as targets as the
projected extended nature of the emission implies that variations in the RM
from the large-scale field across the emission region are likely to be
significant. Medium-resolution interferometers such as the SKA-MID would also
be interesting for such studies, to obtain a synthesized beam of $\approx
50\text{--}100$~mas at GHz-frequencies, which would allow one to resolve out
any extended radio emission and study the RM towards the radio core.

\subsubsection{Contribution of turbulent fields to $\rmgal$} \label{sec:turb_field}

We argued qualitatively above that the RM contributed by turbulent fields
along each sightline, with rms $b_\|$, will be negligible compared to that
produced by large-scale fields, $B_\|$, for illumination beams of size $\approx
100\text{--}250$~pc at the absorber redshift. Here, we derive this size
quantitatively.

The RM produced by $b_\|$ over a three-dimensional volume probed by each
line of sight has zero mean, and a dispersion ($\sigma_{\rm RM, 3D}$) given by
\begin{equation}
\sigma_{\rm RM,3D} = 0.81\,\langle n_{\rm e} \rangle\,b_\|\,\sqrt{l_0\,L}~{\rm rad\, m^{-2}}.
\label{eqn:sigmaRM}
\end{equation}
Here, $l_0$ is the size of the turbulent cells in parsecs, i.e., the
correlation scale of the product $n_{\rm e}\,b_\|$, while $L = h_{\rm
ion}/\cos\,i$ is the path length through the ionized medium. The dispersion of
RM in the plane of the sky ($\sigma_{\rm RM, 2D}$), averaged over a beam of
spatial size $W$ (in pc), is given by $\sigma_{\rm RM, 2D}\approx \sigma_{\rm
RM, 3D}/\sqrt{N}$ \citep[e.g.][]{fletc11}, where $N\approx(W/l_0)^2$ is the
number of turbulent cells within the beam area. The effect of turbulent fields
is negligible when $\sigma_{\rm RM, 2D}/|\rmgal| \ll 1$. To satisfy this
condition, by combining Eqs.~\eqref{eq_rm} and \eqref{eqn:sigmaRM}, we have
\begin{equation}
W \gg \dfrac{b_\|}{|\langle B_\|\rangle|}\,l_0\,\sqrt{\dfrac{l_0\,\cos i}{h_{\rm ion}}}.
\end{equation}
For a sample of disc galaxies with uniform distribution of inclination angles,
$i$ between 0 and $\upi/2$, the mean value of $\sqrt{\cos i}$ is,
\begin{equation}
\begin{split}
\langle \sqrt{\cos i}\rangle  & = \dfrac{2}{\upi}\int_{0}^{\upi/2} \sqrt{\cos i}\,\mathrm{d} i\\
 & = \left(\dfrac{2}{\upi}\right)^{3/2}\Gamma^2\left(\dfrac{3}{4}\right) \approx 0.76\\
\end{split}
\end{equation}
For typical parameter values, $b_\|/|\langle B_\|\rangle| \simeq 2$, $l_0
\approx 50\text{--}100$~pc and $h_{\rm ion} \approx 500$~pc, we find $W \gg
25\text{--}50$ pc. An illumination beam of size $\approx 100\text{--}250$~pc
clearly satisfies this condition. For a larger illumination beam, the variation
of the large-scale field within the beam will give rise to $>10$ per cent
variation of RM. Hence, to avoid significant contamination from RM arising from
the turbulent fields, the background quasars should have polarized emission
with a spatial extent of $\approx 100\text{--}250$~pc.

\subsection{Intervening objects}

Lastly, as noted in Section~\ref{sec:modelPDF}, the intervening galaxy sample
needs to be carefully chosen so that one does not obtain a large $\sigma_{B_0}$ 
from different populations of galaxies in the sample (as seen in Local Volume
galaxies). In order to probe magnetic field evolution in the framework of dynamo 
action, it is best to use sightlines passing through the discs of galaxies. It 
is still not clear what galaxy type or ISM conditions are probed by the various 
absorption lines. In the CGM, mostly probed by systems 
that show C{\sc iv} absorption and/or Lyman-limit systems (LLSs), the magnetic 
field could be of non-dynamo origin, e.g., the primordial seed field amplified 
by magneto-rotational instability (MRI), tidal interactions, and/or gas inflow 
or outflow.

In the local Universe, H{\sc i} column densities within the $R_{25}$ radius
have values $N_{\rm HI} \gtrsim 2\times 10^{20}$ cm$^{-2}$. This threshold
value for $N_{\rm HI}$ is hence a good indicator whether the absorbing gas lies
within a galaxy \citep[e.g.][]{wolfe05}. However, the galaxy type, and even
whether the absorber is a disc galaxy or a dwarf, cannot be directly inferred
from the H{\sc i} column density. Additional information on the gas phase
metallicities ([M/H]; e.g. \citealt{pettini94, rafelski12}), kinematics, and/or
gas temperatures is essential to glean information on the nature of the
absorber host. Here, we discuss briefly the properties of two main absorber
classes that are used as tracers of galaxies at high redshifts, namely, Mg{\sc
ii} absorbers and DLAs.

{\it Intervening Mg{\sc ii} absorbers:} Strong Mg{\sc ii}$\lambda2796\AA$
absorbers at intermediate redshifts have long been associated with the presence
of galaxies close to the quasar sightline \citep[e.g.][]{bergeron91}. Today,
more than 40,000 strong Mg{\sc ii}$\lambda2796\AA$ absorbers are known, mostly at
$0.4 < z < 2.3$, from studies based on the SDSS \citep[e.g.][]{nesto05,zhu13}.
Follow-up Lyman-$\alpha$ absorption studies of Mg{\sc ii} absorbers have shown
that damped Lyman-$\alpha$ absorption (which is expected to arise from
sightlines intersecting galaxy discs) is only seen for Mg{\sc ii}$\lambda2796\AA$
rest equivalent widths $W_0^{\lambda2796} \gtrsim 0.5$~\AA\ \citep{rao06}.
Intermediate-strength Mg{\sc ii} absorbers (with $W_0^{\lambda2796} \approx
0.3\text{--}0.5$~\AA) are expected to trace the outskirts of galaxies or high
velocity clouds \citep{nesto05,nesto07}. It should be emphasized, however, that
a high rest equivalent width $W_0^{\lambda2796} > 0.5$~\AA\ does not guarantee
the presence of damped Lyman-$\alpha$ absorption. \citet{rao06} found that only
$\approx 35$ per cent of Mg{\sc ii} absorbers with $W_0^{\lambda2796} >
0.5$~\AA\ have $N_{\rm HI} \geq 2\times 10^{20}$~cm$^{-2}$. Two-thirds of
strong Mg{\sc ii} absorber sightlines thus appear to trace clouds in the
circumgalactic medium, galactic superwinds, etc., rather than galaxy discs
\citep[e.g.][]{bond01,zibet07,neele16}. The magnetic fields in these regions
could be of completely different origin from those in the discs, and, at least
as important, is likely to have diverse origins as mentioned above. A target
sample of galaxies chosen only by the presence of strong Mg{\sc ii} absorption
is thus unlikely to be suitable as a probe of dynamo action in high-$z$
galaxies. We note that most backlit-experiment studies of magnetic fields in
high-$z$ galaxies have so far been based on such strong Mg{\sc ii} absorbers
\citep[e.g.][]{berne08,berne13,joshi13,farne14,kim16}. But, they don't
necessarily treat them as dynamo origin.

{\it Intervening damped Lyman-$\alpha$ absorbers:} DLAs have the highest H{\sc
i} column densities of all absorbers, $N_{\rm HI} \geq 2 \times
10^{20}$~cm$^{-2}$ \citep{wolfe05}, column densities that only arise in
galaxies in the local Universe. The presence of damped Lyman-$\alpha$
absorption in a quasar spectrum has hence long been used to infer the presence
of a galaxy along the quasar sightline \citep{wolfe86}. Unfortunately, such
high H{\sc i} column densities can arise in a wide range of galaxy types,
ranging from massive disc galaxies through small dwarfs. Additional selection
criteria must hence be imposed on DLA samples to reduce the heterogeneity of
the underlying population. 

Unfortunately, the presence of the bright quasar close to the DLA host galaxy
has meant that it has been very difficult to use optical imaging studies to
characterize the nature of DLA host galaxies. Only around a dozen DLAs at $z
\gtrsim 2$ have identified host galaxies \citep[e.g.][]{krogager12}, and even
this small sample is heavily biased towards high-metallicity absorbers.
Recently, imaging of a sample of foreground DLAs at wavelengths shortward of
the Lyman break produced by higher-redshift absorbers on the same sightline has
shown that DLAs at $ z\approx 2.7$ appear to typically have very low star
formation rates, $\lesssim 0.5\, \rm M_\odot\,yr^{-1}$ \citep{fumagalli14b,
fumagalli15}. This suggests that the DLA host galaxy population at $z \gtrsim
2$ may be dominated by dwarf galaxies. Unfortunately, dwarf galaxies can
only host weak mean-field dynamos and hence should be separated from the target
sample when studying the evolution of the large-scale magnetic fields in disc
galaxies.

One hence needs a spectroscopic indicator of the nature of the host galaxy,
that might allow us to select disc galaxies out of the general DLA population.
There are two obvious possibilities: (1)~the H{\sc i} spin temperature, and
(2)~the gas-phase metallicity.

The H{\sc i} spin temperature ($T_{\rm s}$) provides insights on the
temperature distribution of neutral gas in the ISM
\citep{heiles03,roy13a,chengalur00}. This allows one to statistically
distinguish between large disc galaxies (which contain significant amounts of
cold gas, and hence have a low $T_{\rm s}$) and dwarf galaxies (where the gas
is mostly warm, and which hence have a high $T_{\rm s}$).  In order to infer
$T_{\rm s}$, it is necessary to carry out redshifted H{\sc i} 21-cm absorption
studies of all the target absorption-selected galaxies. A combination of poor
low-frequency frequency coverage and radio frequency interference have meant
that there are only $\approx 50$ DLAs in the literature with searches for
redshifted H{\sc i} 21-cm absorption \citep[e.g.][]{wolfe79,
wolfe85,kanekar06,kanekar07,kanekar14}.

While H{\sc i} 21-cm absorption studies of large DLA samples are likely to be
possible in the future with new low-frequency telescopes
\citep[e.g.][]{kanekar14c}, especially towards the target absorbers of our
proposed RM studies (due to the compactness of radio structure of the
background quasars), we suggest that the more easily-measured gas-phase
metallicity would be a better tool to distinguish between disc and dwarfs
galaxies. Metallicity estimates are now available for more than 300 DLAs at a
wide range of redshifts, using tracers that are relatively insensitive to dust
depletion \citep[e.g.][]{pettini94,pettini97,pettini99,
prochaska03a,prochaska07,kulkarni05,akerman05,rafelski12}. Evidence has been
found for a mass-metallicity relation in DLAs \cite[e.g.][]{moller13,
neeleman13}, similar to that in emission-selected galaxies
\citep[e.g.][]{erb06}, indicating that high-metallicity absorbers are
statistically likely to trace the more-massive disc galaxies. 

High-metallicity DLAs at high redshifts have been found to show higher star
formation rates than the typical DLA population \citep[e.g.][]{fynbo11,
fynbo13, kroga13}. H{\sc i} 21-cm absorption studies of DLA samples have shown
that the gas phase metallicity is anti-correlated with the spin temperature,
with high-metallicity DLAs having lower spin temperatures \citep{kanekar01a,
kanekar09b}. Recently, \citet{neele17} obtained the first detection of [C{\sc
ii}] 158-$\umu$m emission in high-metallicity DLAs at $z \approx 4$, confirming
that the absorbers are massive star-forming galaxies with rotating discs. Thus,
choosing target DLAs based on a high metallicity ([M/H]~$\gtrsim -0.5$; e.g.
\citealt{rafelski12}), appears to be the best way to target disc galaxies at $z
\gtrsim 2$. One might also use comparisons of RM between DLA samples at
different mean metallicities (e.g. [M/H]~$\approx -0.5$, $\approx -1.5$ and
$\approx -2.5$) to study the cosmic evolution of magnetic fields in very
different galaxy environments. This could help to assess the importance and
efficacy of dynamo action as a function of galaxy mass.

Further, we note that it is straightforward to detect the presence of a
DLA along a quasar sightline with even low-resolution spectroscopy, with $R
\lesssim 2000$.  However, measuring DLA metallicities usually requires
follow-up spectroscopy with both high resolution ($R \gtrsim 10,000$) and high
sensitivity. This may prove difficult for very large target galaxy samples.
However, a relatively tight correlation has been detected between the rest
equivalent width of the Si{\sc ii}$\lambda$1526\AA\ line and the DLA
metallicity, and it should be possible to detect this Si{\sc ii} transition in
high-metallicity DLAs via low-resolution spectroscopy \citep{prochaska08}. One
might hence instead use the threshold $W_0^{\lambda 1526} \gtrsim 2$\AA\ based
on the detection spectra to identify high-metallicity DLAs in the sample.

\section{Future work} \label{sec:future}

Our calculations and results are based on the assumption that magnetic fields
in galaxies are confined to the disc with axisymmetric large-scale field. This
assumption is unlikely to be strictly correct. A detailed treatment of the
full three-dimensional magnetic field structure including the vertical magnetic
field, e.g. dipolar or quadrupolar configurations \citep{ruzma88, sokol90,
beck96}, will be considered in a forthcoming paper.

In addition to the dynamo-generated vertical magnetic fields, the effects of
galactic outflows due to star formation activity also should be considered.
Star formation drives magnetized outflows on galactic scales \citep{chyzy16,
damas16, wiene17} which can affect the form of our derived PDF of RM and
thereby affect the interpretations. A careful treatment of the magnetized
outflow is necessary in order to study galactic magnetic fields during the peak
epoch of cosmic star formation history.

The effects of the RM contributed by turbulent magnetic fields can be minimized
by choosing background quasars such that the three-dimensional volume probed by
the illumination beam through the intervening galaxy contains several turbulent
cells.  This simplifies our calculations of $\rmgal$ and helps to measure the
properties of the large-scale disc field. Turbulent fields are themselves
important for a complete understanding of the magneto-ionic medium in galaxies.
The energy density in the turbulent fields are significantly larger than that
in the large-scale field \citep{beck96,beck13book,beck16} and contributes
substantially to the pressure balance in the ISM \citep{beck07,basu13,beck16}.
Further, the ratio of the strength of the random field to that of the
large-scale field can provide additional constraints on the magnetic field
geometry \citep{shuku07,mao17}.  The effects of turbulent fields in the
intervening galaxies will appear as additional wavelength-dependent
depolarization of the linearly polarized signal of the background quasars in
the target sample, as compared to quasars in the control sample. This scenario
is being investigated in another paper of this series.

The success of these backlit-experiments depends crucially on how well one can
model the contribution of $\rmc$, and how accurately it can be isolated in
the observed distribution of $\rmt$ to obtain the distribution of $\rmgal$. In
order to achieve the best results we suggest the following modification to
Eq.~\eqref{rmobs}: 
\begin{equation} \label{rmobs1}
\begin{split}
{\rm RM_t^\prime} & =  (1 + \zgal)^2 \, \rmt \\
 & = \rmgal + {(1 + \zgal)^2}\, {\rm RM_{qso}^\prime}.
\end{split}
\end{equation}
This approach would minimize the redshift dilution of $\rmgal$. Similarly, one
could modify $\rmc$ to:
\begin{equation} \label{rmc1}
{\rm RM_{c}^\prime}  =  (1 + z_{\rm r})^2 \, \rmc.
\end{equation}
Here, $z_{\rm r}$ are values randomly drawn from the set of the sample values
of $\zgal$. This operation is possible for a large enough sample and it can be
shown that ${\rm PDF}[{(1 + \zgal)^2}\, {\rm RM_{qso}^\prime}] = {\rm PDF}[(1 +
z_{\rm r})^2 \, \rmc]$. However, the number of sightlines required depends on
the decomposition of PDFs (see Eq.~\eqref{rmobs1}). We are investigating the
points raised in this section in a series of forthcoming papers.

\section{Summary} \label{sec:summary}

To infer the properties of large-scale magnetic fields in high-$z$ disc
galaxies through backlit-experiments, we have derived analytical expressions
for the probability distributions of $B_\|$ and RM for a galaxy with an
axisymmetric spiral field geometry. We extend this to a sample of disc galaxies
and present an empirical model of the RM distribution when random lines of
sight are shot through galaxies with a random distribution of inclination
angles, impact radii, strengths of the large-scale magnetic field, and
free electron densities. Our study is applicable to future
backlit-experiments where the RM produced in the galaxies has been isolated
from other RM contributions along the line of sight, i.e., RM arising from the
background quasar, the IGM and the Milky Way. The main findings of this study
are:

\begin{enumerate}[(i)]

\item We demonstrate that the methods used in the literature to statistically
infer the large-scale magnetic field strength in samples of high-redshift
galaxies are likely to suffer from significant biases. The biases arise from
using the absolute value of RMs, the different redshift coverages of background
quasars in the target and control samples, and incomplete redshift coverage of
absorber systems in the currently available data.

\item Under our assumed model of the magneto-ionic medium for a galaxy, the
distributions of $B_\|$ and RM for a single galaxy observed along randomly
chosen lines of sight have distinctive features that are dependent on the
strength and radial scale-length of the large-scale magnetic field, and the
radius range probed by the sight lines.

\item The distributions of $B_\|$ and RM are independent of the pitch angle of
the magnetic field when the lines of sight through a galaxy sample all
azimuthal angles. However, within segments of azimuthal angles, the
distributions depend on the magnetic pitch angle.

\item The dispersion in the Faraday rotation arising only from large-scale
magnetic fields in intervening galaxies can give rise to a significant spread
in the distribution of the RM for a galaxy sample, of a magnitude comparable to
that found in previous studies.

\item For a sample of galaxies, where each galaxy has random $i$, $B_0$, $n_0$
and the lines of sight probe random distances from the centre, the distribution
of RM can be empirically modelled as a sum of one Lorentzian and two Gaussian
functions. 

\item The width of the Lorentzian function ($w_1$) gives an estimate of the
mean magnetic field strength, $\langle B_0 \rangle$, of the sample, provided
the sample dispersion, $\sigma_{B_0}$, is sufficiently small, i.e., $\langle
B_0 \rangle/\sigma_{B_0} \gtrsim 1$. This emphasizes the importance of
selecting absorber galaxies carefully so as to avoid heterogeneous galaxy
samples.

\item Choosing disc galaxies as a target sample is critical to study the
evolution of large-scale magnetic fields in the future work on disc dynamo
action. This selection criterion is best achieved when the galaxies are
identified as damped Lyman-$\alpha$ absorbers with high metallicities.
Comparing Faraday rotation in spiral and dwarf galaxies (the latter are not
expected to host mean-field dynamos) can shed light on conditions for, and the
consequences, of galactic dynamo action.

\end{enumerate}

\section*{Acknowledgements}

We thank the referee, Prof. Lawrence Rudnick, for critical and insightful
comments which have improved the presentation of the paper. We thank Luiz F.
S. Rodrigues and Luke Chamandy for helpful discussions on dynamo action in
galaxies. We also thank David J. Champion for critical comments on the
manuscript and Rainer Beck for helpful discussions. AB would like to thank the
warm hospitality at Newcastle University during his visit there. AB
acknowledges the online service provided by WolframAlpha
(\url{https://www.wolframalpha.com}) which extensively helped to reduce the
complexity of the mathematical functions. AF and AS are grateful to the STFC
(ST/N000900/1, Project 2) and the Leverhulme Trust (RPG-2014-427) for partial
financial support. NK acknowledges support from the Department of Science and
Technology via a Swarnajayanti Fellowship (DST/SJF/PSA-01/2012-13).

Funding for SDSS-III has been provided by the Alfred P. Sloan Foundation, the
Participating Institutions, the National Science Foundation, and the U.S.
Department of Energy Office of Science. The SDSS-III web site is
\url{http://www.sdss3.org/}.

SDSS-III is managed by the Astrophysical Research Consortium for the
Participating Institutions of the SDSS-III Collaboration including the
University of Arizona, the Brazilian Participation Group, Brookhaven National
Laboratory, Carnegie Mellon University, University of Florida, the French
Participation Group, the German Participation Group, Harvard University, the
Instituto de Astrofisica de Canarias, the Michigan State/Notre Dame/JINA
Participation Group, Johns Hopkins University, Lawrence Berkeley National
Laboratory, Max Planck Institute for Astrophysics, Max Planck Institute for
Extraterrestrial Physics, New Mexico State University, New York University,
Ohio State University, Pennsylvania State University, University of Portsmouth,
Princeton University, the Spanish Participation Group, University of Tokyo,
University of Utah, Vanderbilt University, University of Virginia, University
of Washington, and Yale University.

\bibliographystyle{mnras}

\bibliography{abasu_etal_mnr.bbl} 

\begin{thebibliography}{}
\makeatletter
\relax
\def\mn@urlcharsother{\let\do\@makeother \do\$\do\&\do\#\do\^\do\_\do\%\do\~}
\def\mn@doi{\begingroup\mn@urlcharsother \@ifnextchar [ {\mn@doi@}
  {\mn@doi@[]}}
\def\mn@doi@[#1]#2{\def\@tempa{#1}\ifx\@tempa\@empty \href
  {http://dx.doi.org/#2} {doi:#2}\else \href {http://dx.doi.org/#2} {#1}\fi
  \endgroup}
\def\mn@eprint#1#2{\mn@eprint@#1:#2::\@nil}
\def\mn@eprint@arXiv#1{\href {http://arxiv.org/abs/#1} {{\tt arXiv:#1}}}
\def\mn@eprint@dblp#1{\href {http://dblp.uni-trier.de/rec/bibtex/#1.xml}
  {dblp:#1}}
\def\mn@eprint@#1:#2:#3:#4\@nil{\def\@tempa {#1}\def\@tempb {#2}\def\@tempc
  {#3}\ifx \@tempc \@empty \let \@tempc \@tempb \let \@tempb \@tempa \fi \ifx
  \@tempb \@empty \def\@tempb {arXiv}\fi \@ifundefined
  {mn@eprint@\@tempb}{\@tempb:\@tempc}{\expandafter \expandafter \csname
  mn@eprint@\@tempb\endcsname \expandafter{\@tempc}}}

\bibitem[\protect\citeauthoryear{{Ahn} et~al.,}{{Ahn} et~al.}{2012}]{ahn12}
{Ahn} C.~P.,  et~al., 2012, \mn@doi [\apjs] {10.1088/0067-0049/203/2/21}, \href
  {http://adsabs.harvard.edu/abs/2012ApJS..203...21A} {203, 21}

\bibitem[\protect\citeauthoryear{{Akahori}, {Ryu}  \& {Gaensler}}{{Akahori}
  et~al.}{2016}]{akaho16}
{Akahori} T.,  {Ryu} D.,   {Gaensler} B.~M.,  2016, \mn@doi [\apj]
  {10.3847/0004-637X/824/2/105}, \href
  {http://adsabs.harvard.edu/abs/2016ApJ...824..105A} {824, 105}

\bibitem[\protect\citeauthoryear{{Akerman}, {Ellison}, {Pettini}  \&
  {Steidel}}{{Akerman} et~al.}{2005}]{akerman05}
{Akerman} C.~J.,  {Ellison} S.~L.,  {Pettini} M.,   {Steidel} C.~C.,  2005,
  A\&A, 440, 499

\bibitem[\protect\citeauthoryear{{Aller}, {Aller}, {Hughes}  \&
  {Latimer}}{{Aller} et~al.}{1999}]{aller99}
{Aller} M.~F.,  {Aller} H.~D.,  {Hughes} P.~A.,   {Latimer} G.~E.,  1999,
  \mn@doi [\apj] {10.1086/306799}, \href
  {http://adsabs.harvard.edu/abs/1999ApJ...512..601A} {512, 601}

\bibitem[\protect\citeauthoryear{{Arshakian}, {Beck}, {Krause}  \&
  {Sokoloff}}{{Arshakian} et~al.}{2009}]{arsha09}
{Arshakian} T.~G.,  {Beck} R.,  {Krause} M.,   {Sokoloff} D.,  2009, \mn@doi
  [\aap] {10.1051/0004-6361:200810964}, \href
  {http://adsabs.harvard.edu/abs/2009A%26A...494...21A} {494, 21}

\bibitem[\protect\citeauthoryear{{Basu} \& {Roy}}{{Basu} \&
  {Roy}}{2013}]{basu13}
{Basu} A.,  {Roy} S.,  2013, \mn@doi [\mnras] {10.1093/mnras/stt845}, \href
  {http://adsabs.harvard.edu/abs/2013MNRAS.433.1675B} {433, 1675}

\bibitem[\protect\citeauthoryear{Beck}{Beck}{2007}]{beck07}
Beck R.,  2007, \mn@doi [\aap] {10.1051/0004-6361:20066988}, 470, 539

\bibitem[\protect\citeauthoryear{{Beck}}{{Beck}}{2015}]{beck15}
{Beck} R.,  2015, \mn@doi [\aap] {10.1051/0004-6361/201425572}, \href
  {http://adsabs.harvard.edu/abs/2015A%26A...578A..93B} {578, A93}

\bibitem[\protect\citeauthoryear{{Beck}}{{Beck}}{2016}]{beck16}
{Beck} R.,  2016, \mn@doi [\aapr] {10.1007/s00159-015-0084-4}, \href
  {http://adsabs.harvard.edu/abs/2016A%26ARv..24....4B} {24, 4}

\bibitem[\protect\citeauthoryear{Beck \& Wielebinski}{Beck \&
  Wielebinski}{2013}]{beck13book}
Beck R.,  Wielebinski R.,  2013, {Planets, Stars and Stellar Systems Vol. 5.
  Springer, Dordrecht}.
p.~641

\bibitem[\protect\citeauthoryear{Beck, Brandenburg, Moss, Shukurov  \&
  Sokoloff}{Beck et~al.}{1996}]{beck96}
Beck R.,  Brandenburg A.,  Moss D.,  Shukurov A.,   Sokoloff D.,  1996, \mn@doi
  [\araa] {10.1146/annurev.astro.34.1.155}, 34, 155

\bibitem[\protect\citeauthoryear{{Bergeron} \& {Boiss{\'e}}}{{Bergeron} \&
  {Boiss{\'e}}}{1991}]{bergeron91}
{Bergeron} J.,  {Boiss{\'e}} P.,  1991, A\&A, 243, 344

\bibitem[\protect\citeauthoryear{{Berkhuijsen}, {Horellou}, {Krause},
  {Neininger}, {Poezd}, {Shukurov}  \& {Sokoloff}}{{Berkhuijsen}
  et~al.}{1997}]{berkh97}
{Berkhuijsen} E.~M.,  {Horellou} C.,  {Krause} M.,  {Neininger} N.,  {Poezd}
  A.~D.,  {Shukurov} A.,   {Sokoloff} D.~D.,  1997, \aap, \href
  {http://adsabs.harvard.edu/abs/1997A%26A...318..700B} {318, 700}

\bibitem[\protect\citeauthoryear{{Berkhuijsen}, {Urbanik}, {Beck}  \&
  {Han}}{{Berkhuijsen} et~al.}{2016}]{berkh16}
{Berkhuijsen} E.~M.,  {Urbanik} M.,  {Beck} R.,   {Han} J.~L.,  2016, \mn@doi
  [\aap] {10.1051/0004-6361/201527322}, \href
  {http://adsabs.harvard.edu/abs/2016A%26A...588A.114B} {588, A114}

\bibitem[\protect\citeauthoryear{Bernet, Miniati, Lilly, Kronberg  \&
  Dessauges-Zavadsky}{Bernet et~al.}{2008}]{berne08}
Bernet M.,  Miniati F.,  Lilly S.,  Kronberg P.,   Dessauges-Zavadsky M.,
  2008, \mn@doi [\nat] {10.1038/nature07105}, 454, 302

\bibitem[\protect\citeauthoryear{{Bernet}, {Miniati}  \& {Lilly}}{{Bernet}
  et~al.}{2013}]{berne13}
{Bernet} M.~L.,  {Miniati} F.,   {Lilly} S.~J.,  2013, \mn@doi [\apjl]
  {10.1088/2041-8205/772/2/L28}, \href
  {http://adsabs.harvard.edu/abs/2013ApJ...772L..28B} {772, L28}

\bibitem[\protect\citeauthoryear{{Bond}, {Churchill}, {Charlton}  \&
  {Vogt}}{{Bond} et~al.}{2001}]{bond01}
{Bond} N.~A.,  {Churchill} C.~W.,  {Charlton} J.~C.,   {Vogt} S.~S.,  2001,
  \mn@doi [\apj] {10.1086/323876}, \href
  {http://esoads.eso.org/abs/2001ApJ...562..641B} {562, 641}

\bibitem[\protect\citeauthoryear{Bondi, Dallacasa, Stanghellini  \& Ceca}{Bondi
  et~al.}{1996}]{bondi96book}
Bondi M.,  Dallacasa D.,  Stanghellini C.,   Ceca R.~D.,  1996, Extended
  Emission in BL Lac Objects.
Springer Netherlands, Dordrecht, p.~53

\bibitem[\protect\citeauthoryear{{Bondi}, {March{\~a}}, {Polatidis},
  {Dallacasa}, {Stanghellini}  \& {Ant{\'o}n}}{{Bondi} et~al.}{2004}]{bondi04}
{Bondi} M.,  {March{\~a}} M.~J.~M.,  {Polatidis} A.,  {Dallacasa} D.,
  {Stanghellini} C.,   {Ant{\'o}n} S.,  2004, \mn@doi [\mnras]
  {10.1111/j.1365-2966.2004.07903.x}, \href
  {http://adsabs.harvard.edu/abs/2004MNRAS.352..112B} {352, 112}

\bibitem[\protect\citeauthoryear{{Burkhart}, {Lazarian}, {Ossenkopf}  \&
  {Stutzki}}{{Burkhart} et~al.}{2013}]{burkh13}
{Burkhart} B.,  {Lazarian} A.,  {Ossenkopf} V.,   {Stutzki} J.,  2013, \mn@doi
  [\apj] {10.1088/0004-637X/771/2/123}, \href
  {http://adsabs.harvard.edu/abs/2013ApJ...771..123B} {771, 123}

\bibitem[\protect\citeauthoryear{{Chamandy}}{{Chamandy}}{2016}]{chama16}
{Chamandy} L.,  2016, \mn@doi [\mnras] {10.1093/mnras/stw1941}, \href
  {http://adsabs.harvard.edu/abs/2016MNRAS.462.4402C} {462, 4402}

\bibitem[\protect\citeauthoryear{Chamandy, Subramanian  \& Shukurov}{Chamandy
  et~al.}{2013}]{chama13}
Chamandy L.,  Subramanian K.,   Shukurov A.,  2013, \mn@doi [\mnras]
  {10.1093/mnras/sts297}, 428, 3569

\bibitem[\protect\citeauthoryear{Chengalur \& Kanekar}{Chengalur \&
  Kanekar}{2000}]{chengalur00}
Chengalur J.~N.,  Kanekar N.,  2000, MNRAS, 318, 303

\bibitem[\protect\citeauthoryear{{Chy{\.z}y}, {Drzazga}, {Beck}, {Urbanik},
  {Heesen}  \& {Bomans}}{{Chy{\.z}y} et~al.}{2016}]{chyzy16}
{Chy{\.z}y} K.~T.,  {Drzazga} R.~T.,  {Beck} R.,  {Urbanik} M.,  {Heesen} V.,
  {Bomans} D.~J.,  2016, \mn@doi [\apj] {10.3847/0004-637X/819/1/39}, \href
  {http://adsabs.harvard.edu/abs/2016ApJ...819...39C} {819, 39}

\bibitem[\protect\citeauthoryear{{Cordes} \& {Lazio}}{{Cordes} \&
  {Lazio}}{2002}]{ne2001}
{Cordes} J.~M.,  {Lazio} T.~J.~W.,  2002, ArXiv Astrophysics e-prints, \href
  {http://adsabs.harvard.edu/abs/2002astro.ph..7156C} {}

\bibitem[\protect\citeauthoryear{{Damas-Segovia} et~al.,}{{Damas-Segovia}
  et~al.}{2016}]{damas16}
{Damas-Segovia} A.,  et~al., 2016, \mn@doi [\apj] {10.3847/0004-637X/824/1/30},
  \href {http://adsabs.harvard.edu/abs/2016ApJ...824...30D} {824, 30}

\bibitem[\protect\citeauthoryear{{Erb}, {Shapley}, {Pettini}, {Steidel},
  {Reddy}  \& {Adelberger}}{{Erb} et~al.}{2006}]{erb06}
{Erb} D.~K.,  {Shapley} A.~E.,  {Pettini} M.,  {Steidel} C.~C.,  {Reddy} N.~A.,
    {Adelberger} K.~L.,  2006, ApJ, 644, 813

\bibitem[\protect\citeauthoryear{{Farnes}, {O'Sullivan}, {Corrigan}  \&
  {Gaensler}}{{Farnes} et~al.}{2014}]{farne14}
{Farnes} J.~S.,  {O'Sullivan} S.~P.,  {Corrigan} M.~E.,   {Gaensler} B.~M.,
  2014, \mn@doi [\apj] {10.1088/0004-637X/795/1/63}, \href
  {http://adsabs.harvard.edu/abs/2014ApJ...795...63F} {795, 63}

\bibitem[\protect\citeauthoryear{{Federrath}, {Chabrier}, {Schober},
  {Banerjee}, {Klessen}  \& {Schleicher}}{{Federrath} et~al.}{2011}]{feder11}
{Federrath} C.,  {Chabrier} G.,  {Schober} J.,  {Banerjee} R.,  {Klessen}
  R.~S.,   {Schleicher} D.~R.~G.,  2011, \mn@doi [\prl]
  {10.1103/PhysRevLett.107.114504}, \href
  {http://adsabs.harvard.edu/abs/2011PhRvL.107k4504F} {107, 114504}

\bibitem[\protect\citeauthoryear{{Fletcher}}{{Fletcher}}{2010}]{fletc10}
{Fletcher} A.,  2010, in {Kothes} R.,  {Landecker} T.~L.,   {Willis} A.~G.,
  eds,  ASP Conf. Series Vol. 438, The Dynamic Interstellar Medium: A
  Celebration of the Canadian Galactic Plane Survey. p.~197

\bibitem[\protect\citeauthoryear{Fletcher, Beck, Shukurov, Berkhuijsen  \&
  Horellou}{Fletcher et~al.}{2011}]{fletc11}
Fletcher A.,  Beck R.,  Shukurov A.,  Berkhuijsen E.,   Horellou C.,  2011,
  \mn@doi [\mnras] {10.1111/j.1365-2966.2010.18065.x}, 412, 2396

\bibitem[\protect\citeauthoryear{{Fumagalli}, {O'Meara}, {Prochaska}, {Kanekar}
   \& {Wolfe}}{{Fumagalli} et~al.}{2014}]{fumagalli14b}
{Fumagalli} M.,  {O'Meara} J.~M.,  {Prochaska} J.~X.,  {Kanekar} N.,   {Wolfe}
  A.~M.,  2014, MNRAS, 444, 1282

\bibitem[\protect\citeauthoryear{{Fumagalli}, {O'Meara}, {Prochaska},
  {Rafelski}  \& {Kanekar}}{{Fumagalli} et~al.}{2015}]{fumagalli15}
{Fumagalli} M.,  {O'Meara} J.~M.,  {Prochaska} J.~X.,  {Rafelski} M.,
  {Kanekar} N.,  2015, MNRAS, 446, 3178

\bibitem[\protect\citeauthoryear{{Fynbo} et~al.,}{{Fynbo}
  et~al.}{2011}]{fynbo11}
{Fynbo} J.~P.~U.,  et~al., 2011, \mn@doi [\mnras]
  {10.1111/j.1365-2966.2011.18318.x}, \href
  {http://adsabs.harvard.edu/abs/2011MNRAS.413.2481F} {413, 2481}

\bibitem[\protect\citeauthoryear{{Fynbo} et~al.,}{{Fynbo}
  et~al.}{2013}]{fynbo13}
{Fynbo} J.~P.~U.,  et~al., 2013, \mn@doi [\mnras] {10.1093/mnras/stt1579},
  \href {http://adsabs.harvard.edu/abs/2013MNRAS.436..361F} {436, 361}

\bibitem[\protect\citeauthoryear{Gent, Shukurov, Sarson, Fletcher  \&
  Mantere}{Gent et~al.}{2013a}]{gent13b}
Gent F.,  Shukurov A.,  Sarson G.,  Fletcher A.,   Mantere M.,  2013a, \mn@doi
  [\mnras] {10.1093/mnrasl/sls042}, 430, L40

\bibitem[\protect\citeauthoryear{Gent, Shukurov, Fletcher, Sarson  \&
  Mantere}{Gent et~al.}{2013b}]{gent13a}
Gent F.,  Shukurov A.,  Fletcher A.,  Sarson G.,   Mantere M.,  2013b, \mn@doi
  [\mnras] {10.1093/mnras/stt560}, 432, 1396

\bibitem[\protect\citeauthoryear{Glen, Leemis  \& Drew}{Glen
  et~al.}{2004}]{glen04}
Glen A.~G.,  Leemis L.~M.,   Drew J.~H.,  2004, \mn@doi [Computational
  Statistics \& Data Analysis]
  {http://dx.doi.org/10.1016/S0167-9473(02)00234-7}, 44, 451

\bibitem[\protect\citeauthoryear{{Hanasz}, {Otmianowska-Mazur}, {Kowal}  \&
  {Lesch}}{{Hanasz} et~al.}{2009}]{hanas09}
{Hanasz} M.,  {Otmianowska-Mazur} K.,  {Kowal} G.,   {Lesch} H.,  2009, \mn@doi
  [\aap] {10.1051/0004-6361/200810279}, \href
  {http://adsabs.harvard.edu/abs/2009A%26A...498..335H} {498, 335}

\bibitem[\protect\citeauthoryear{Heiles \& Troland}{Heiles \&
  Troland}{2003}]{heiles03}
Heiles C.,  Troland T.~H.,  2003, ApJS, 145, 329

\bibitem[\protect\citeauthoryear{{Joshi} \& {Chand}}{{Joshi} \&
  {Chand}}{2013}]{joshi13}
{Joshi} R.,  {Chand} H.,  2013, \mn@doi [\mnras] {10.1093/mnras/stt1277}, \href
  {http://adsabs.harvard.edu/abs/2013MNRAS.434.3566J} {434, 3566}

\bibitem[\protect\citeauthoryear{{Kacprzak}, {Churchill}, {Evans}, {Murphy}  \&
  {Steidel}}{{Kacprzak} et~al.}{2011}]{kacpr11}
{Kacprzak} G.~G.,  {Churchill} C.~W.,  {Evans} J.~L.,  {Murphy} M.~T.,
  {Steidel} C.~C.,  2011, \mn@doi [\mnras] {10.1111/j.1365-2966.2011.19261.x},
  \href {http://adsabs.harvard.edu/abs/2011MNRAS.416.3118K} {416, 3118}

\bibitem[\protect\citeauthoryear{{Kanekar}}{{Kanekar}}{2014}]{kanekar14c}
{Kanekar} N.,  2014, ApJL, 797, L20

\bibitem[\protect\citeauthoryear{{Kanekar} \& {Chengalur}}{{Kanekar} \&
  {Chengalur}}{2001}]{kanekar01a}
{Kanekar} N.,  {Chengalur} J.~N.,  2001, A\&A, 369, 42

\bibitem[\protect\citeauthoryear{{Kanekar}, {Subrahmanyan}, {Ellison}, {Lane}
  \& {Chengalur}}{{Kanekar} et~al.}{2006}]{kanekar06}
{Kanekar} N.,  {Subrahmanyan} R.,  {Ellison} S.~L.,  {Lane} W.~M.,
  {Chengalur} J.~N.,  2006, MNRAS, 370, L46

\bibitem[\protect\citeauthoryear{{Kanekar}, {Chengalur}  \& {Lane}}{{Kanekar}
  et~al.}{2007}]{kanekar07}
{Kanekar} N.,  {Chengalur} J.~N.,   {Lane} W.~M.,  2007, MNRAS, 375, 1528

\bibitem[\protect\citeauthoryear{{Kanekar}, {Smette}, {Briggs}  \&
  {Chengalur}}{{Kanekar} et~al.}{2009}]{kanekar09b}
{Kanekar} N.,  {Smette} A.,  {Briggs} F.~H.,   {Chengalur} J.~N.,  2009, ApJ,
  705, L40

\bibitem[\protect\citeauthoryear{{Kanekar} et~al.,}{{Kanekar}
  et~al.}{2014}]{kanekar14}
{Kanekar} N.,  et~al., 2014, MNRAS, 438, 2131

\bibitem[\protect\citeauthoryear{{Kim}, {Lilly}, {Miniati}, {Bernet}, {Beck},
  {O'Sullivan}  \& {Gaensler}}{{Kim} et~al.}{2016}]{kim16}
{Kim} K.~S.,  {Lilly} S.~J.,  {Miniati} F.,  {Bernet} M.~L.,  {Beck} R.,
  {O'Sullivan} S.~P.,   {Gaensler} B.~M.,  2016, \mn@doi [\apj]
  {10.3847/0004-637X/829/2/133}, \href
  {http://adsabs.harvard.edu/abs/2016ApJ...829..133K} {829, 133}

\bibitem[\protect\citeauthoryear{{Krogager}, {Fynbo}, {M{\o}ller}, {Ledoux},
  {Noterdaeme}, {Christensen}, {Milvang-Jensen}  \& {Sparre}}{{Krogager}
  et~al.}{2012}]{krogager12}
{Krogager} J.-K.,  {Fynbo} J.~P.~U.,  {M{\o}ller} P.,  {Ledoux} C.,
  {Noterdaeme} P.,  {Christensen} L.,  {Milvang-Jensen} B.,   {Sparre} M.,
  2012, MNRAS, 424, L1

\bibitem[\protect\citeauthoryear{{Krogager} et~al.,}{{Krogager}
  et~al.}{2013}]{kroga13}
{Krogager} J.-K.,  et~al., 2013, \mn@doi [\mnras] {10.1093/mnras/stt955}, \href
  {http://adsabs.harvard.edu/abs/2013MNRAS.433.3091K} {433, 3091}

\bibitem[\protect\citeauthoryear{{Kronberg}, {Perry}  \& {Zukowski}}{{Kronberg}
  et~al.}{1992}]{kronb92}
{Kronberg} P.~P.,  {Perry} J.~J.,   {Zukowski} E.~L.~H.,  1992, \mn@doi [\apj]
  {10.1086/171104}, \href {http://adsabs.harvard.edu/abs/1992ApJ...387..528K}
  {387, 528}

\bibitem[\protect\citeauthoryear{{Kulkarni}, {Fall}, {Lauroesch}, {York},
  {Welty}, {Khare}  \& {Truran}}{{Kulkarni} et~al.}{2005}]{kulkarni05}
{Kulkarni} V.~P.,  {Fall} S.~M.,  {Lauroesch} J.~T.,  {York} D.~G.,  {Welty}
  D.~E.,  {Khare} P.,   {Truran} J.~W.,  2005, ApJ, 618, 68

\bibitem[\protect\citeauthoryear{{Lazarian} \& {Pogosyan}}{{Lazarian} \&
  {Pogosyan}}{2000}]{lazar00}
{Lazarian} A.,  {Pogosyan} D.,  2000, \mn@doi [\apj] {10.1086/309040}, \href
  {http://adsabs.harvard.edu/abs/2000ApJ...537..720L} {537, 720}

\bibitem[\protect\citeauthoryear{{Lee} et~al.,}{{Lee} et~al.}{2013}]{lee13}
{Lee} K.-G.,  et~al., 2013, \mn@doi [\aj] {10.1088/0004-6256/145/3/69}, \href
  {http://adsabs.harvard.edu/abs/2013AJ....145...69L} {145, 69}

\bibitem[\protect\citeauthoryear{Leroy, Walter, Brinks, Bigiel, de Blok, Madore
   \& Thornley}{Leroy et~al.}{2008}]{leroy08}
Leroy A.,  Walter F.,  Brinks E.,  Bigiel F.,  de Blok W.,  Madore B.,
  Thornley M.,  2008, \mn@doi [\aj] {10.1088/0004-6256/136/6/2782}, 136, 2782

\bibitem[\protect\citeauthoryear{Lister}{Lister}{2001}]{lister01}
Lister M.~L.,  2001, \apj, 562, 208

\bibitem[\protect\citeauthoryear{Lister et~al.,}{Lister
  et~al.}{2016}]{lister16}
Lister M.~L.,  et~al., 2016, \aj, 152, 12

\bibitem[\protect\citeauthoryear{{Mao}, {Gaensler}, {Haverkorn}, {Zweibel},
  {Madsen}, {McClure-Griffiths}, {Shukurov}  \& {Kronberg}}{{Mao}
  et~al.}{2010}]{mao10}
{Mao} S.~A.,  {Gaensler} B.~M.,  {Haverkorn} M.,  {Zweibel} E.~G.,  {Madsen}
  G.~J.,  {McClure-Griffiths} N.~M.,  {Shukurov} A.,   {Kronberg} P.~P.,  2010,
  \mn@doi [\apj] {10.1088/0004-637X/714/2/1170}, \href
  {http://adsabs.harvard.edu/abs/2010ApJ...714.1170M} {714, 1170}

\bibitem[\protect\citeauthoryear{{Mao} et~al.,}{{Mao} et~al.}{2017}]{mao17}
{Mao} S.~A.,  et~al., 2017, \mn@doi [Nature Astronomy]
  {https://doi.org/10.1038/s41550-017-0218-x}, 1, 621

\bibitem[\protect\citeauthoryear{{M{\o}ller}, {Fynbo}, {Ledoux}  \&
  {Nilsson}}{{M{\o}ller} et~al.}{2013}]{moller13}
{M{\o}ller} P.,  {Fynbo} J.~P.~U.,  {Ledoux} C.,   {Nilsson} K.~K.,  2013,
  MNRAS, 430, 2680

\bibitem[\protect\citeauthoryear{{Neeleman}, {Wolfe}, {Prochaska}  \&
  {Rafelski}}{{Neeleman} et~al.}{2013}]{neeleman13}
{Neeleman} M.,  {Wolfe} A.~M.,  {Prochaska} J.~X.,   {Rafelski} M.,  2013, ApJ,
  769, 54

\bibitem[\protect\citeauthoryear{{Neeleman} et~al.,}{{Neeleman}
  et~al.}{2016}]{neele16}
{Neeleman} M.,  et~al., 2016, \mn@doi [\apjl] {10.3847/2041-8205/820/2/L39},
  \href {http://adsabs.harvard.edu/abs/2016ApJ...820L..39N} {820, L39}

\bibitem[\protect\citeauthoryear{{Neeleman}, {Kanekar}, {Prochaska},
  {Rafelski}, {Carilli}  \& {Wolfe}}{{Neeleman} et~al.}{2017}]{neele17}
{Neeleman} M.,  {Kanekar} N.,  {Prochaska} J.~X.,  {Rafelski} M.,  {Carilli}
  C.~L.,   {Wolfe} A.~M.,  2017, \mn@doi [Science] {10.1126/science.aal1737},
  \href {http://adsabs.harvard.edu/abs/2017Sci...355.1285N} {355, 1285}

\bibitem[\protect\citeauthoryear{{Nestor}, {Turnshek}  \& {Rao}}{{Nestor}
  et~al.}{2005}]{nesto05}
{Nestor} D.~B.,  {Turnshek} D.~A.,   {Rao} S.~M.,  2005, \mn@doi [\apj]
  {10.1086/427547}, \href {http://adsabs.harvard.edu/abs/2005ApJ...628..637N}
  {628, 637}

\bibitem[\protect\citeauthoryear{{Nestor}, {Turnshek}, {Rao}  \&
  {Quider}}{{Nestor} et~al.}{2007}]{nesto07}
{Nestor} D.~B.,  {Turnshek} D.~A.,  {Rao} S.~M.,   {Quider} A.~M.,  2007,
  \mn@doi [\apj] {10.1086/511411}, \href
  {http://esoads.eso.org/abs/2007ApJ...658..185N} {658, 185}

\bibitem[\protect\citeauthoryear{{Noterdaeme}, {Petitjean}, {Ledoux}  \&
  {Srianand}}{{Noterdaeme} et~al.}{2009}]{noter09}
{Noterdaeme} P.,  {Petitjean} P.,  {Ledoux} C.,   {Srianand} R.,  2009, \mn@doi
  [\aap] {10.1051/0004-6361/200912768}, \href
  {http://adsabs.harvard.edu/abs/2009A%26A...505.1087N} {505, 1087}

\bibitem[\protect\citeauthoryear{{Noterdaeme} et~al.,}{{Noterdaeme}
  et~al.}{2012}]{noter12}
{Noterdaeme} P.,  et~al., 2012, \mn@doi [\aap] {10.1051/0004-6361/201220259},
  \href {http://adsabs.harvard.edu/abs/2012A%26A...547L...1N} {547, L1}

\bibitem[\protect\citeauthoryear{{Ohno} \& {Shibata}}{{Ohno} \&
  {Shibata}}{1993}]{ohno93}
{Ohno} H.,  {Shibata} S.,  1993, \mn@doi [\mnras] {10.1093/mnras/262.4.953},
  \href {http://adsabs.harvard.edu/abs/1993MNRAS.262..953O} {262, 953}

\bibitem[\protect\citeauthoryear{{Oppermann} et~al.,}{{Oppermann}
  et~al.}{2015}]{opper15}
{Oppermann} N.,  et~al., 2015, \mn@doi [\aap] {10.1051/0004-6361/201423995},
  \href {http://adsabs.harvard.edu/abs/2015A%26A...575A.118O} {575, A118}

\bibitem[\protect\citeauthoryear{{Oren} \& {Wolfe}}{{Oren} \&
  {Wolfe}}{1995}]{oren95}
{Oren} A.~L.,  {Wolfe} A.~M.,  1995, \mn@doi [\apj] {10.1086/175726}, \href
  {http://adsabs.harvard.edu/abs/1995ApJ...445..624O} {445, 624}

\bibitem[\protect\citeauthoryear{{Pakmor}, {Marinacci}  \& {Springel}}{{Pakmor}
  et~al.}{2014}]{pakmo14}
{Pakmor} R.,  {Marinacci} F.,   {Springel} V.,  2014, \mn@doi [\apjl]
  {10.1088/2041-8205/783/1/L20}, \href
  {http://adsabs.harvard.edu/abs/2014ApJ...783L..20P} {783, L20}

\bibitem[\protect\citeauthoryear{{P{\^a}ris} et~al.,}{{P{\^a}ris}
  et~al.}{2012}]{paris12}
{P{\^a}ris} I.,  et~al., 2012, \mn@doi [\aap] {10.1051/0004-6361/201220142},
  \href {http://adsabs.harvard.edu/abs/2012A%26A...548A..66P} {548, A66}

\bibitem[\protect\citeauthoryear{{Pettini}, {Smith}, {Hunstead}  \&
  {King}}{{Pettini} et~al.}{1994}]{pettini94}
{Pettini} M.,  {Smith} L.~J.,  {Hunstead} R.~W.,   {King} D.~L.,  1994, ApJ,
  426, 79

\bibitem[\protect\citeauthoryear{{Pettini}, {Smith}, {King}  \&
  {Hunstead}}{{Pettini} et~al.}{1997}]{pettini97}
{Pettini} M.,  {Smith} L.~J.,  {King} D.~L.,   {Hunstead} R.~W.,  1997, ApJ,
  486, 665

\bibitem[\protect\citeauthoryear{{Pettini}, {Ellison}, {Steidel}  \&
  {Bowen}}{{Pettini} et~al.}{1999}]{pettini99}
{Pettini} M.,  {Ellison} S.~L.,  {Steidel} C.~C.,   {Bowen} D.~V.,  1999, ApJ,
  510, 576

\bibitem[\protect\citeauthoryear{{Prochaska}, {Gawiser}, {Wolfe}, {Castro}  \&
  {Djorgovski}}{{Prochaska} et~al.}{2003}]{prochaska03a}
{Prochaska} J.~X.,  {Gawiser} E.,  {Wolfe} A.~M.,  {Castro} S.,   {Djorgovski}
  S.~G.,  2003, ApJ, 595, L9

\bibitem[\protect\citeauthoryear{Prochaska, Herbert-Fort  \& Wolfe}{Prochaska
  et~al.}{2005}]{prochaska05}
Prochaska J.~X.,  Herbert-Fort S.,   Wolfe A.~M.,  2005, ApJ, 635, 123

\bibitem[\protect\citeauthoryear{Prochaska, Wolfe, Howk, Gawiser, Burles  \&
  Cooke}{Prochaska et~al.}{2007}]{prochaska07}
Prochaska J.~X.,  Wolfe A.~M.,  Howk J.~C.,  Gawiser E.,  Burles S.~M.,   Cooke
  J.,  2007, ApJS, 171, 29

\bibitem[\protect\citeauthoryear{{Prochaska}, {Chen}, {Wolfe},
  {Dessauges-Zavadsky}  \& {Bloom}}{{Prochaska} et~al.}{2008}]{prochaska08}
{Prochaska} J.~X.,  {Chen} H.-W.,  {Wolfe} A.~M.,  {Dessauges-Zavadsky} M.,
  {Bloom} J.~S.,  2008, ApJ, 672, 59

\bibitem[\protect\citeauthoryear{{Rafelski}, {Wolfe}, {Prochaska}, {Neeleman}
  \& {Mendez}}{{Rafelski} et~al.}{2012}]{rafelski12}
{Rafelski} M.,  {Wolfe} A.~M.,  {Prochaska} J.~X.,  {Neeleman} M.,   {Mendez}
  A.~J.,  2012, ApJ, 755, 89

\bibitem[\protect\citeauthoryear{{Rao}, {Turnshek}  \& {Nestor}}{{Rao}
  et~al.}{2006}]{rao06}
{Rao} S.~M.,  {Turnshek} D.~A.,   {Nestor} D.~B.,  2006, \mn@doi [\apj]
  {10.1086/498132}, \href {http://adsabs.harvard.edu/abs/2006ApJ...636..610R}
  {636, 610}

\bibitem[\protect\citeauthoryear{{Rodrigues}, {Shukurov}, {Fletcher}  \&
  {Baugh}}{{Rodrigues} et~al.}{2015}]{rodri15}
{Rodrigues} L.~F.~S.,  {Shukurov} A.,  {Fletcher} A.,   {Baugh} C.~M.,  2015,
  \mn@doi [\mnras] {10.1093/mnras/stv816}, \href
  {http://adsabs.harvard.edu/abs/2015MNRAS.450.3472R} {450, 3472}

\bibitem[\protect\citeauthoryear{Rohatgi}{Rohatgi}{1976}]{rohat76}
Rohatgi V.~K.,  1976, {An Introduction to Probability Theory Mathematical
  Statistics}.
Wiley, New York

\bibitem[\protect\citeauthoryear{{Roy}, {Kanekar}, {Braun}  \&
  {Chengalur}}{{Roy} et~al.}{2013}]{roy13a}
{Roy} N.,  {Kanekar} N.,  {Braun} R.,   {Chengalur} J.~N.,  2013, MNRAS, 436,
  2352

\bibitem[\protect\citeauthoryear{Ruzmaikin, Sokoloff  \& Shukurov}{Ruzmaikin
  et~al.}{1988}]{ruzma88}
Ruzmaikin A.,  Sokoloff D.,   Shukurov A.,  1988, \mn@doi [Nature]
  {http://dx.doi.org/10.1038/336341a0}, 336, 341

\bibitem[\protect\citeauthoryear{{Schnitzeler}}{{Schnitzeler}}{2010}]{schni10}
{Schnitzeler} D.~H.~F.~M.,  2010, \mn@doi [\mnras]
  {10.1111/j.1745-3933.2010.00957.x}, \href
  {http://adsabs.harvard.edu/abs/2010MNRAS.409L..99S} {409, L99}

\bibitem[\protect\citeauthoryear{Shukurov}{Shukurov}{2007}]{shuku07}
Shukurov A.,  2007, {Introduction to galactic dynamos, in ``Mathematical
  Aspects of Natural Dynamo''}.
eds. E. Dormy and B. Desjardins, CRC Press

\bibitem[\protect\citeauthoryear{Sokoloff \& Shukurov}{Sokoloff \&
  Shukurov}{1990}]{sokol90}
Sokoloff D.,  Shukurov A.,  1990, \mn@doi [Nature]
  {http://dx.doi.org/10.1038/347051a0}, 347, 51

\bibitem[\protect\citeauthoryear{{Subramanian}}{{Subramanian}}{1999}]{kandu99}
{Subramanian} K.,  1999, \mn@doi [Physical Review Letters]
  {10.1103/PhysRevLett.83.2957}, \href
  {http://adsabs.harvard.edu/abs/1999PhRvL..83.2957S} {83, 2957}

\bibitem[\protect\citeauthoryear{Sveshnikov}{Sveshnikov}{1968}]{svesh68}
Sveshnikov A.~A.,  1968, Problems in Probability Theory, Mathematical
  Statistics, and Theory of Random Functions, Edited by A.A. Sveshnikov.
  Translated by Scripta Technica, Inc. Edited by Bernard R. Gelbaum.
Saunders Mathematics Books

\bibitem[\protect\citeauthoryear{{Wiener}, {Pfrommer}  \& {Peng Oh}}{{Wiener}
  et~al.}{2017}]{wiene17}
{Wiener} J.,  {Pfrommer} C.,   {Peng Oh} S.,  2017, \mn@doi [\mnras]
  {10.1093/mnras/stx127}, \href
  {http://adsabs.harvard.edu/abs/2017MNRAS.467..906W} {467, 906}

\bibitem[\protect\citeauthoryear{{Wolfe} \& {Davis}}{{Wolfe} \&
  {Davis}}{1979}]{wolfe79}
{Wolfe} A.~M.,  {Davis} M.~M.,  1979, \mn@doi [\aj] {10.1086/112470}, \href
  {http://adsabs.harvard.edu/abs/1979AJ.....84..699W} {84, 699}

\bibitem[\protect\citeauthoryear{{Wolfe}, {Briggs}, {Turnshek}, {Davis},
  {Smith}  \& {Cohen}}{{Wolfe} et~al.}{1985}]{wolfe85}
{Wolfe} A.~M.,  {Briggs} F.~H.,  {Turnshek} D.~A.,  {Davis} M.~M.,  {Smith}
  H.~E.,   {Cohen} R.~D.,  1985, ApJ, 294, L67

\bibitem[\protect\citeauthoryear{Wolfe, Turnshek, Smith  \& Cohen}{Wolfe
  et~al.}{1986}]{wolfe86}
Wolfe A.~M.,  Turnshek D.~A.,  Smith H.~E.,   Cohen R.~D.,  1986, ApJS, 61, 249

\bibitem[\protect\citeauthoryear{Wolfe, Gawiser  \& Prochaska}{Wolfe
  et~al.}{2005}]{wolfe05}
Wolfe A.~M.,  Gawiser E.,   Prochaska J.~X.,  2005, ARA\&A, 43, 861

\bibitem[\protect\citeauthoryear{{Zensus}}{{Zensus}}{1997}]{zensus97}
{Zensus} J.~A.,  1997, \mn@doi [\araa] {10.1146/annurev.astro.35.1.607}, \href
  {http://adsabs.harvard.edu/abs/1997ARA%26A..35..607Z} {35, 607}

\bibitem[\protect\citeauthoryear{{Zhu} \& {M{\'e}nard}}{{Zhu} \&
  {M{\'e}nard}}{2013}]{zhu13}
{Zhu} G.,  {M{\'e}nard} B.,  2013, \mn@doi [\apj]
  {10.1088/0004-637X/770/2/130}, \href
  {http://adsabs.harvard.edu/abs/2013ApJ...770..130Z} {770, 130}

\bibitem[\protect\citeauthoryear{{Zibetti}, {M{\'e}nard}, {Nestor}, {Quider},
  {Rao}  \& {Turnshek}}{{Zibetti} et~al.}{2007}]{zibet07}
{Zibetti} S.,  {M{\'e}nard} B.,  {Nestor} D.~B.,  {Quider} A.~M.,  {Rao} S.~M.,
    {Turnshek} D.~A.,  2007, \mn@doi [\apj] {10.1086/511300}, \href
  {http://adsabs.harvard.edu/abs/2007ApJ...658..161Z} {658, 161}

\bibitem[\protect\citeauthoryear{de Avillez \& Breitschwerdt}{de~Avillez \&
  Breitschwerdt}{2005}]{deavi05}
de Avillez M.,  Breitschwerdt D.,  2005, \mn@doi [\aap]
  {10.1051/0004-6361:20042146}, 436, 585

\makeatother
\end{thebibliography}

\appendix

\section{Probability distribution functions} \label{pdfcalc}

\begin{figure*}
\begin{tabular}{cc}
{\mbox{\includegraphics[width=6cm]{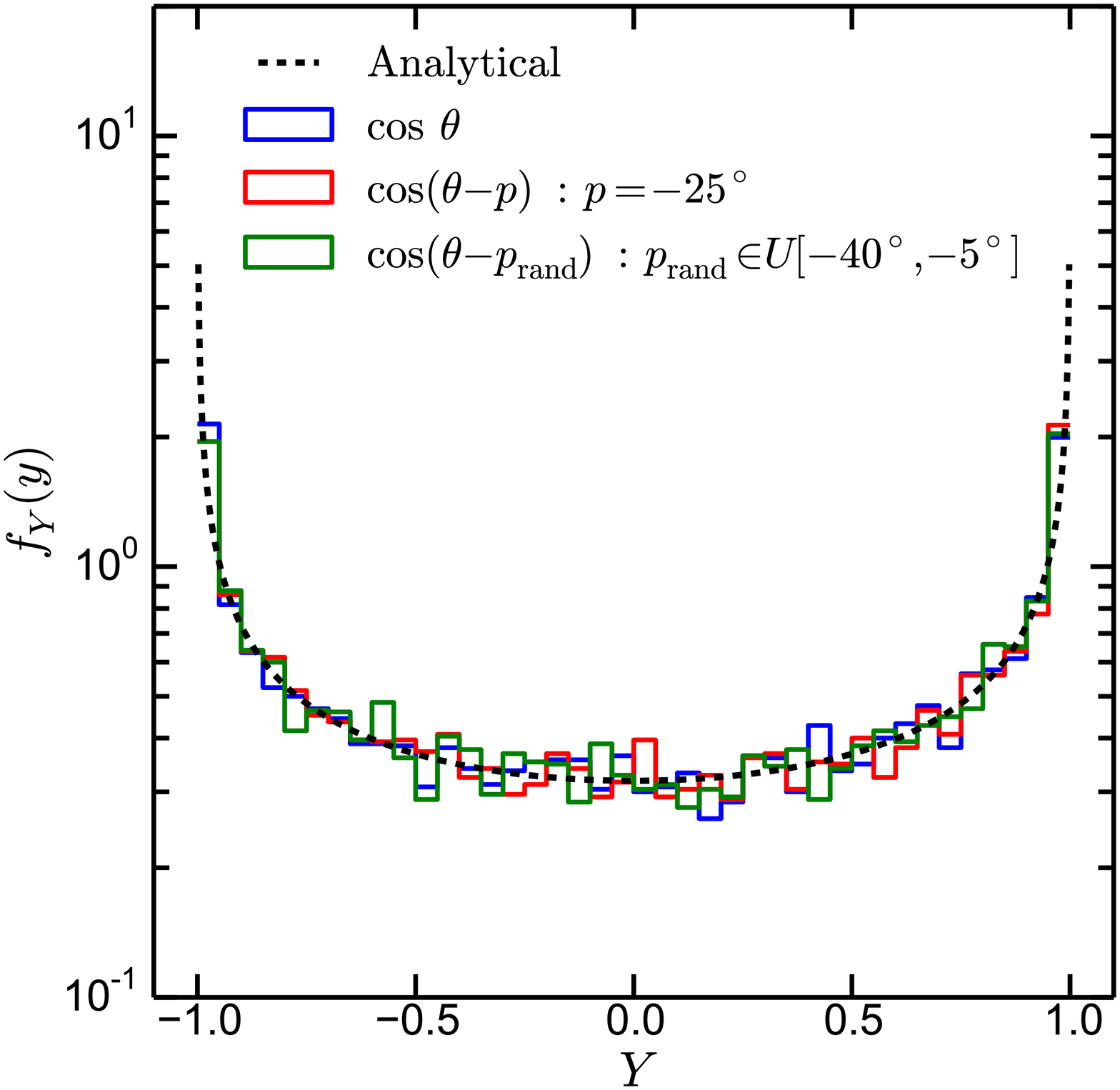}}} &
{\mbox{\includegraphics[width=6cm]{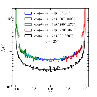}}} \\
\end{tabular}
\caption{{\it Left:} Distribution of $\cos \theta$ (blue histogram),
$\cos(\theta - p)$ with constant $p = -25^\circ$ (red histogram) and
$\cos(\theta - p_{\rm rand})$ with $p_{\rm rand}$ uniformly distributed, such
that, $p_{\rm rand} \in U[-40^\circ, -5^\circ]$ (green histogram). Here,
$\theta$ is distributed uniformly between $0^\circ$ and $360^\circ$. The black
dashed curve is the analytical PDF given in Eq.~\eqref{eq:pdf_cos}. {\it Right:}
Distribution of $Y = \cos (\theta - p)$ for $p = -25^\circ$. Each colour is for
for uniform distribution of $\theta$ lying in the range $n\times 90^\circ$ and
$(n+1)\times 90^\circ$, where $n = 0, 1, 2, 3$. The corresponding curves are
the analytical functions given by Eqns.~\eqref{eq:pdf_cos_seg1} and
\eqref{eq:pdf_cos_seg2}. For comparison, we show the distribution for $\theta$ in
the range $0-360^\circ$ in black, same as the left-hand panel.}
\label{distr} 
\end{figure*}

Under assumptions described in Section~\ref{sec:method}, the magnetic field
along the line of sight ($B_{\rm \parallel}$) in Eq.~\eqref{eq_bparallel} is
given by,
\begin{equation}
B_\parallel = -B_0\, {\rm e}^{-r/r_B}\, \cos(\theta - p)\, \sin\,i. \label{eq_bparallel_app}
\end{equation}

The probability distribution function, $f_Y(y)$, of function of a random
variable $X$, where $Y = f(X)$ and $X = f^{-1}(Y)$ is single valued, is given
by \citep{svesh68},
\begin{equation}
f_Y(y) = f_X[f^{-1}(y)] \left|\frac{d f^{-1}(y)}{d y}\right|.
\end{equation}
Here, $f^{-1}$ denotes the inverse function.

Thus, we compute the PDF of $Y = \cos (\Theta - p)$, i.e., $f_Y(y)$, as:
\begin{equation}
f_Y(y) =
\begin{cases}
~\dfrac{1}{\upi \sqrt{1 - y^2}}, & \text{for } -1 \leq y \leq 1,\\
~0, & \text{otherwise.}\\
\end{cases}
\label{eq:pdf_cos}
\end{equation}
Interestingly, the PDF of $\cos(\theta - p)$ is independent of the value of the
pitch angle $p$ because of the periodicity of the function in $0 \le \theta <
2\upi$. In Fig.~\ref{distr} (left-hand panel), we show the distribution of the
functions $\cos(\theta)$ and $\cos(\theta - p)$ with constant $p$ and $\cos(\theta
- p_{\rm rand})$ with uniform distribution of $p_{\rm rand}$ from $-40^\circ$ to $-5^\circ$.
Clearly, as the PDF is independent of $p$, all the
distributions can be represented by a single analytical form.

\begin{figure}
\begin{tabular}{c}
\end{tabular}
{\mbox{\includegraphics[width=6cm]{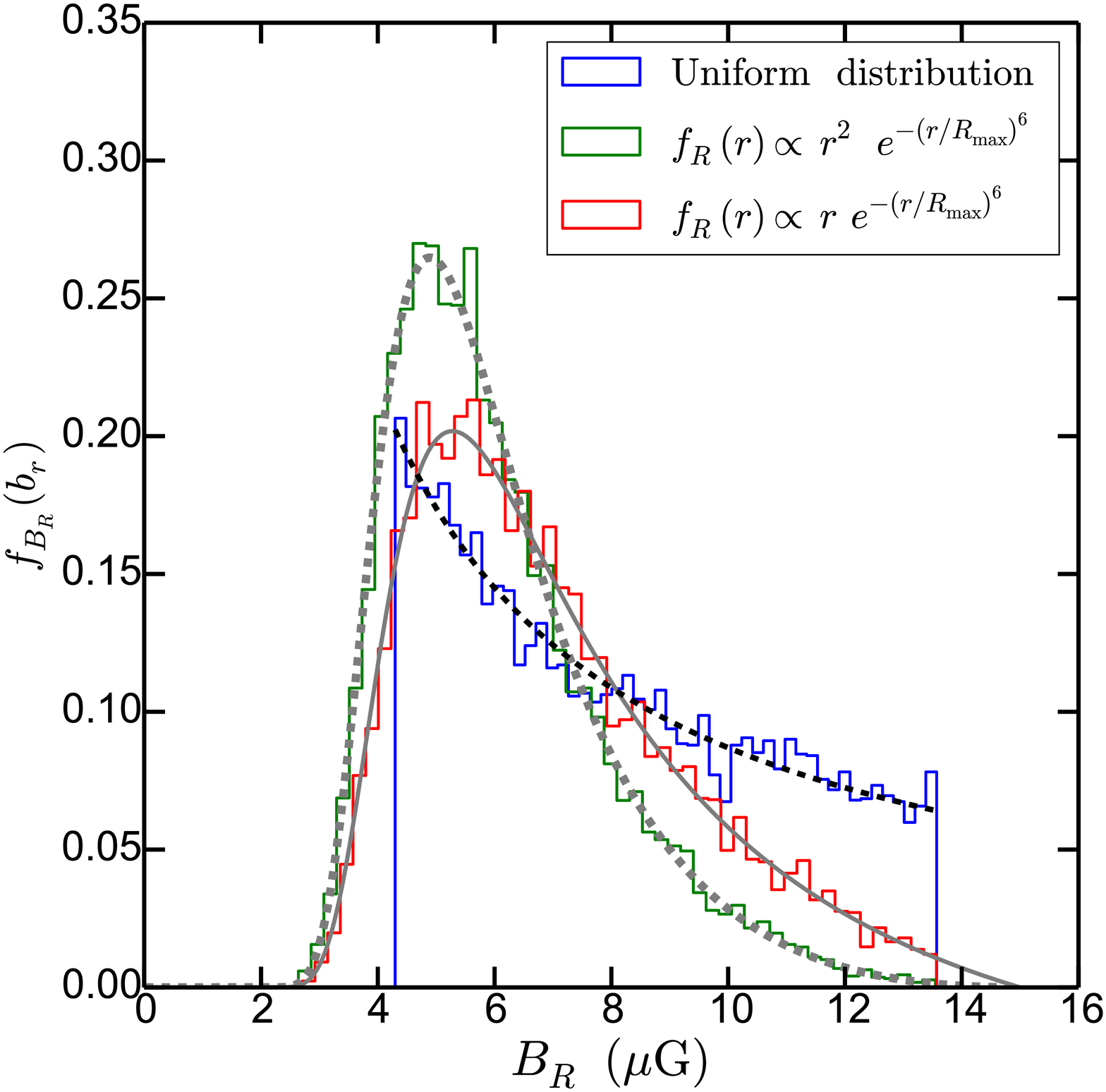}}} \\
\caption{Distribution of the function $B_R = B_0\, {\rm e}^{-r/r_B}$ for
a uniform distribution of $r$ given by $U[R_{\rm min}, R_{\rm max}]$ (blue
histogram) and for the case when $r$ is distributed as per Eq.~\eqref{r_pdf2}
with $\eta=1$ (red histogram) and $\eta=2$ (green histogram). The corresponding
curves are the analytical PDFs given in Eqs.~\eqref{eq:pdf_r} and
\eqref{eq:pdf_r1}.} 
\label{theta_segment}
\end{figure}

However, within segments of a galaxy, i.e., 
\begin{equation}
f_\Theta(\theta)=
\begin{cases}
~\dfrac{1}{\theta_{\rm max} - \theta_{\rm min}}, & \text{for } \theta_{\rm min} \leq \theta \leq \theta_{\rm max},
\\
~0, & \text{otherwise,}
\\
\end{cases}
\end{equation}
where $\theta_{\rm max} - \theta_{\rm min} < \upi$ and $0, \upi, 2\upi \not \in
(\theta_{\rm min} - p, \theta_{\rm max} - p)$, the PDF of $\cos(\theta - p)$ is
given by
\begin{equation}
f_Y(y) = 
\begin{cases}
~\left(\dfrac{1}{\theta_{\rm max} - \theta_{\rm min}}\right) \left(\dfrac{1}{\sqrt{1 - y^2}}\right), &~\\
~~~~~~~~~~~y \in \, [\cos(\theta_{\rm min} - p), \, \cos(\theta_{\rm max} - p)], & \\
~0, ~~~~~~\text{otherwise.}& ~
\\
\end{cases}
\label{eq:pdf_cos_seg1}
\end{equation}
In the cases where $0, \upi, 2\upi \in (\theta_{\rm min} - p, \theta_{\rm max} -
p)$, the distribution is given by,
\begin{equation}
f_Y(y) =
\begin{cases}
~\left(\dfrac{1}{\theta_{\rm max} - \theta_{\rm min}}\right) \left(\dfrac{1}{\sqrt{1 - y^2}}\right), &~\\
~~~~~~~~~~~~~~y \in \, [\cos(\theta_{\rm min} - p), \, \cos(\theta_{\rm max} - p)], & ~\\
~\left(\dfrac{2}{\theta_{\rm max} - \theta_{\rm min}}\right) \left(\dfrac{1}{\sqrt{1 - y^2}}\right)\,\Pi_{\cos(\theta_{\rm max} - p), (+,-)1}(y), &~\\
~~~~~~~~~~~~~~~~~~~ y \in [\cos(\theta_{\rm max} - p), (+,-)1],& ~\\
~0, ~~~~~~~~~~~\text{otherwise.}& ~
\\
\end{cases}
\label{eq:pdf_cos_seg2}
\end{equation}
Here, $\Pi_{a,(+,-)1}(y)$ represents the boxcar function in the range
$\cos(\theta_{\rm max} - p)$ and $(+,-)1$ if $2\upi, \upi \in (\theta_{\rm min} -
p, \theta_{\rm max} - p)$. The distribution of $\cos(\theta - p)$ within
segments of $\upi/2$ in azimuthal angles is shown in Fig.~\ref{theta_segment}.
Note that the above analytical description is valid for segments of length
$\upi/2$ with $\theta_{\rm min} = n\times\upi/2$ and $|p| < \upi/4$. For a general
treatment, the segments where $\cos(\theta - p)$ no longer remain single valued
need to be considered appropriately.

\begin{figure*}
\begin{tabular}{ccc}
{\mbox{\includegraphics[width=6cm]{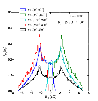}}} &
{\mbox{\includegraphics[width=6cm]{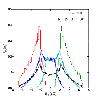}}} &
{\mbox{\includegraphics[width=6cm]{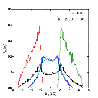}}} \\
\end{tabular}
\caption{Distribution of $B_\parallel$ for a single galaxy which is inclined at
$30^\circ$ and $B_0=15~\umu$G. Different colours are for different different
segments of azimuthal angle $\theta$. For comparison we also show the PDF of
$B_\parallel$ for the entire galaxy as the black histogram. The different
panels are for different values of the pitch angle $p$. It is evident from the
distributions, the PDF of $B_\parallel$ is independent of $p$ for the entire
galaxy. However, within the different segments, the distribution depends on the
pitch angle.}
\label{distr_bparallel_seg}
\end{figure*}

Similarly, the PDF of $B_R = B_0\,{\rm e}^{-r/r_B}$, $f_{B_R}(b_r)$, for
a uniform distribution of $r$ (Case 2 in Section~\ref{sec:rpdf}) is given by,
\begin{equation}
f_{B_R}(b_r)=
\begin{cases}
~\dfrac{r_B}{(R_{\rm max} - R_{\rm min})\,b_r}, &   B_0 {\rm e}^{-R_{\rm max}/r_B} \le b_r \le B_0 {\rm e}^{-R_{\rm min}/r_B}, \\
~0, & \text{otherwise.}
\end{cases}
\label{eq:pdf_r}
\end{equation}
In the case where $r$ is distributed according to Eq.~\eqref{r_pdf2} (Case 1 in
Section~\ref{sec:rpdf}), the PDF of $B_R$ has the form,
\begin{equation} \label{eq:pdf_r1}
\begin{split}
f_{B_R}(b_r)& = \frac{6}{\Gamma\left(\frac{\eta+1}{6}\right)}\,\left(\frac{r_B}{R_{\rm max}}\right)^{\eta+1} \,\frac{1}{b_r}\,
\left[-\ln\left(\frac{b_r}{B_0}\right)\right]^\eta\,{\rm e}^{-\left[\frac{r_B}{R_{\rm max}}\,\ln\left(\frac{b_r}{B_0}\right)\right]^6},\\
 & ~~~~~~~~~~~~~~~~~~~~~~~~~~~~~~~~~~~~~~~~~~0 \,\leq\, b_r\, \leq \, B_0.
\end{split}
\end{equation}
The distributions of $B_0\,{\rm e}^{-r/r_B}$ for the above two cases are shown
in Fig.~\ref{distr} (right panel) for $B_0 = 15~\umu$G, $r_B=20$ kpc, and
$R_{\rm min}$ and $R_{\rm max}$ equal to 2 and 25 kpc, respectively.

\subsection{Distribution function of $B_\|$ and RM for a single galaxy} \label{appendix_RM}

The probability density of the product of two continuous random variables
($V=XY$) is given by \citep{rohat76, glen04},
\begin{equation}
f_V(v) = \int_{-\infty}^{+\infty} f_{X,Y}\left(x, v/x\right)\frac{1}{|x|}dx.
\label{eq:pdf_2func}
\end{equation}
Here, $f_{X,Y}(x,y)$ is the joint PDF of the continuous variables $X$ and $Y$.
Applying the above relation to Eq.~\eqref{eq_bparallel_app}, we compute the PDF of
$B_\parallel$, $f_{B_\parallel}(b_\parallel)$, for a galaxy as a function of
$B_0$, $i$ and $r_B$ as,
\begin{equation}
f_{B_\parallel}(b_\parallel) = 
\begin{cases}
~\dfrac{k_1}{|b_\parallel|}\left[ \arcsin \left(\dfrac{|b_\parallel|}{a} \right)-  \arcsin\left(\dfrac{|b_\parallel|}{b}\right) \right], & -a \le b_\parallel \le a,\\
~\dfrac{k_1}{|b_\parallel|} \, \arccos\left(\dfrac{|b_\parallel|}{b} \right), & b_\parallel \in [-b,-a) \cup (a,b].\\
\end{cases}
\end{equation}
Here, $k_1=r_B/[\upi(R_{\rm max} - R_{\rm min})]$, $a=B_0 \sin i\, {\rm
e}^{-R_{\rm max}/r_B}$ and $b=B_0 \sin i\, {\rm e}^{-R_{\rm min}/r_B}$. The
distribution of $B_\parallel$ is shown in the right panel of
Fig.~\ref{distr_bparallel}.

As discussed above, the distribution function is independent of the pitch angle
because of the periodicity of $\cos(\theta- p)$ in the range from $0$ to
$2\upi$. However, within segments of azimuthal angles in a galaxy, the symmetry
no longer holds true, making the PDF within each segment dependent on the pitch
angle. In Fig.~\ref{distr_bparallel_seg} we show the segment-wise PDF of
$B_\parallel$ for various values of $p$. In this case, the locations of the
characteristic peaks of the PDF (at $\pm a$) are modified to $a\,
\cos(\theta_{\rm min} - p)$ and $a\, \cos(\theta_{\rm max} - p)$ in the
segments $[0, 90^\circ]$ and $[180^\circ, 270^\circ]$, respectively.

Since the RM along each line of sight through a galaxy as given by
Eq.~\eqref{eq_rm} has the same form as that of Eq.~\eqref{eq_bparallel_app},
the analytical form of the PDF of RM can be similarly written, replacing $r_B$
by $r_0^\prime$ and $B_0\, \sin i$ by $0.81\,B_0\, n_0\,h_{\rm ion}\, \tan i$.

\section{Variation of $\rmgal$ distribution} \label{appendix_variation}

\begin{figure*}
\begin{tabular}{ccc}
{\mbox{\includegraphics[width=6cm]{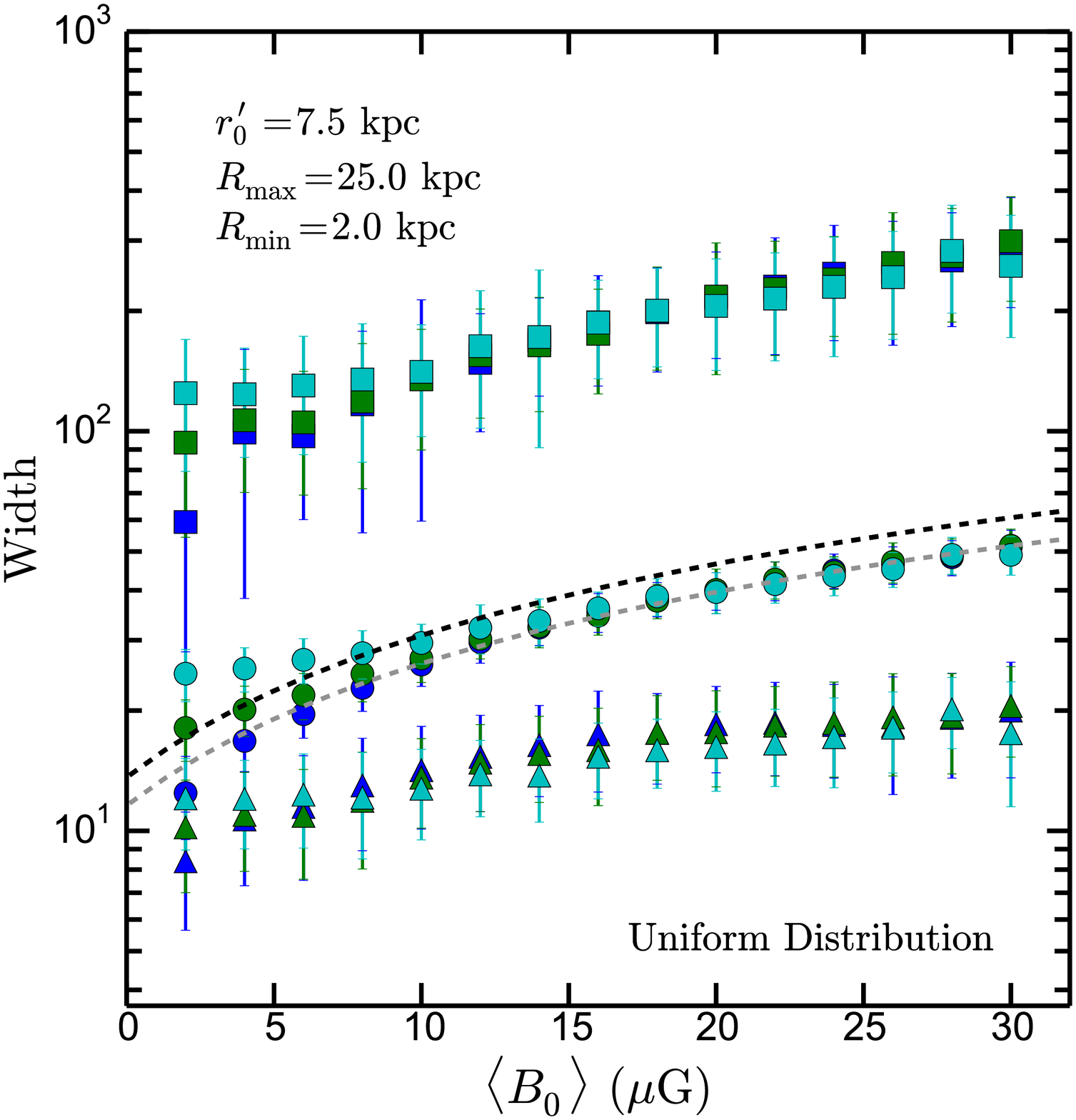}}} &
{\mbox{\includegraphics[width=6cm]{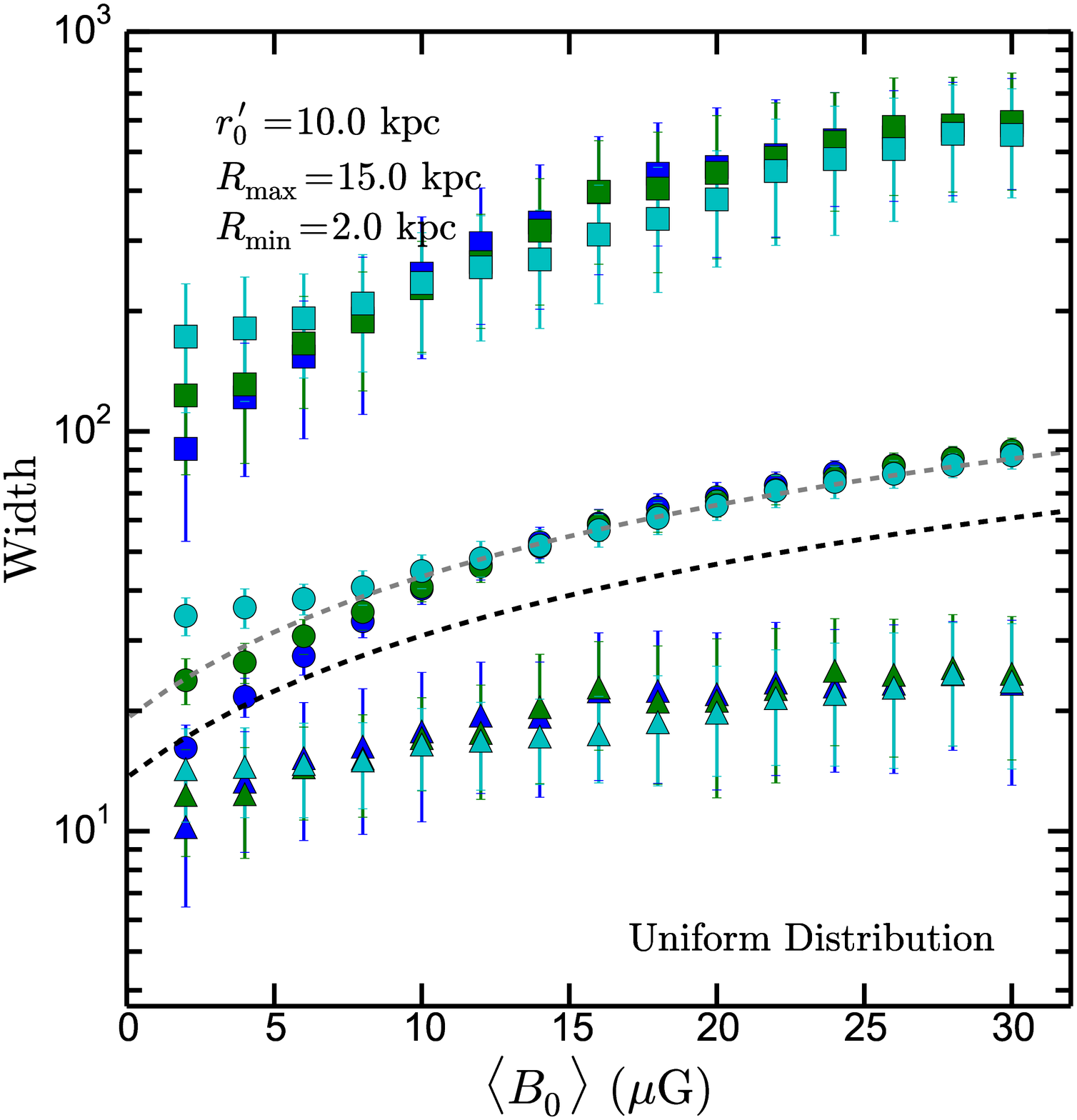}}} &
{\mbox{\includegraphics[width=6cm]{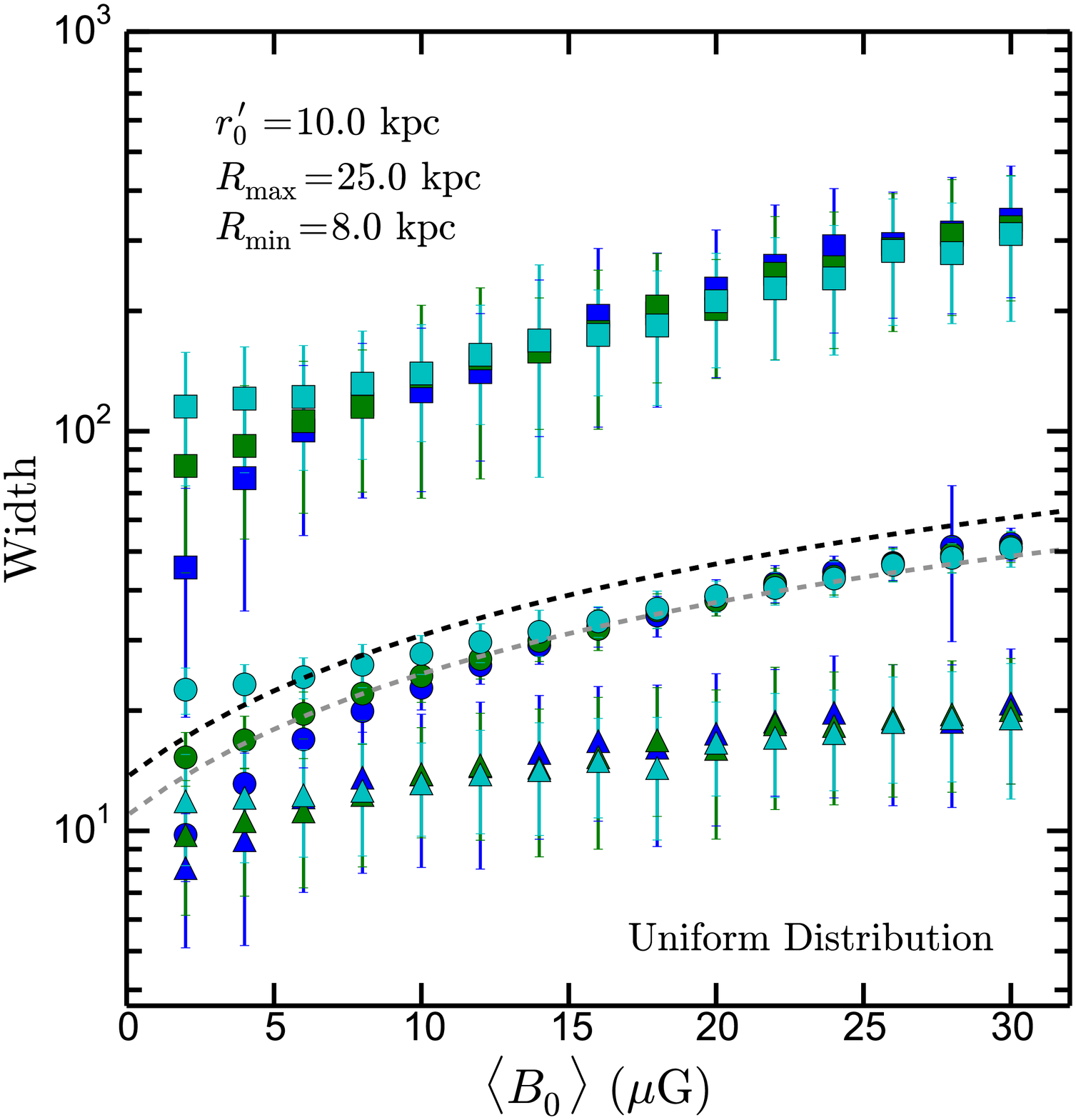}}} \\
{\mbox{\includegraphics[width=6cm]{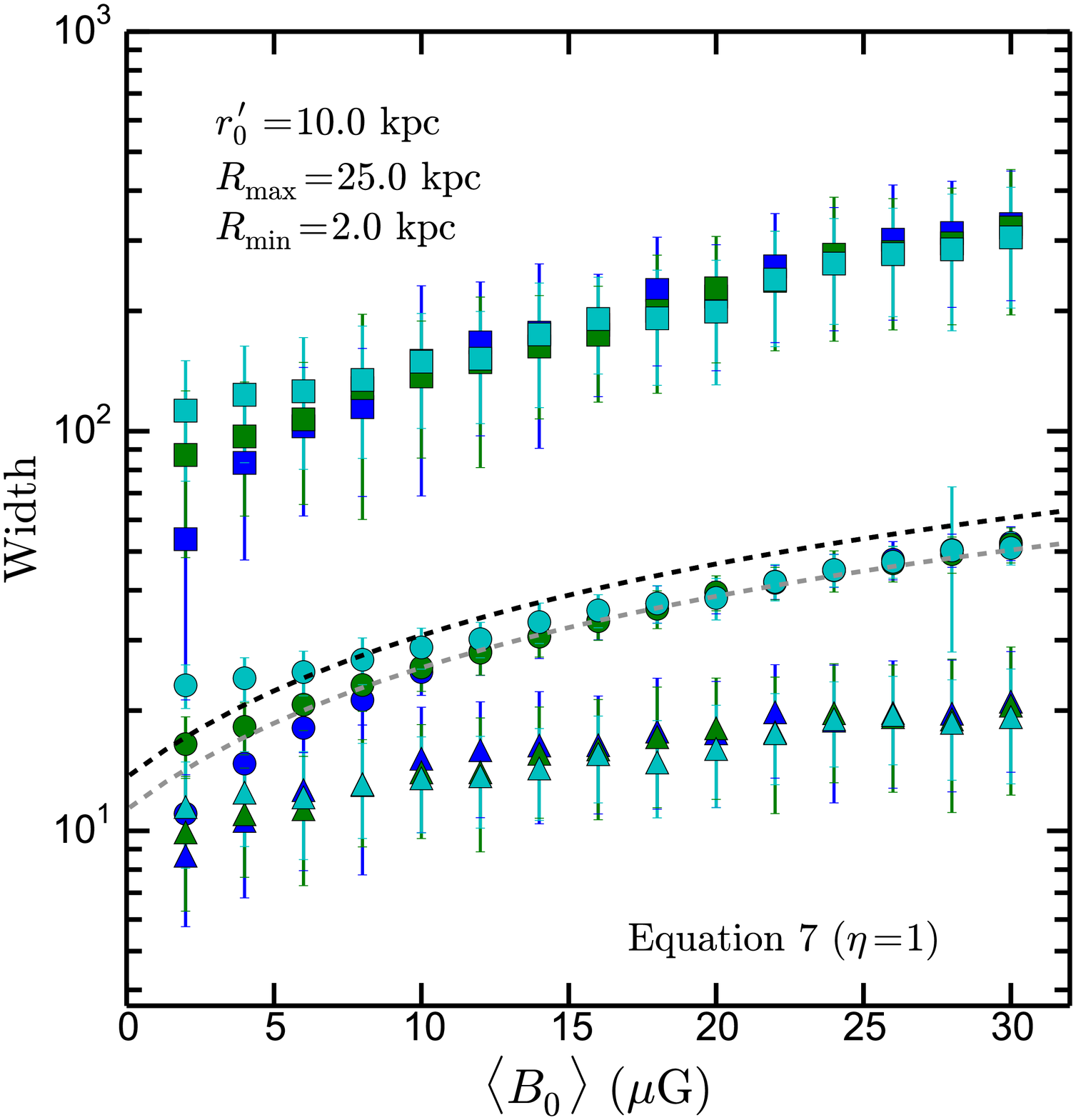}}} &
{\mbox{\includegraphics[width=6cm]{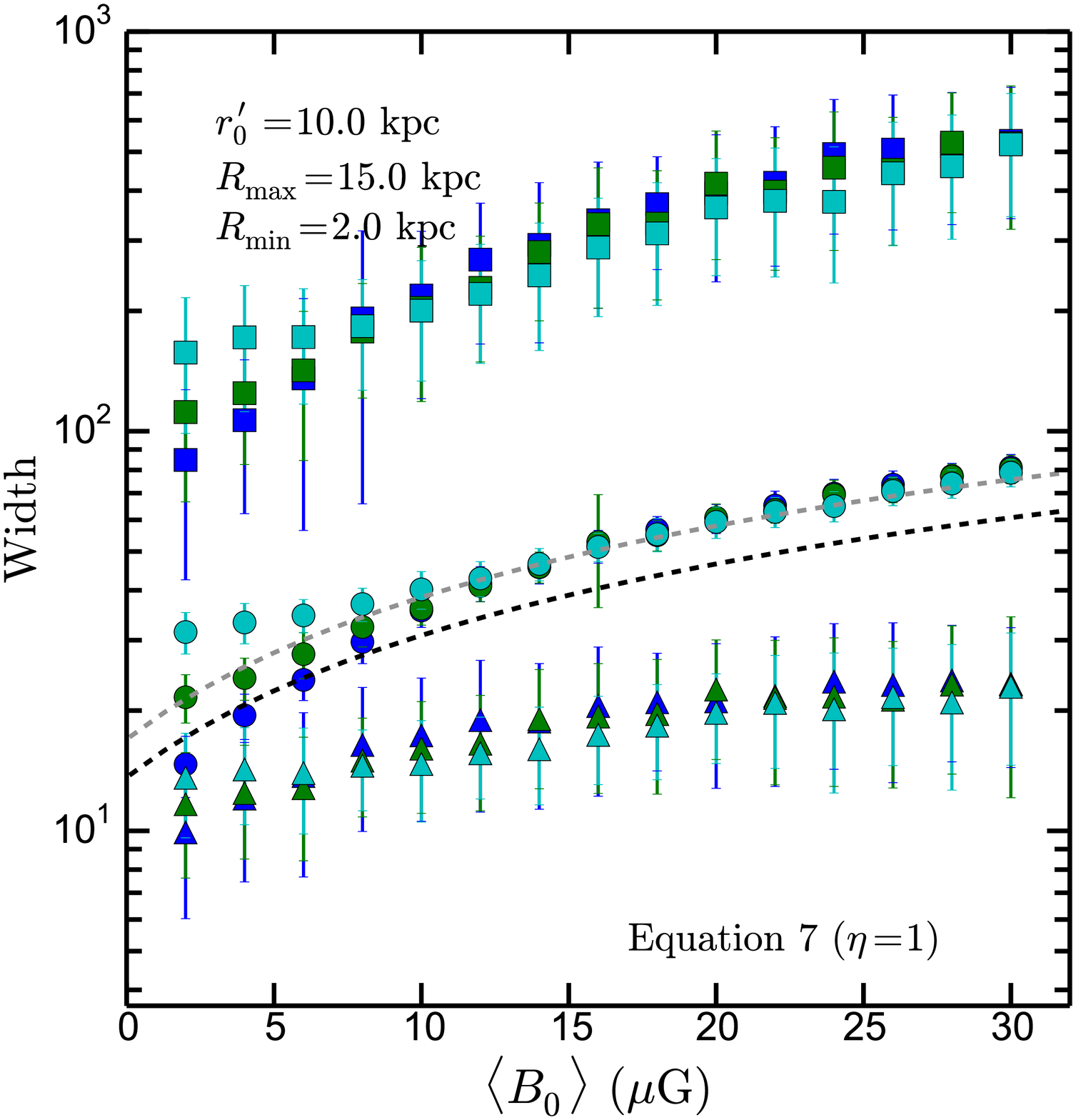}}} &
{\mbox{\includegraphics[width=6cm]{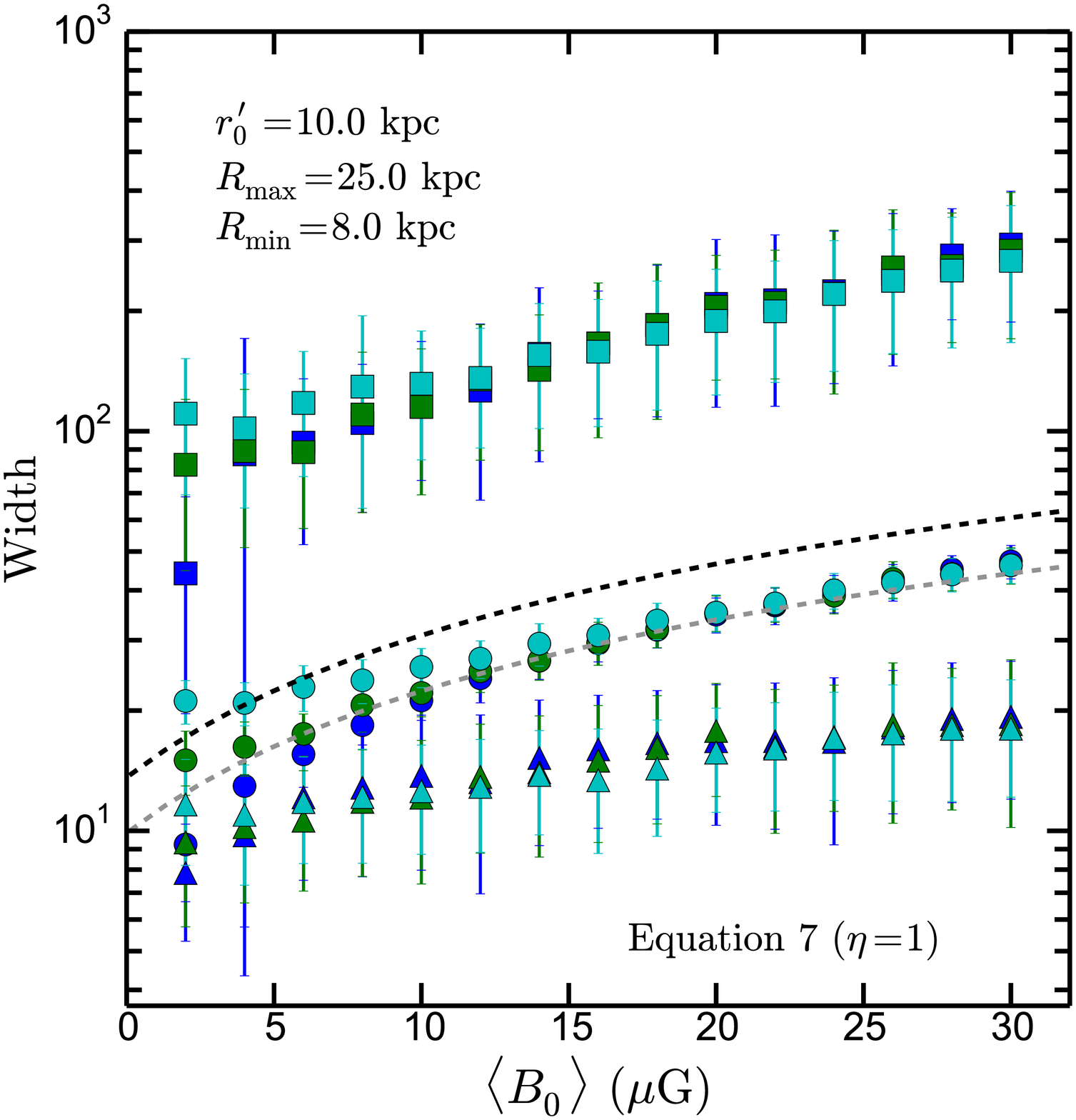}}} \\
\end{tabular}
\caption{Variation of the widths of the modelled components to fit the PDF
of $\rmgal$ with $\langle B_0 \rangle$ for radii distribution as per Case 2
(top panels) and Case 1 with $\eta=1$ (bottom panels). The different columns
are for different choices of the fixed parameters listed in the top left corner
of the plots. The symbols and colours have the same meaning as in
Fig.~\ref{width}. Black dashed curve is the empirical fit given by
Eq.~\eqref{wlor_model} and the grey dashed curve is obtained by scaling the
black curve with factors listed in Table~\ref{tab:correction}.}
\label{width_appendix}
\end{figure*}

\begin{table} \centering 
 \caption{Scaling factor ($a_{\rm s}$) to be applied to Eq.~\eqref{wlor_model}
for different choices of fixed parameters.} 
  \begin{tabular}{@{}lcccc@{}} 
 \hline 
Radial distribution & $R_{\rm min}$, $R_{\rm max}$  & $r_0^\prime$ & $r_B$, $r_{\rm e}$  & $a_{\rm s}$\\
                    & (kpc) & (kpc) &  (kpc) & \\
\hline
Uniform          & 2, 25 & 12.5  & 25, 25 & 1.18 \\
(Case 2)         &  & 10 & 20, 20 & 1.00 \\
                 &  & 7.5  & 15, 15 & 0.85 \\
                 &  & 7.5  & 20, 12 & 0.85 \\
                 &  & 5  & 20, 6.67 & 0.69 \\
                 &  & 5  & 10, 10 & 0.69 \\
                 &  & 3.75  & 15, 5 & 0.64 \\
                 & 0, 25 & 10  & 20, 20 & 1.12  \\
                 & 2, 35 & 10  & 20, 20 & 0.86  \\
                 & 2, 45 & 10  & 20, 20 & 0.78  \\
                 & 8, 25 & 10  & 20, 20  & 0.78 \\
                 & 2, 15 & 10  & 20, 20  & 1.34 \\
                 & 8, 35 & 10  & 20, 20  & 0.64 \\
                 & 8, 35 & 7.5  & 15, 15  & 0.51 \\
                 &  &   &   &  \\
Eq.~\eqref{r_pdf2}; $\eta=1$ & 2, 25 & 10  & 20, 20 & 0.80 \\
(Case 1)         &  & 7.5  & 15, 15 & 0.64 \\
                 &  & 5  & 10, 10 & 0.51 \\
                 & 8, 25  & 10  & 20, 20 & 0.71 \\
                 & 2, 15  & 10  & 20, 20 & 1.21 \\
\hline
\end{tabular}
\label{tab:correction} 
\end{table}

\begin{figure*}
\begin{tabular}{cc}
{\mbox{\includegraphics[height=6cm]{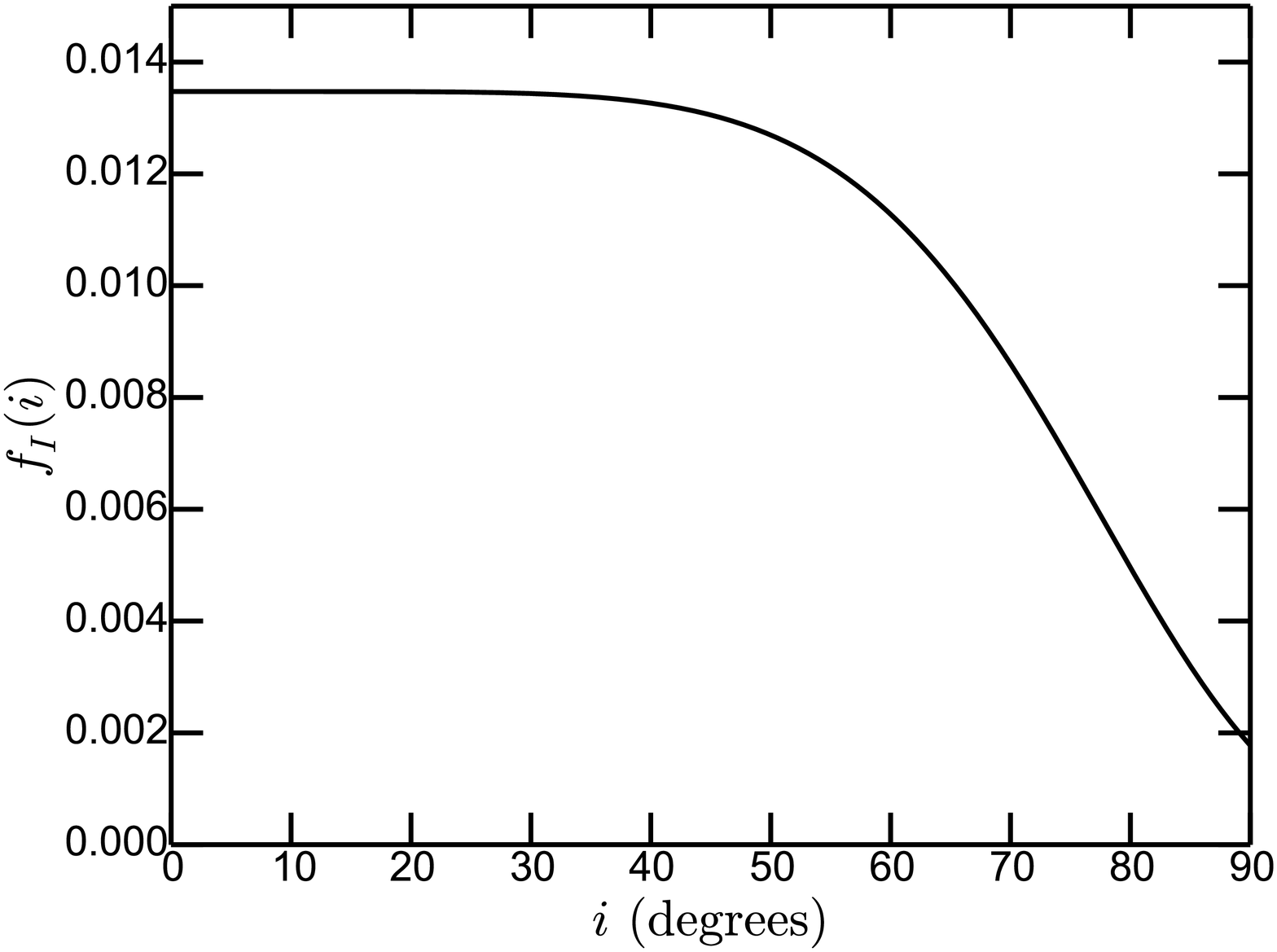}}} &
{\mbox{\includegraphics[width=6cm]{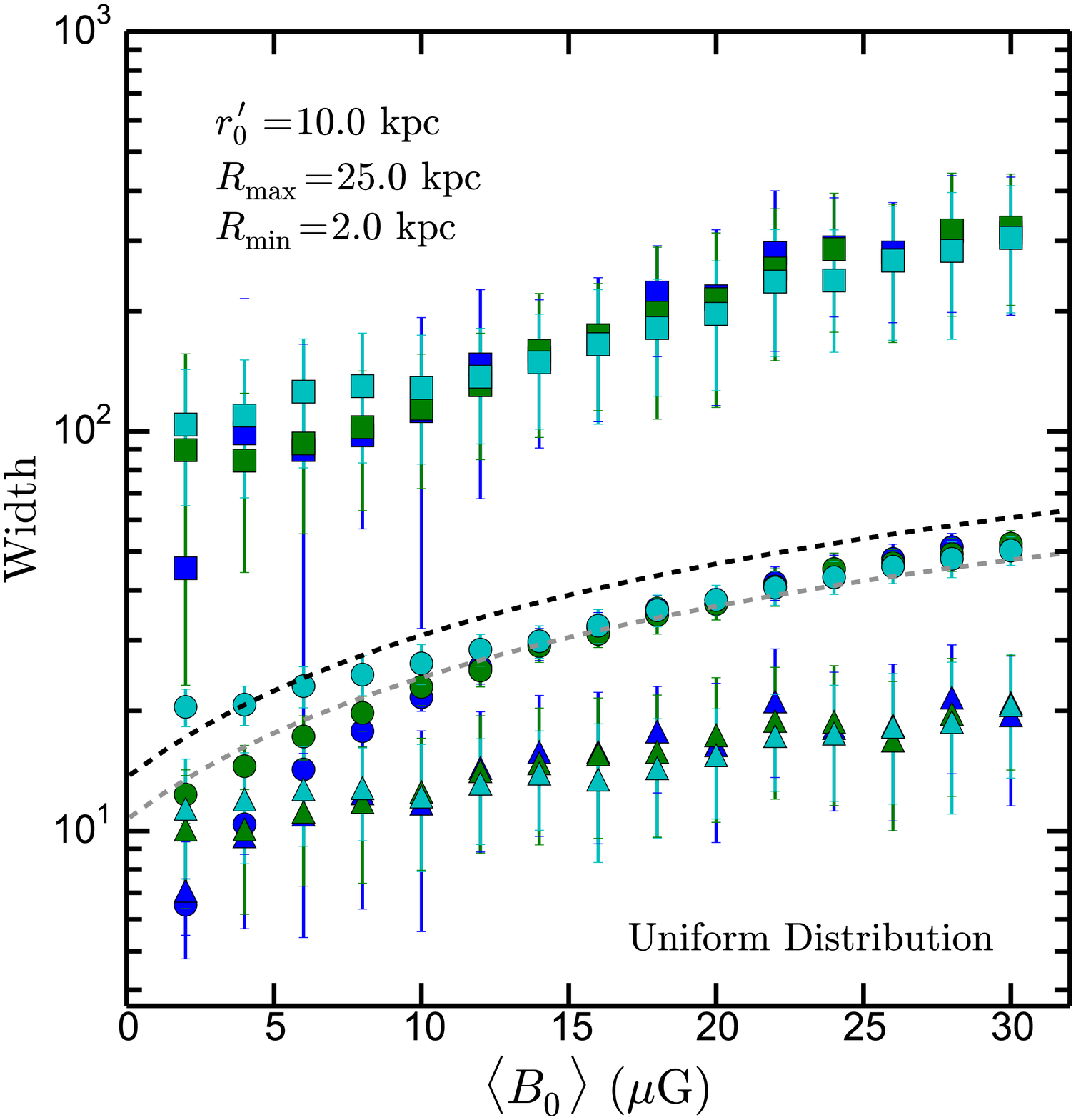}}} \\
\end{tabular}
\caption{{\it Left}: Modified distribution of inclination angle of
galaxies given by Eq.~\eqref{eq:mod_pdf_i}. {\it Right}: Variation of the
widths of the components to model the PDF of $\rmgal$ with $\langle B_0
\rangle$ for uniform distribution of radii and inclination angles distributed
as shown in the left panel. The symbols and colours have the same meaning as in
Fig.~\ref{width}. The black dashed curve is the empirical model given by
Eq.~\eqref{wlor_model} and the grey curve is obtained by scaling the black
curve by a factor 0.78.}
\label{width_incl_appendix}
\end{figure*}

Here, we assess how the empirical modelling describing the variation of
$w_1$ with $\langle B_0 \rangle$ in Eq.~\eqref{wlor_model} changes for
different assumptions for the fixed parameters, such as the distribution of
radii, radial scale-lengths, and inclination angles. 

In Fig.~\ref{width_appendix}, we show the variation of $w_1$, $w_2$ and $w_3$
with $\langle B_0 \rangle$ for different choices of the fixed parameters. For
the top panels, the radii of intersection are distributed uniformly, i.e.,
Case~2 in Section~\ref{sec:rpdf}, while, for the bottom panels, the radii are
distributed as per Case~1 with $\eta=1$. The different columns are for
different choices of the fixed parameters. For all the cases, the PDF of
$\rmgal$ is modelled as a sum of one Lorentzian and two Gaussian functions
described in Section~\ref{sec:RMsample}. We find that $w_1$ is typically
$\sim20$ per cent lower for Case~1 as compared to Case~2, for the same set of
parameters (see the bottom panel of Fig.~\ref{width_appendix}). The black
dashed curve is the empirical model given by Eq.~\eqref{wlor_model} and
Fig.~\ref{width}, i.e., for Case 1 only. We find that the overall dependence of
$w_1$, $w_2$ and $w_3$ on $\langle B_0 \rangle$ does not significantly depend
on the assumed values of the fixed parameters. The empirical model for the
variation of $w_1$ with $\langle B_0 \rangle$ can be made to fit the data by
scaling Eq.~\eqref{wlor_model} as, 
\begin{equation}
w_1\left(\langle B_0 \rangle\right) = a_{\rm s}\,\left(p_0 + p_1\, \langle B_0 \rangle + p_2\,\langle B_0 \rangle\right),
\end{equation}
where $a_{\rm s}$ is a scaling factor whose values for different choices of
the fixed parameters are listed in Table~\ref{tab:correction}. The scaled
empirical models are shown as the grey dashed curves in
Fig.~\ref{width_appendix}.

\subsection{Effect of radial scale length}

The location of the cusps and the width of the PDF of $\rmgal$ for a
single galaxy (Eq.~\eqref{eq:pdf_rm}) depends on the effective radial scale
length, $r_0^\prime$, of the product $n_{\rm e}\,B_\|$ as $\exp(-R_{\rm
min}/r_0^\prime)$ and $\exp(-R_{\rm max}/r_0^\prime)$, respectively. Thus, we
expect the width of the PDF of $\rmgal$ for a sample of galaxies to depend on
$r_0^\prime$ such that $w_1$ should decrease with decreasing $r_0^\prime$.
Such a behaviour can indeed be seen in both Table~\ref{tab:correction} and the
top-left panel of Fig.~\ref{width_appendix}. In fact, as expected, the overall
width of the PDF of $\rmgal$ does not depend on the individual radial scale
lengths of $B_\|$ or $n_{\rm e}$, i.e., $r_B$ or $r_{\rm e}$, respectively. 
The scale factor changes by $\lesssim 40$ per cent within the typical
range of values of $r_0^\prime$. This results in $\langle B_0 \rangle$ being
constrained within $\approx 50$ per cent of its true value, while
estimating it using the set of fixed parameters in Eq.~\eqref{wlor_model}.

\subsection{Effect of $R_{\rm min}$ and $R_{\rm max}$}

$R_{\rm min}$ and $R_{\rm max}$ will have an opposite effect on the width of
the distribution of $\rmgal$ for a sample of galaxies as compared to
$r_0^\prime$, i.e., we expect the widths to decrease with increasing $R_{\rm
min}$ and $R_{\rm max}$. In the middle panels of Fig.~\ref{width_appendix}, we
show the variation of the widths of the fitted components with $\langle B_0
\rangle$ for $R_{\rm max}=15$ kpc. Clearly, the values of $w_1$ in this case
are larger by a factor of $\approx 1.3$ (see Table~\ref{tab:correction}) than
those for the case $R_{\rm max}=25$~kpc, which was used to model
Eq.~\eqref{wlor_model}.

In the right-hand panels of Fig.~\ref{width_appendix} show that $w_1$ decreases
by a factor of $\approx 0.8$ on increasing $R_{\rm min}$ from 2~kpc to 8~kpc.
We note that $R_{\rm min}$ is not expected to vary widely for a given
absorption line species. For example, for the case of DLAs, $R_{\rm min}$ is
determined by the radius within which the H{\sc i} gas is sufficiently ionized
such that $N_{\rm HI}$ drops below the DLA threshold column density. We expect
that such an effect will be dominant for radius $\lesssim2$ kpc \citep[see,
e.g.,][for nearby galaxies]{leroy08}. $R_{\rm max}$, however, can have a larger
range of values depending on the mass and the evolutionary stage of the
galaxies. We find that the width of the PDF of $\rmgal$ changes less than the
other parameters on varying $R_{\rm min}$ and $R_{\rm max}$. For example,
changing $R_{\rm max}$ from 25~kpc to 45~kpc causes $w_1$ to reduce by only
$\approx 20$ per cent. Hence, within the typical range of possible values for
$R_{\rm min}$ and $R_{\rm max}$, $\langle B_0 \rangle$ can be constrained to
within $\approx 20$ per cent.

\subsection{Effect of inclination angle distribution} \label{sec:inclination}

We also tested how the empirical modeling of $w_1$ changes with $\langle B_0
\rangle$ for a different assumption of the distribution of inclination angles.
As pointed out in Section~\ref{rand_distr}, our assumption of a uniform
distribution for the inclination angles of the galaxies is likely to be
inadequate. Because of comparatively larger projected area on the sky for
relatively face-on galaxies, probability of finding a quasar behind them will
be higher than those for highly inclined galaxies.  We therefore model the
distribution of inclination angles based on Eq.~\eqref{r_pdf2} with $\eta = 0$
as,
\begin{equation}
f_I(i) = \dfrac{6}{i_{\rm max}\,\Gamma(1/6)}\,{\rm e}^{-(i/i_{\rm max})^6} \;.
\label{eq:mod_pdf_i}
\end{equation}
The left panel of Fig.~\ref{width_incl_appendix} shows a modified form of the
distribution of inclination angles for $i_{\rm max} = 80^\circ$, such that, low
inclined galaxies are preferred. For such an inclination angle distribution,
$w_1$ is $\approx 20$ per cent lower than that for a uniform distribution.
However, in this case, simple scaling of Eq.~\eqref{wlor_model} shows slight
deviations, unlike the earlier cases, especially for lower values of $\langle
B_0 \rangle$ and high $\sigma_{B_0}$. A more realistic modelling of the
combined effects of the distributions of inclination angles, impact radii and
column densities of the absorbing gas (as pointed out in
Section~\ref{rand_distr}) is required to study how they affect on our results.

In summary, within the typical ranges of values for the parameters
$r_0^\prime$, $R_{\rm min}$ and $R_{\rm max}$, $\langle B_0 \rangle$ can be
constrained to within $\approx 50$ per cent. In other words, the possible
lower- and higher-end of the true value of $\langle B_0 \rangle$ would be
within $\approx 50$ per cent of $\langle B_0 \rangle$ estimated using
Eq.~\eqref{wlor_model}. To estimate $\langle B_0 \rangle$ to a better
accuracy, additional information on these parameters are necessary.

\section{List of variables}

\begin{table*} \centering 
 \caption{Notation and definitions of variables used in the text.} 
  \begin{tabular}{@{}lcl@{}} 
 \hline 
Variable & Definition/Typical values  & Description\\
\hline
$B_r$ & $\dotsb$  & Radial component of the large-scale field \\
$B_\theta$ & $\dotsb$  & Azimuthal component of the large-scale field \\
$B_z$ & $\dotsb$  & Magnetic field component perpendicular to the disc \\
$i$ & 0--90$^\circ$ (30$^\circ$)  & Inclination angle ($i = 0^\circ$ is face-on) \\
$\theta$ & 0--360$^\circ$  & Azimuthal angle \\ 
$p$ & $\arctan(B_r/B_\theta) $ & Pitch angle of the axisymmetric spiral disc field \\
$B_0$ & $2\text{--}30\,\umu$G ($15\,\umu$G) & Coherent field strength at the center of the galaxy \\
$r_B$ & $\sim 15\text{--}25$ kpc (20 kpc) & Radial scale-length of the large-scale field \\
$R_{\rm min}$ & ($\sim2$ kpc) & Minimum impact parameter  \\
$R_{\rm max}$ & ($\sim 25$ kpc) & Maximum impact parameter \\
$B(r)$ & $\sqrt{B_r(r)^2 + B_\theta(r)^2} = B_0\,{\rm e}^{-r/r_B}$ & Radial variation of the large-scale field \\
$B_\parallel$ & $-(B_r \sin \theta + B_\theta \cos \theta) \sin i + B_z \cos i$ & Magnetic field component along the line of sight \\
              & $\equiv -B_0\, {\rm e}^{-r/r_B}\,\cos (\theta - p) \sin i$ & Assuming $B_z$ is negligible\\
$\langle B_\|\rangle$ & $\dotsb$ & Average $B_\|$ along the line of sight\\
$b_\|$ & $\dotsb$ & rms of turbulent magnetic fields along the line of sight\\
\hline
\hline
$n_{\rm 0}$ & $0.01\text{--}0.05$\,cm$^{-3}$ (0.03\,cm$^{-3}$) & Free electron density at the center of the galaxy \\
$\langle n_{\rm e}\rangle$ & $\dotsb$ & Average free electron density along the line of sight \\
$r_{\rm e}$ & $\sim 15\text{--}20$ kpc (20 kpc) & Radial scale-length of free electron density in the thick ionized medium\\
$h_{\rm ion}$ & ($\sim 500$ pc)  & Thickness of the ionized medium \\
 \hline 
 \hline 
$\rm RM$  & $0.81\, \langle n_{\rm e}\rangle\,\langle B_\parallel\rangle\, h_{\rm ion}$  & Rotation Measure \\
$\rmgal$ & $-0.81\, n_0\, B_0\, {\rm e}^{-r/r_0^\prime}\,\cos (\theta - p) \tan i\, h_{\rm ion}$  & RM of galaxy or an absorber species for a single line-of-sight \\ 
$\rmqso$ & $10\text{--}50$ rad m$^{-2}$  & Intrinsic RM of quasar in a target sample \\
$\rm RM_{qso, c}$ & $\dotsb$  & Intrinsic RM of quasar in a control sample \\
$\zgal$ & $\dotsb$  & Redshift of the absorber galaxy \\
$\zqso$ &  $\dotsb$ & Redshift of the quasar in the target sample \\
$z_{\rm qso, c}$ & $\dotsb$  & Redshift of the quasar in the control sample \\
$\Delta_{\rm RM}$ & $\rm RM_{MW} + RM_{IGM} + \delta_{\rm RM}$ & Net RM contributed along the path \\
$\rmt$ & $\dfrac{\rmgal}{(1 + \zgal)^2} + \dfrac{\rmqso}{(1 + \zqso)^2} + \Delta_{\rm RM}$  & Total RM along a target sightline\\
$\rm RM_{qso}^\prime$ & $\dfrac{\rmqso}{(1 + \zqso)^2} + \Delta_{\rm RM}$  & Total RM along a target sightline except the contribution of the absorber\\
$\rmc$ & $\rm \dfrac{RM_{qso, c}}{(1 + z_{\rm qso, c})^2} + \Delta_{RM}$ & Total RM along a control sightline  \\
\hline
\hline
$\sigma_{\rm gal}$ & $\dotsb$  & Dispersion of $\rmgal$ in the target sample \\
$\sigma_{\rm qso}$ & ($\sim 50\,\umu$G)  & Dispersion of $\rm RM_{qso}^\prime$ in the target sample and $\rmc$ in the control sample\\
$\sigma_{\rm t}$ & $\dotsb$  & Dispersion of $\rmt$ of the target sample \\
$\langle B_0 \rangle$ & $1\text{--}30\,\umu$G ($15\,\umu$G) & Mean $B_0$ for a sample of galaxies\\
$\sigma_{B_0}$    & $1\text{--}10\,\umu$G ($5\,\umu$G) & Dispersion of $B_0$ for a sample of galaxies \\
\hline
\end{tabular}
\begin{flushleft}
The values in parenthesis represent the fixed values for the parameters used to generate the 
plots in the text unless specified otherwise.
\end{flushleft}
\label{variables} 
\end{table*}

\bsp

\label{lastpage}

\end{document}